\documentclass[
10pt,
]{achemso}

\usepackage{graphicx}
\usepackage{amsmath, amsthm, amssymb, amsfonts}

\usepackage{microtype}
\usepackage{booktabs}
\usepackage{soul}
\usepackage[para]{threeparttable}
 \usepackage{lineno}
\usepackage[english]{babel}
\usepackage[utf8]{inputenc}
\usepackage[pdftex]{hyperref, color}
\usepackage[figure, figure*]{hypcap}

\usepackage{upgreek}
\usepackage{lmodern}
\usepackage{wrapfig}
\usepackage{dcolumn}
\usepackage{bm}

\usepackage{bbold}
\usepackage[T1]{fontenc}
\usepackage{color,bm, braket}
\usepackage{siunitx}
\usepackage[skip=0pt plus1pt, indent=20pt]{parskip}
\usepackage{gensymb}

\usepackage[table]{xcolor}
\renewcommand{\arraystretch}{1}
\newcolumntype{s}{ p{2.5cm}}
\arrayrulecolor[HTML]{DB5800}

\usepackage{xcolor}

\newcommand {\va}{VS$_2$}
\newcommand {\ve}{VSe$_2$}
\newcommand {\vx}{VS$_{x}$}
\newcommand {\vai}{V$_{5}$S$_{8}$}
\newcommand {\vas}{V$_{4}$S$_{7}$}
\newcommand {\ro}{($\sqrt{3}\times\sqrt{3})R30\degree$}

\usepackage{caption}
\DeclareCaptionFont{12pt}{\fontsize{12pt}{10pt}\selectfont}
	\DeclareGraphicsExtensions{.pdf}

\author{Camiel van Efferen}
\email{efferen@ph2.uni-koeln.de}
\affiliation{II. Physikalisches Institut, Universit\"{a}t zu K\"{o}ln, Z\"{u}lpicher Stra\ss e 77, 50937 K\"{o}ln, Germany}

\author{Joshua Hall}
\affiliation{II. Physikalisches Institut, Universit\"{a}t zu K\"{o}ln, Z\"{u}lpicher Stra\ss e 77, 50937 K\"{o}ln, Germany}

\author{Nicolae Atodiresei} 
\affiliation{Peter Grünberg Institut and Institute for Advanced Simulation, Forschungszentrum Jülich, Wilhelm-Johnen Stra\ss e, 52428 J\"{u}lich, Germany}

\author{Virg\'inia Boix}
\affiliation{ Division of Synchrotron Radiation Research, Department of Physics, Lund University, P.O. Box 118, SE-221 00 Lund, Sweden}

\author{Affan Safeer}
\affiliation{II. Physikalisches Institut, Universit\"{a}t zu K\"{o}ln, Z\"{u}lpicher Stra\ss e 77, 50937 K\"{o}ln, Germany}

\author{Tobias Wekking}
\affiliation{II. Physikalisches Institut, Universit\"{a}t zu K\"{o}ln, Z\"{u}lpicher Stra\ss e 77, 50937 K\"{o}ln, Germany}

\author{Nikolay Vinogradov}
\affiliation{MAX IV Laboratory, Lund University, P.O Box 118, SE-221 00 Lund, Sweden}

\author{Alexei Preobrajenski}
\affiliation{MAX IV Laboratory, Lund University, P.O Box 118, SE-221 00 Lund, Sweden}

\author{Jan Knudsen}
\affiliation{ Division of Synchrotron Radiation Research, Department of Physics, Lund University, P.O. Box 118, SE-221 00 Lund, Sweden}
\affiliation{MAX IV Laboratory, Lund University, P.O Box 118, SE-221 00 Lund, Sweden}

\author{Jeison Fischer}
\affiliation{II. Physikalisches Institut, Universit\"{a}t zu K\"{o}ln, Z\"{u}lpicher Stra\ss e 77, 50937 K\"{o}ln, Germany}

\author{Wouter Jolie} 
\affiliation{II. Physikalisches Institut, Universit\"{a}t zu K\"{o}ln, Z\"{u}lpicher Stra\ss e 77, 50937 K\"{o}ln, Germany}

\author{Thomas Michely} 
\affiliation{II. Physikalisches Institut, Universit\"{a}t zu K\"{o}ln, Z\"{u}lpicher Stra\ss e 77, 50937 K\"{o}ln, Germany}

\title{Novel 2D vanadium sulphides: synthesis, atomic structure engineering and charge density waves}


\begin{document}

\begin{abstract}
Two new ultimately thin vanadium rich 2D materials based on \va{} are created via molecular beam epitaxy and investigated using scanning tunneling microscopy, X-ray photoemission spectroscopy and density-functional theory calculations. The controlled synthesis of stoichiometric single-layer \va{} or either of the two vanadium-rich materials is achieved by varying the sample coverage and the sulphur pressure during annealing. Through annealing of small stoichiometric single-layer \va{} islands without S pressure, S-vacancies spontaneously order in 1D arrays, giving rise to patterned adsorption. Via the comparison of density-functional theory calculations with scanning tunneling microscopy data, the atomic structure of the S-depleted phase, with a stoichiometry of \vas, is determined. By depositing larger amounts of vanadium and sulphur, which are subsequently annealed in a S-rich atmosphere, self-intercalated ultimately thin \vai{}-derived layers are obtained, which host $2 \times 2$ V-layers between sheets of \va{}. We provide atomic models for the thinnest \vai{}-derived structures. Finally, we use scanning tunneling spectroscopy to investigate the charge density wave observed in the 2D \vai{}-derived islands. 
\end{abstract}

\maketitle

Atomic structure engineering of two-dimensional (2D) materials in order to tailor their electronic and chemical properties or to create novel phases has been on the forefront of recent research. Various methods to achieve this have been explored, like the introduction of point or line defects~\cite{Ma2014, Lin2015, Coelho2018}, creating horizontal or vertical heterostructures~\cite{Huang2014, Vano2021}, doping or gating~\cite{Lin2014a, Sutter2016, Wang2017, VanEfferen2022}, as well as intercalation of native or foreign atoms between the layers~\cite{Friend1987, Wan2016}. 

The creation of defects has been particularly successful in engineering transition metal dichalcogenides (TMDCs)~\cite{Hu2018}. Chalcogen vacancies alone can be used to dope TMDCs~\cite{Komsa2012a}, to develop or enhance a magnetic ground state~\cite{Yu2019, Chua2020} or to gain increased surface reactivity~\cite{Liu2019, Wang2020}. Vacancies can be created by extrinsic means like electron-beam irradiation~\cite{Komsa2012a, Lin2015a}, but can also spontaneously form under suitable conditions through thermal annealing~\cite{Liu2019, Chua2020}. 

Since TMDCs are layered materials, they are also easily intercalated, with additional metal atoms placed between the TMDC sheets~\cite{Wan2016}. Intercalated atomic layers have been used to enhance electronic conductivity, induce ferromagnetism, phase transitions or Ising superconductivity in TMDCs~\cite{Liu2023, Zhao2020, Wang2014, Tan2017, Zhang2022}. While these intercalants have been used extensively to intercalate bulk TMDCs~\cite{Friend1987, Wang2014, Tan2017, Liu2023}, recently the focus has shifted to bilayer TMDC intercalation, being the thinnest intercalated material~\cite{Kanetani2012, Lasek2020, Zhao2020}. 

Among TMDCs, vanadium based compounds like \va{} have attracted substantial theoretical research interest due to their predicted electronically correlated and magnetic ground states when thinned down to a single layer~\cite{Ma2012, Zhang2013, Isaacs2016, Zhuang2016}. However, due to the lack of a stable bulk polymorph~\cite{Mulazzi2010, Gauzzi2014}, \va{} is a particularly challenging material to synthesize as a few-layer system and it was a late addition of the single-layer TMDCs when it was synthesized on Au(111)~\cite{Arnold2018} and on quasi-freestanding on Gr/Ir(111), where a $9 \times \sqrt{3}R30\degree$ charge density wave (CDW) was found~\cite{VanEfferen2021}. Owing to the wealth of experiments performed on \va{} and its sister compound \ve{} showing the absence of net magnetic moment in these materials, it is commonly accepted that the theoretically predicted ferromagnetic ground state~\cite{Ma2012, Zhang2013, Isaacs2016, Zhuang2016} is not realized in pristine single-layers of these materials~\cite{Coelho2019, Chua2020, VanEfferen2021}. 

In contrast, the stable compound \vai{} exhibits layer-dependent magnetism, which transforms with decreasing thickness from anti-ferromagnetic to ferromagnetic~\cite{Hardy2016, Niu2017, Zhang2020}. These results have so far been limited to samples down to \SI{3}{\nm} thickness since this compound has not yet been realized in its minimum thickness configuration of two \va{} layers sandwiching a $2\times2$ V intercalation layer. Transport measurements of bulk \vai{} show no increase in the resistivity at low temperatures, suggesting it does not have a CDW~\cite{Nozaki1975, Moutaabbid2016}, though upon cooling down to \SI{100}{\K}, an anisotropic contraction of V-V bonds was observed using X-ray diffraction~\cite{Bensch1993}. Several studies also note that the magnetic moment of the intercalated V atoms can couple to the itinerant electrons in the \va{} sheets, leading to a Kondo effect~\cite{Niu2020, Zhou2022}.

\begin{figure*}
	\centering
		\includegraphics[width=0.9\textwidth]{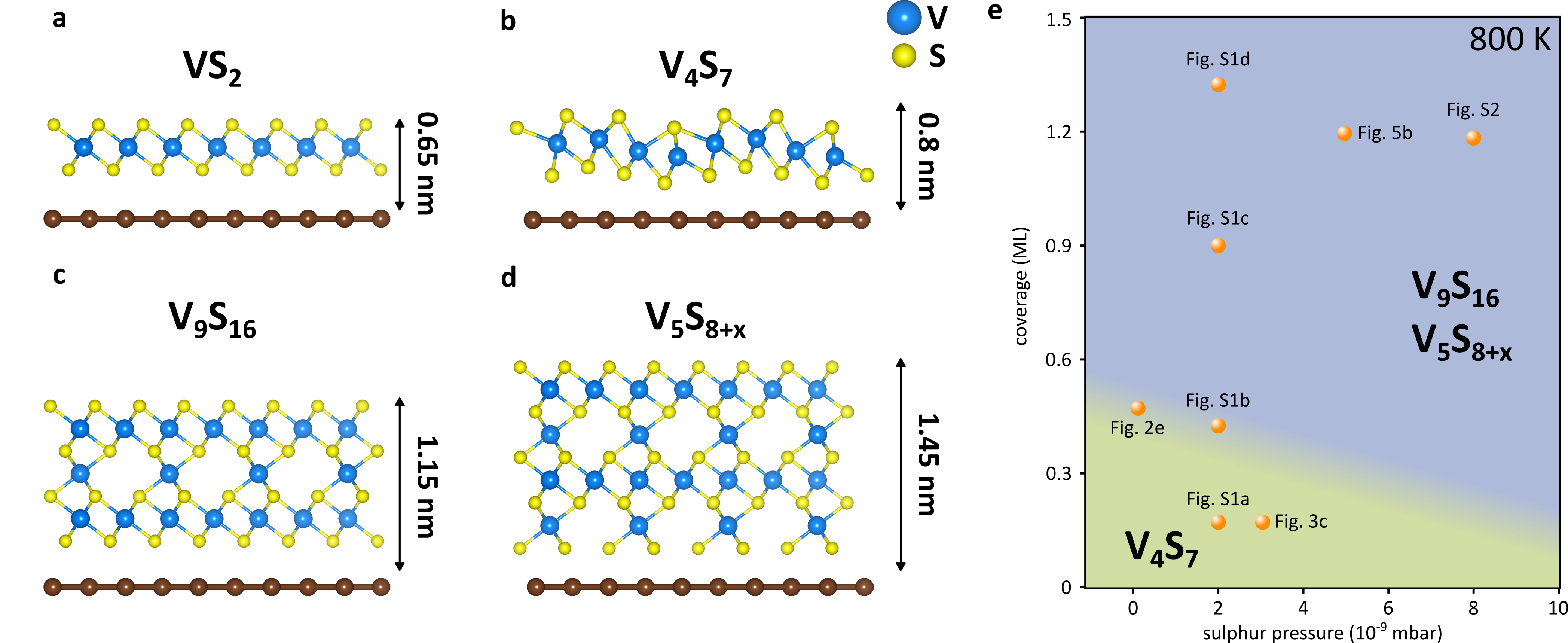}
			\caption{ \textbf{Atomic structures and phase diagram of VS$_x$ compounds. a-d}~Atomic structure models of \va, \vas, V$_{9}$S$_{16}$ and V$_5$S$_{8+x}$, respectively. Below the structures, a Gr substrate layer is drawn; besides the images, the apparent height measured with STM is indicated. The suggested structure of \vas{} is based on DFT calculations. For V$_5$S$_{8+x}$ the actual configuration of the S atoms in the lowest layer is not known. \textbf{e}~Schematic representation of phases present after annealing temperature of \SI{800}{\K}. The figures indicated next to data points represent the corresponding compounds. Fig. S1 and Fig. S2 are in the Supplementary Information. With increasing initial coverage and annealing S pressure, the transition of \va{} into \vas{} is suppressed. Instead, a $2\times2$ self-intercalated material is obtained, with stoichiometry V$_{9}$S$_{16}$ or V$_5$S$_{8+x}$ depending on the number of intercalation layers.}
 \label{fig:Fig0}
\end{figure*}

Here we explore two novel 2D vanadium-sulphide compounds, created using a two-step molecular beam epitaxy process. The synthesis consists of an initial growth step at room temperature and an annealing step at elevated temperature to enhance island shape and alignment. Depending on the amount of deposited material during growth, the temperature, and the S pressure during annealing, three materials of different V:S stoichiometry are obtained. Since V deposition always takes place in large S excess, the deposited amount is characterized through the amount of V deposited - the V atoms stay on the surface, while excess S evaporates. The unit used is monolayer (ML), which characterizes the V amount needed to grow a full single layer of stoichiometric \va, \textit{i.e.} an amount of $1.12 \times 10^{19}$ V atoms per m$^2$. The synthesis is performed in each case on the inert graphene (Gr)/Ir(111) substrate, which preserves the intrinsic properties of the material under investigation~\cite{Hall2018, Murray2019}. 

Single-layer islands of stoichiometric and phase pure \va{} are obtained when less than $0.5$ ML are deposited at room temperature and annealed to a temperature not exceeding $\SI{600}{\K}$~\cite{VanEfferen2021}. Under these conditions, the resulting single-layer \va{} is always present in the 1T phase, as depicted in Fig.~\ref{fig:Fig0}a. When stoichiometric \va{} is exposed to higher annealing temperatures, it gradually transforms into \vas{}, a process that is completed at an annealing temperature of $\approx \SI{800}{\K}$. \vas{} has periodically arranged rows of S vacancies and an accompanying buckling of the lattice, see Fig.~\ref{fig:Fig0}b. However, when the initial coverage during growth is larger, specifically when the islands resulting from the initial growth carry second layer islands of significant size, a different compound is formed upon annealing. It consists of layers of \va{} self-intercalated with V atoms in a $2\times2$ pattern. The bulk stoichiometry of this phase would be \vai, which has a NiAs structure with ordered V vacancies every second layer (leaving a quarter of the V atoms to form the $2\times2$)~\cite{Kawada1975}. The V:S ratio of bulk \vai{} is a limit reached only when the number of layers goes to infinity. We obtain \vai-derived islands of minimal thickness, consisting of two layers of \va{} with a $2\times2$ layer of V atoms intercalated in the van der Waals (vdWs) gap between the layer. Consequently the stoichiometry is V$_{9}$S$_{16}$, see Fig.~\ref{fig:Fig0}c. We also find evidence for the presence of a doubly intercalated structure. In that case, a second $2\times2$ layer of V, passivated by S atoms, intercalated in the vdWs gap between the bottom \va{} sheet and Gr, as depicted in Fig.~\ref{fig:Fig0}d. We denote this structure as V$_5$S$_{8+x}$, since the precise amount of S saturating the V is unknown. To show the different growth regimes in S pressure and coverage needed to obtain either the vacancy row structure \vas{} or the 2D derivatives of \vai{}, a schematic representation is shown in Fig.~\ref{fig:Fig0}e, wherein the structures obtained after room temperature growth and subsequent annealing at a temperature of \SI{800}{\K} are collected. 

In this study, the materials shown schematically in Fig~\ref{fig:Fig0} are investigated using scanning tunneling microscopy (STM), X-ray photoemission spectroscopy (XPS) and density-functional theory (DFT) calculations. Furthermore, using low-temperature scanning tunneling spectroscopy (STS), a \ro{} CDW is found in the 2D \vai{}-derived islands, with a transition temperature below \SI{110}{\K}. An overview of our results is given in Table~\ref{Table1}.

\arrayrulecolor[HTML]{eaa51f}
\renewcommand{\arraystretch}{2}
\begin{table}
\centering
\fontsize{10pt}{10pt}\selectfont
\begin{tabular}{|p{2.75cm}|p{2.75cm}|p{2.75cm}|p{2.75cm}|p{2.75cm}| }
\hline
\rowcolor[HTML]{eaa51f} \multicolumn{5}{|c|}{Growth parameters and properties of VS$_x$ compounds} \\
\hline
\text{stoichiometry}&VS$_2$& V$_4$S$_7$ & V$_9$S$_{16}$ & V$_5$S$_{8+x}$ \\
\hline
\text{coverage (ML)} & $< 0.5$ & $< 0.5$ & \multicolumn{2}{c|}{$> 0.5$} \\
\hline
annealing temperature (K) & $\le 600$ & $\approx 800$ & \multicolumn{2}{c|}{$\approx 800$} \\
\hline
annealing S pressure (mbar) &$0-8 \times 10^{-9}$ & $0-8 \times 10^{-9}$ & \multicolumn{2}{c|}{$5 \times 10^{-9}$} \\
\hline
\text{purity}  & \text{phase pure} & \text{phase pure} & not phase pure, mixed with other \vai-derived structures & not phase pure, dominant with excess V\\
\hline
\text{CDW, $T_c$} & $9\times\sqrt{3}R30\degree$, $> 300$K & none, none & \multicolumn{2}{c|}{\ro, $\approx 110$K} \\
\hline
\text{structure} & T-phase TMDC & S-vacancy rows & \multicolumn{2}{c|}{$2\times2$ - intercalation, \vai-derived} \\
\hline
apparent height (nm) & 0.65 & 0.8 & 1.15 & 1.45 \\
\hline
\text{figure} & 2d & 2f & 5b & 6d \\
\hline
\end{tabular}
\caption{\textbf{Overview of VS$_x$ compounds and their synthesis parameters.}
\label{Table1}}
\end{table}

\section{Results}
 \subsection{Creating 1D-patterned \vas{} from single-layer \va}

\begin{figure*}
	\centering
		\includegraphics[width=0.9\textwidth]{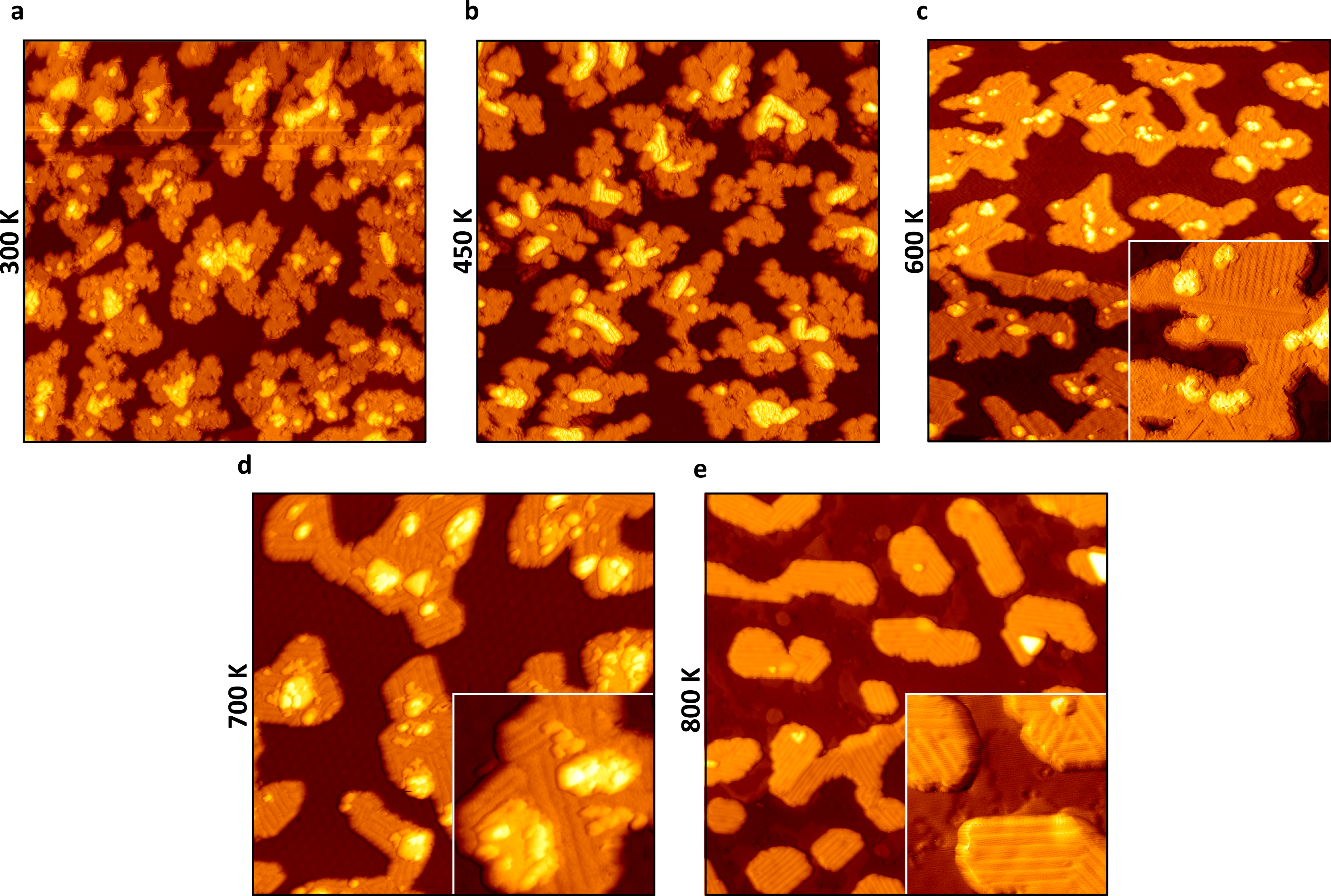}
			\caption{\textbf{Evolution of stoichiometric \va{} with temperature. a}~STM topograph of \va{} islands directly after growth at \SI{300}{\K}. \textbf{b-e}~STM topographs after annealing to the indicated temperatures. The insets in \textbf{c-e} are close-ups of the topographs, highlighting the different superstructures. All annealing was performed without S background pressure. Images taken at \SI{300}{\K}. Image information (image size, sample bias, tunneling current): main panels~\SI{80 \times 80 }{\nano \meter^2}, inset~\SI[parse-numbers=false]{20 \times 20}{\nano \meter^2}. \textbf{a}~\SI{-1.0}{\V}, \SI{210}{\pico \ampere}; \textbf{b}~\SI{-1.0}{\V}, \SI{90}{\pico \ampere}; \textbf{c}~\SI{-1.0}{\V}, \SI{100}{\pico \ampere}, inset: \SI{-0.7}{\V}, \SI{550}{\pico \ampere}; \textbf{d}~\SI{-1.2}{\V}, \SI{110}{\pico \ampere}, inset: \SI{-1.2}{\V}, \SI{110}{\pico \ampere}; \textbf{e}~\SI{-1.3}{\V}, \SI{10}{\pico \ampere}, inset: \SI{-1.3}{\V}, \SI{10}{\pico \ampere}.}
 \label{fig:Fig1}
\end{figure*}

In Fig.~\ref{fig:Fig1}a the sample topography directly after growth at room temperature is shown. Dendritic first layer islands of  \SI{0.65}{\nm} apparent height are visible, covered by small second layer islands. After annealing in vacuum to \SI{450}{\K}, see Fig.~\ref{fig:Fig1}b, the single-layer islands are less rough, but dendritic shapes are still prevalent. After annealing to \SI{600}{\K}, in Fig.~\ref{fig:Fig1}c, the islands are more compact with smoother edges. A superstructure, identified in previous work as a unidirectional charge density wave with an approximate unit cell of $9 \times \sqrt{3}\text{R}30\degree$~\cite{VanEfferen2021}, covers all but the smallest regions of the islands. The CDW appears only after annealing to \SI{600}{\K}, as its formation requires sufficiently large islands of stoichiometric \va. The instability of the CDW on small islands of \va{} has been documented in previous work~\cite{VanEfferen2021}.

Further annealing to \SI{700}{\K}, see Fig.~\ref{fig:Fig1}d, leads to the dominant presence of a new type of striped superstructure (note that already at \SI{600}{\K} some stripes are present, compare Fig.~\ref{fig:Fig1}c). The dark stripes have non-uniform spacing with a minimal width of $\approx \SI{1}{\nm}$ and several domains of different orientations, as can be seen in the inset. After annealing to \SI{800}{\K}, in Fig.~\ref{fig:Fig1}e, the \va{} islands are elongated but compact and display a well-ordered pattern of stripes. The stripes are generally aligned with the island edges and exhibit only three domains, rotated by $120 \degree$. The islands have an altered apparent height of \SI{0.8}{\nm} (see Note\,3 of the Supporting Information (SI) for line profiles of the relevant topographies). The area of the islands has decreased - correspondingly, patches of what are presumably S and V atoms intercalated under the Gr are visible. Annealing beyond \SI{800}{\K} leads to the formation of higher structures, discussed below, and elongated crystallites which are not studied in the present manuscript, see Note\,4 of the SI. 

\begin{figure*}
	\centering
		\includegraphics[width=0.45\textwidth]{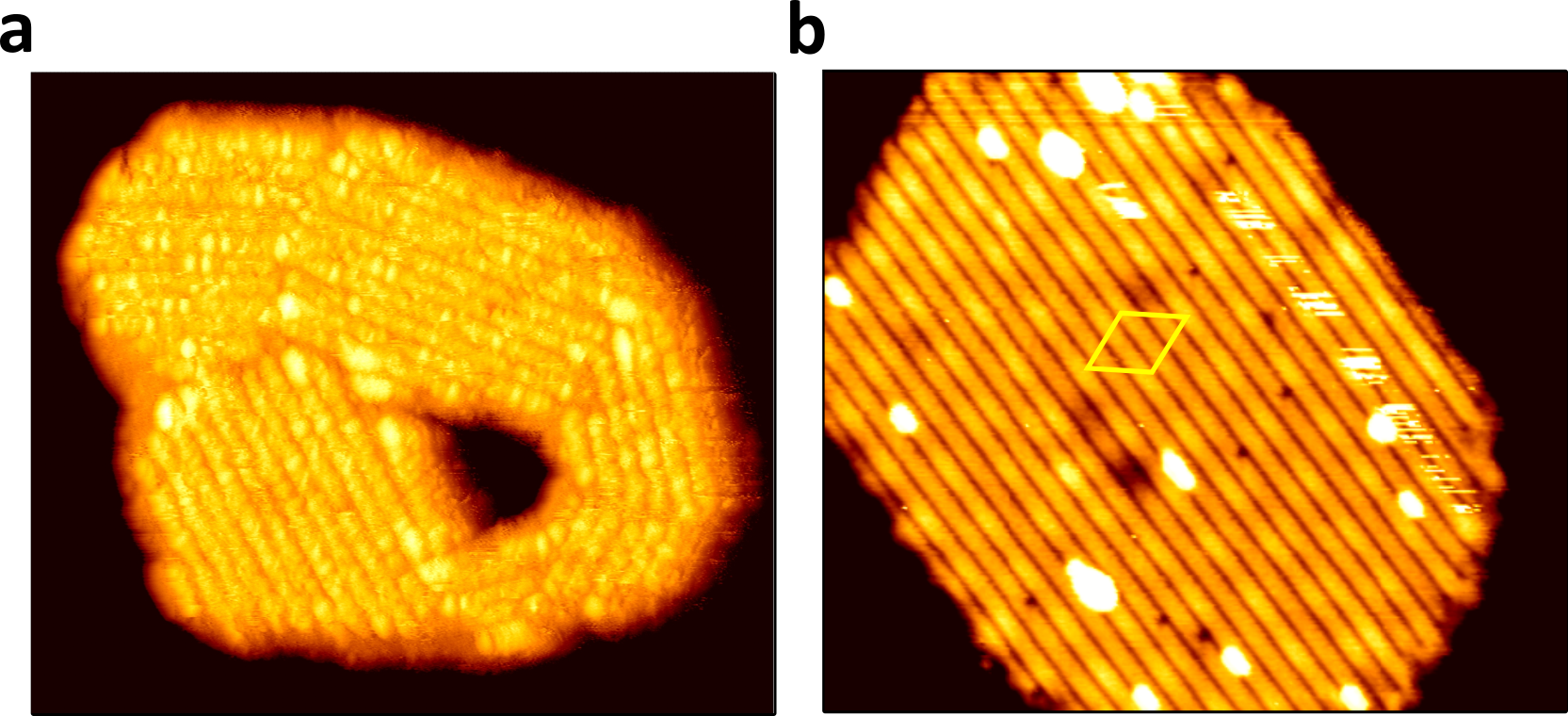}
			\caption{\textbf{Patterned adsorption of S on striped superstructure. a}~STM topograph of a \vas{} island with dark stripes covered by adsorbates after annealing in S-rich environment to \SI{800}{\K}. \textbf{b}~STM topograph of island surface after quick heating to \SI{600}{\K}. Most of the adsorbates have desorbed. The yellow rhombus indicates the Gr/Ir(111) moiré unit cell shining through the \vas{} island. Note that example island shown in \textbf{b} is not the same island as in \textbf{a}.  
Images taken at \SI{300}{\K}. Image information (image size, sample bias, tunneling current): \textbf{a}~\SI{35 \times 30 }{\nano \meter^2}, \SI{-0.5}{\V}, \SI{10}{\pico \ampere}; \textbf{b}~\SI{30 \times 30 }{\nano \meter^2}, \SI{-0.3}{\V}, \SI{20}{\pico \ampere}.}

 \label{fig:Fig2a}
\end{figure*}

Focussing on the stripe superstructure, Fig.~\mbox{\ref{fig:Fig2a}a} shows the result of annealing the stripes to \SI{800}{\K} in a moderate sulphur background of $P_{\mathrm{sulphur}}^{\mathrm{a}}$ = \SI{2 \times 10^{-9}}{mbar}. Bright stripes of adsorbates are present on the island, with a spacing corresponding to the periodicity of the superstructure. Subsequent annealing largely removes the adsorbates uncovering the superstructure again, as shown in Fig.~\ref{fig:Fig2a}b. Only some bright adsorbate clusters are left. The moiré of Gr/Ir(111) (marked with a yellow rhombus) becomes apparent again after annealing.

We interpret the situation as follows: upon cooldown from the annealing step, residual S, which is hard to pump out, adheres to the islands. The subsequent brief heating in clean UHV causes desorption of these adsorbed S atoms. Irrespective of the details, the observations imply a modulated reactivity of the stripe superstructure. Note that for stoichiometric \va{} islands such structured adsorption was never observed. Note also the reconstructed island edges in \mbox{Fig.~\ref{fig:Fig2a}b}, with the the stripes causing kinked edges in segments parallel and orthogonal to them. The stripe superstructure being imprinted on the step edge highlights that the structure is not of electronic but of structural origin.

\begin{figure*}
	\centering
		\includegraphics[width=0.9\textwidth]{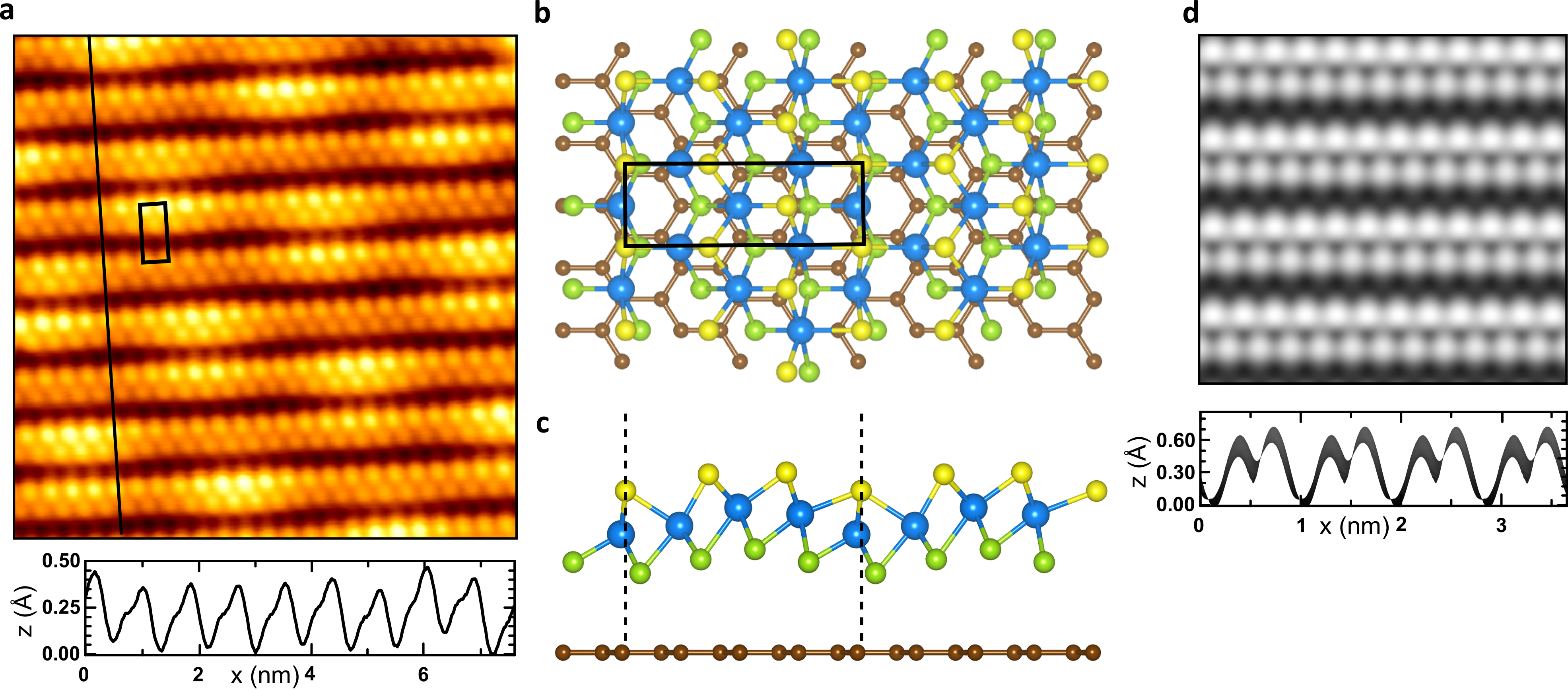}
			\caption{\textbf{Reconstruction of \va{} into \vas{} through defect-row formation. a}~Atomically resolved STM topograph of the stripes. The unit cell of the dominant striped superstructure (black) is indicated. The black line indicates the position of the line profile. \textbf{b,c}~Top and side view of structural model of S-deficient \va{} on graphene obtained by DFT calculations. The black box indicates the unit cell. The bottom S atoms are colored green to allow for easy identification when viewed from the top. \textbf{d}~Simulated STM image (isosurface of the charge density) at \SI{-0.5}{\V}. Below, the isosurface is shown from the side, so the corrugation can be seen.
Image taken at \mbox{\SI{1.7}{\K}}. Image information (image size, sample bias, tunneling current): \textbf{a}~\mbox{\SI{7.5 \times 7.5}{\nano \meter^2}}, \mbox{\SI{-0.5}{\V}}, \mbox{\SI{1.4}{\nano \ampere}}
}

 \label{fig:Fig2}
\end{figure*}

Turning to the atomic structure of the stripes, Fig.~\ref{fig:Fig2}a shows an atomically resolved image of the surface of a striped single-layer island. The rectangular unit cell is shown as a black box. Its dimensions are \mbox{\SI{(3.20 \pm 0.05)}{\angstrom}} along its short axis and \mbox{\SI{(9.1 \pm 0.1)}{\angstrom}} along its long axis. The line profile along the black line shows an apparent height corrugation of about \mbox{\SI{0.4}{\angstrom}}. Depending on the tunneling conditions the apparent height varied in the range from \mbox{\SI{0.4}{\angstrom}} to \mbox{\SI{0.6}{\angstrom}}. The STM contrast allows identification of two rows of atoms (presumably S atoms separated by grooves with a faint substructure.)

Assuming the superstructure to result from S desorption and given the rectangular symmetry of the unit cell with a S atom at each corner of the unit cell, it would be compatible with a VS$_2$ lattice from which either every second or every fourth row of S desorbed. If this would occur without any other changes in the bonding distances of the V and lower S layer, one would expect a periodicity of either \mbox{$2 \times a_\text{VS2} \times \sqrt{3} /2 = \SI{5.5}{\angstrom}$} or \mbox{$4 \times a_\text{VS2} \times \sqrt{3} /2 = \SI{11.1}{\angstrom}$} normal to the removed rows. While the periodicity resulting from the removal of every second S row is inconsistent with the measured unit cell dimension, the somewhat larger unit cell size derived from our assumption of every fourth S row removed could be rationalized by a bond order -- bond length argument:  the remaining atoms strengthen their bonds and consequently shorten them. 

A model involving the removal of every fourth row was also proposed for a similar sized unit cell with same symmetry by Liu et al. \mbox{\cite{Liu2019}} for the sister compound VSe$_2$. They observed a striped phase after annealing monolayer VSe$_2$ grown on graphite.

To clarify the situation we conducted DFT calculations confining VS$_2$ to the size of the experimental unit cell. We placed the \va{} layer on Gr and removed every fourth top layer S row. After relaxation, the structure shown in Fig.~\mbox{\ref{fig:Fig2}b,c} was obtained. The nominal composition of the comound is \vas. The removal of the S atom row has led to a buckling of the structure and changed the coordination of one of the remaining rows of S atoms from threefold to fourfold. The simulated STM image shown in \mbox{Fig.~\ref{fig:Fig2}d}, using precisely the same sample bias of -0.5 V as in the experiment, shows striking agreement with the experimental topograph of \mbox{Fig.~\ref{fig:Fig2}a}. The predicted corrugation of the structure of \mbox{$\SI{\approx 0.6}{\angstrom}$} matches reasonably well with  experiment, as does the asymmetry of the profile.

Giving up the constraint of an initially intact \va{} layer with just S rows removed, alternative models can be considered. If an additional \va{} unit is removed per cell, the compound V$_3$S$_5$ is formed by bond reorganisation. Simulated STM topographs reproduce the experimental STM contrast reasonably well (compare Note~5 of the SI). However, the corrugation of the resulting structure is well above \mbox{\SI{2}{\angstrom}}, at variance with experiment. Its formation would furthermore involve bond breaking and either loss V or expansion of islands, which are also considered to be unlikely. A model similar to this scenario was proposed by Chua et al.~\mbox{\cite{Chua2020}} to explain a striped phase after annealing of monolayer VSe$_2$ on MoS$_2$. However, their unit cell was substantially larger and of different symmetry than the S compound discussed here.

\begin{figure*}
	\centering
		\includegraphics[width=\textwidth]{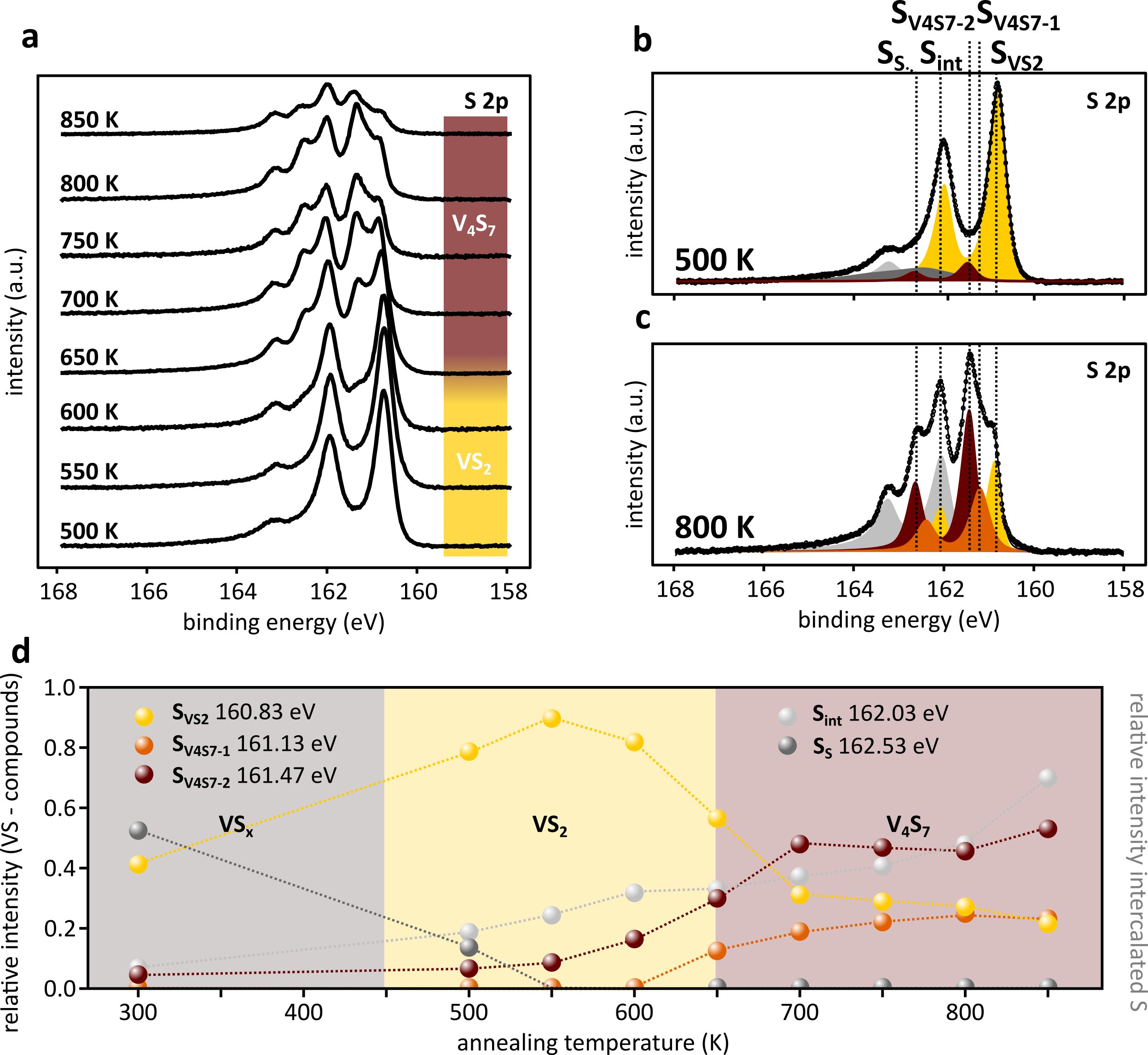}
			\caption{\textbf{XPS of single-layer \va. a}~Temperature resolved X-ray photoemission spectra (XPS) of S 2p$_{3/2}$ and S 2p$_{1/2}$ core-level spectra of \va{} measured at a photon energy of $hv = \SI{260}{\eV}$. For each successive datapoint the sample was annealed to the indicated temperature without additional S and then measured at \SI{300}{\kelvin}. \textbf{b, c}~XPS spectra of S 2p core-level spectra fitted with 5 components. The fitted components are denoted as S$_\text{VS2}$ (yellow), S$_\text{V4S7-1}$ and S$_\text{V4S7-2}$ (orange and brown), S$_\text{int}$ (silver) and S$_\text{S}$ (dark grey), see text for discussion. The data points are represented by black circles and the overall fit by a solid black line. \textbf{d}~Change of the relative intensities of the fit components during the entire annealing sequence as a function of temperature. Dashed lines to guide the eye. Note that the relative intensity of the intercalated S atoms S$_\text{int}$ has been plotted on the right y-axis versus the total S signal of the atoms in the vanadium compound.}
 \label{fig:Fig3}
\end{figure*}

In order to obtain additional information about the effect of annealing stoichiometric \va, we performed high-resolution X-ray photoemission spectroscopy (XPS) of the S 2p core level. Fig.~\ref{fig:Fig3}a shows the S 2p core-level spectra obtained for samples with increasing annealing temperatures in the absence of S pressure from bottom to top. Spin-orbit coupling of the S 2p level gives rise to an S 2p$_{3/2}$ and S 2p$_{1/2}$ doublet separated by \SI{1.19(3)}{\eV}. At \SI{500}{\K} only a single S 2p doublet is visible, with peaks at \SI{160.83(3)}{\eV} and \SI{162.02(3)}{\eV}. As the annealing temperature is increased, more doublets become pronounced. In particular, the S 2p peak intensities shift toward higher binding energies. At the highest annealing temperature of \SI{850}{\K} the total intensity of the signal is significantly reduced.

Fitting the XPS spectra across the whole range of annealing temperatures requires five doublets, which will be discussed by reference to the lower binding energy 2p$_{3/2}$ peak of the doublet. In Fig.~\ref{fig:Fig3}b, a single component (\SI{160.83(3)}{\eV}, yellow) dominates the spectrum, denoted S$_\text{VS2}$. In contrast, the \SI{800}{\K} spectrum in Fig.~\ref{fig:Fig3}c is made up of multiple doublets of similar intensity. Crucially, two new doublets close to the original yellow component are present at \SI{800}{\K}, which we denote S$_\text{V4S7-1}$ (\SI{161.13(3)}{\eV}, orange) and S$_\text{V4S7-2}$ ((\SI{161.47(3)}{\eV}, brown). Comparison to the \SI{500}{\K} spectrum furthermore reveals that the dark grey (\SI{162.53(3)}{\eV}) component S$_\text{S}$ has disappeared, whereas the silver (\SI{162.03(3)}{\eV}) component S$_\text{int}$ is still present and has larger relative intensity. 

From STM images we know that at or below \SI{600}{\K}, mainly unperturbed (though defective) single-layer \va{} islands are present (compare Fig.~\ref{fig:Fig1}a,b,c). We therefore assign S$_\text{VS2}$ to stoichiometric \va{}. S$_\text{S}$, which disappears rapidly when the sample is heated up from room temperature, we assign to sulphur species which react or desorb upon annealing. S$_\text{int}$, on the other hand, can be straightforwardly assigned to S atoms intercalated between Gr and Ir(111)~\cite{Pielic2020}. The presence of intercalated S atoms is inferred from STM images (see Note\,6 of the SI) and low-energy electron diffraction (LEED), an example of which is shown in Fig.~\ref{fig:Fig5}. The assignment is further based on the presence of this peak at the same location for all samples investigated in this study. Its high binding energy furthermore sets it apart from the peaks intrinstic to the vanadium sulphide compound. S$_\text{V4S7-1}$ and S$_\text{V4S7-2}$ finally, are related to the formation of the striped vacancy row \vas{} observed in STM, as these components are present in significant intensity only at higher annealing temperatures.

Plotting the relative intensities of all components against the annealing temperature in Fig.~\ref{fig:Fig3}d, we can broadly distinguish three stages. At \SI{300}{\K}, stoichiometric \va{} is coexisting with unreacted sulphur (compare Fig.~\ref{fig:Fig1}a and Note~7 of the SI). Upon annealing to \SI{500}{\K}, the sulphur desorbs from the surface, leaving a sample consisting mainly of stoichiometric \va{}, with about $90\%$ of the vanadium sulphide intensity stemming from the S$_\text{VS2}$ component. Above \SI{600}{\K}, the sample undergoes a transition to \vas{}, with S$_\text{V4S7-1}$ and S$_\text{V4S7-2}$ becoming prominent with a ratio that tends to $1:2$. Because of this ratio, we assign S$_\text{V4S7-2}$ to the two almost equivalent, threefold coordinated S atom rows in the DFT calculation of Fig~\mbox{\ref{fig:Fig2}b,c}, and S$_\text{V4S7-1}$ to the other S atom row, which has a fourfold coordination, see Note~8 of the SI. While S$_\text{VS2}$ is reduced in intensity and shifts to a slightly higher binding energy by $\approx \SI{100}{\meV}$ when the sample is annealed, it remains present up to \SI{800}{\K}, when the sample surface uniformly exhibits the striped superstructure. We infer that this signal stems from the bottom S atoms, which are in a chemical environment not too different to that of pristine \va. At higher annealing temperatures, the S detached from \va{} intercalates below Gr, which can be seen in Fig.~\ref{fig:Fig1}e, leading to the strong increase in the relative intensity of the S$_\text{int}$ component.

In conclusion, we interpret \vas{} to emerge gradually with increasing annealing temperature through loss of sulphur from the initial \va{} islands and an accompanying reconstruction of the atomic lattice (compare Figs.~\ref{fig:Fig1}c-e). At about \SI{800}{\K}, the density of the vacancy rows has reached a saturation, resulting in uniform \vas{} characterized by every fourth sulphur top layer row missing. The XPS spectra, which show the appearance of new S 2p doublets at considerably higher binding energies, support this interpretation. In particular, two new S 2p components indicate that most of the top layer S atoms are in a different chemical environment compared to pristine \va, which makes a different stoichiometry likely. The loss in S is accompanied with a substantial lattice reconstruction, also consistent with the reshaping of the islands, which reflect the vacancy row periodicity. In contrast, no significant changes are observed in the V 2p XP spectra, see Note~9 of the SI.

The observation of a striped sulphur depleted phase after annealing of a monolayer of \va{} or \ve{} appears to be a common feature of these materials. However, the precise structure that is obtained depends on the substrate. Annealing of \va{} on graphite results in V$_4$Se$_7$, as observed by Liu \textit{et al.}~\mbox{\cite{Liu2019}}; similarly, annealing of \va{} on graphene results in \vas{}, as observed by us. Despite the difference in the chalcogen, the same stoichiometry and very similar structures with a rectangular unit cell and an orthorhombic 2D Bravais lattice are formed. Although annealing of \ve{} on MoS$_2$ as done by Chua \textit{et al.}~\mbox{\cite{Chua2020}} leads to the same stoichiometry V$_4$Se$_7$, the unit cell is oblique and the symmetry is that of a monoclinic 2D Bravais lattice. Finally, annealing of monolayer VS$_2$ on Au(111), as conducted by Arnold \textit{et al.}~\mbox{\cite{Arnold2018}} and Kamber \textit{et al.}~\mbox{\cite{Kamber2021}}, results in a more S-depleted stoichiometry of V$_2$S$_3$, which in addition has a quite different internal structure of the cell. It would display single rows of S atoms rather than the double rows as observed experimentally and simulated by DFT (compare \mbox{Fig.~\ref{fig:Fig2})}.

The formation of chalcogen vacancy rows or line defects due to annealing, electron or laser beam irradiation seems furthermore a general feature in TMDCs~\cite{Komsa2013, Lu2015a, Elibol2018, Zhao2019}. In the case of single-layer \ve, annealing-induced Se-vacancy rows were used to lift the spin-frustration of the material, leading to room-temperature ferromagnetism~\cite{Chua2020}. Since the CDW ground state responsible for the magnetic frustration~\cite{Wong2019} is very similar in both stoichiometric single-layers of \va{} and \ve{}, it would be of great interest to track the effect of annealing \va{} on its magnetic properties. In contrast to \ve{} however, where it is possible to transform the patterned state back to stoichiometric \ve{} \textit{via} low-temperature annealing after deposition of Se atoms~\cite{Liu2019, Chua2020}, low-temperature annealing in sulphur vapour did not recreate stoichiometric \va{} from \vas. Depending on the coverage of the sample, either adsorbed S atoms on top of the patterned material were observed, depicted in Fig.~\ref{fig:Fig2}d, or - in the case of larger coverage - higher structures were obtained, which will be discussed below.

\begin{figure*}
	\centering
		\includegraphics[width=0.9\textwidth]{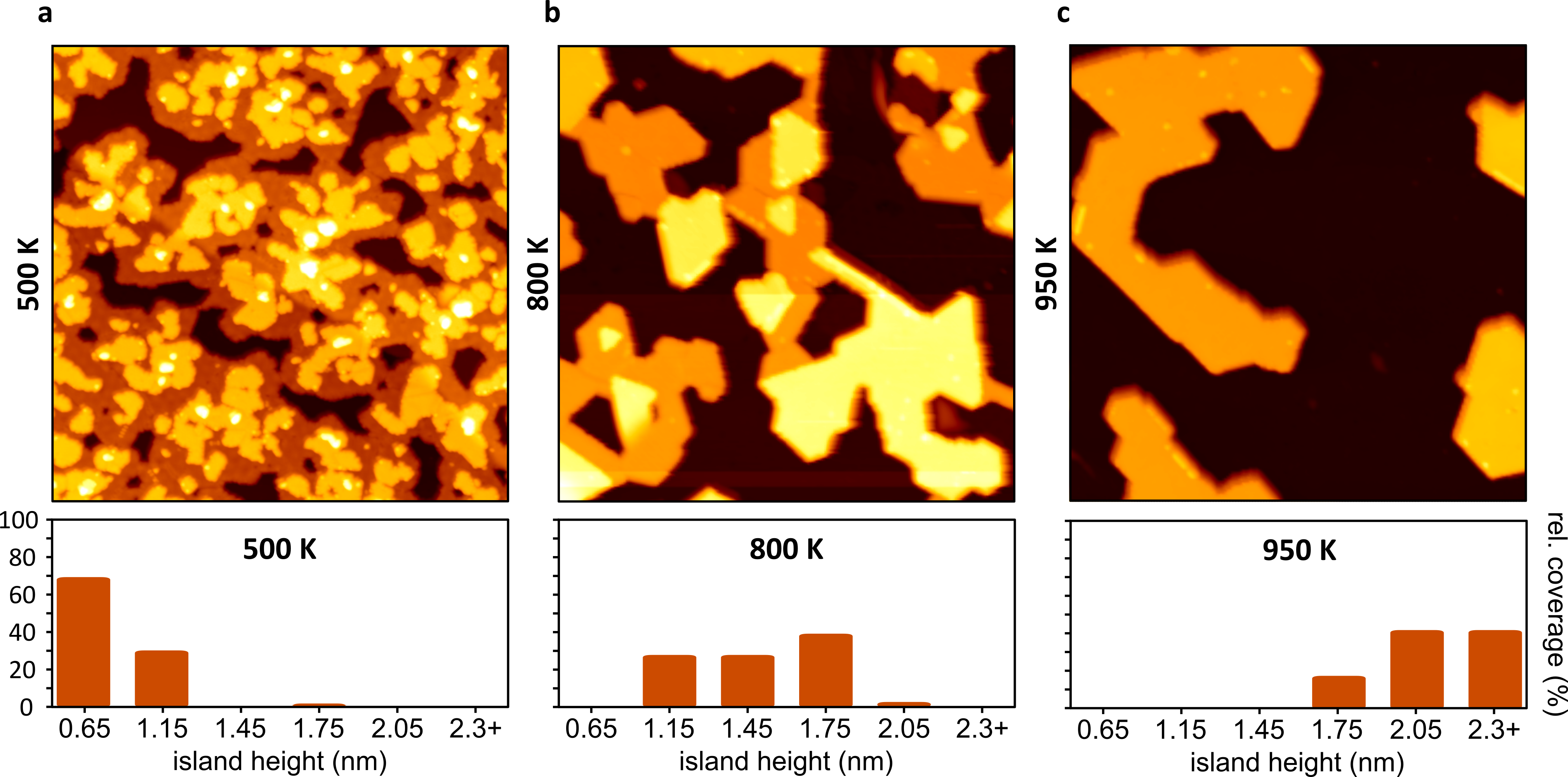}
			\caption{\textbf{Evolution of multiheight VS$_x$ with temperature. a}~STM topograph after deposition of $\approx$ 1.2 ML, annealed at \SI{550}{\kelvin} in a S-rich atmosphere. \textbf{b}~STM topograph after annealing to \SI{800}{\kelvin}. \textbf{c}~STM topograph after annealing to \SI{950}{\kelvin}. Below the STM topographs in \textbf{a-c}, histograms of the relative coverage of islands of different apparent heights is shown.
			Images taken at \SIrange{4}{7}{\K}.\\ STM parameters: \textbf{a-c}~\SI{80 \times 80 }{\nano \meter^2}, \textbf{a}~\SI{1.0}{\V}, \SI{50}{\pico \ampere}, \textbf{b}~\SI{-1.0}{\V}, \SI{50}{\pico \ampere}, \textbf{c}~\SI{1.0}{\V}, \SI{100}{\pico \ampere}.
}
			
 \label{fig:Fig4}
\end{figure*}

\subsection{Self-intercalation of V atoms in multilayer \va}
Instead of single-layer \va{} or \vas{}, we can selectively grow a new 2D material by changing the growth conditions to favor higher structures. Already during growth at \SI{300}{\K} multilayer structures form (see Fig.~\ref{fig:Fig1}a). As we deposit more material, a substantial fraction of the material will grow on top of single-layer \va. When we then anneal in S-rich environment to about \SI{800}{\K}, structures with several height levels are created, which do not manifest S vacancies in their top layer.

In Fig.~\ref{fig:Fig4}a-c the evolution of such a sample is depicted. After annealing $\approx 1.2$ ML of VS$_x$ to \SI{550}{\K}, the apparent height distribution shows the \SI{0.65}{\nm} apparent height characteristic for single-layer \va, a significant area fraction of \SI{1.15}{\nm} height and some small areas of \SI{1.75}{\nm} apparent height, as shown in Fig.~\ref{fig:Fig4}a. Annealing the sample to \SI{800}{\K} leads to a transformation of island shapes and apparent heights, shown in Fig.~\ref{fig:Fig4}b. The island edges are straight, largely aligned to the dense packed directions of Gr. In addition, two new island apparent heights appear (\SI{1.45}{\nm} and \SI{2.05}{\nm}), both \SI{0.3}{\nm} offset from the higher islands already present at \SI{550}{\K}. The lowest islands are now \SI{1.15}{\nm} high. No single-layer \va{} is present on the surface. While the different islands are generally sharply demarcated by step edges from one another, continuous transitions between them are also observed, like in the bottom left of Fig.~\ref{fig:Fig4}b. These transitions will be discussed below. When the sample is further annealed to \SI{950}{\K}, as shown in Fig.~\ref{fig:Fig4}c, only islands with apparent heights of \SI{1.75}{\nm} and up are observed. We thus find that as the annealing temperature increases, the island density goes down, while the average apparent island height increases, with island heights separated by $\approx \SI{0.3}{\nm}$. The resulting sample cannot be understood as multilayer \va, since the height of each \va{} layer is $\approx \SI{0.6}{\nm}$. We therefore refer to these samples as multiheight VS$_x$. Our investigation will focus on the lower two levels of \SI{1.15}{\nm} and \SI{1.45}{\nm} apparent height, which will be identified as 2D derivates of \vai.

\begin{figure*}
	\centering
		\includegraphics[width=0.9\textwidth]{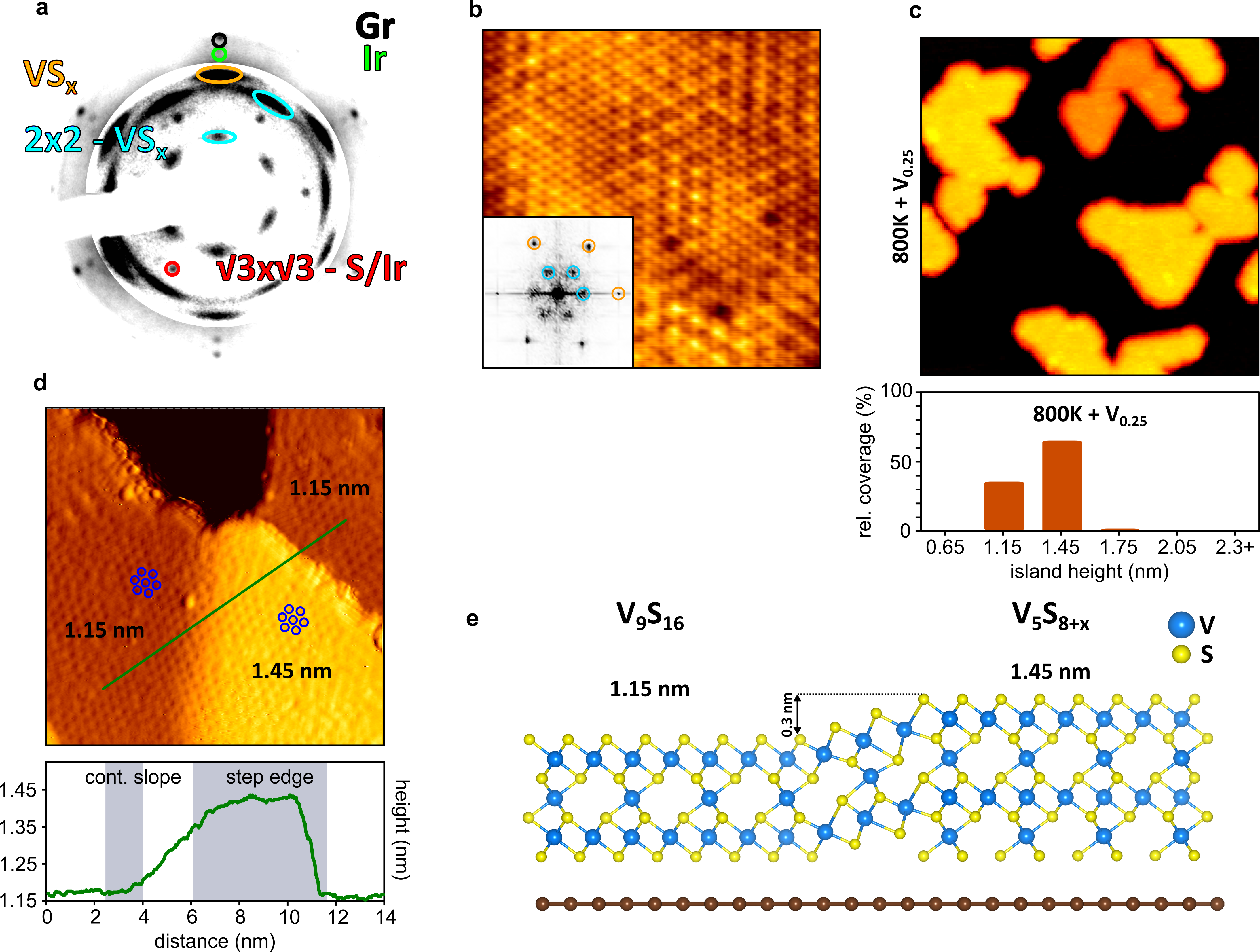}
			\caption{\textbf{$2\times2$ self-intercalation in multiheight VS$_x$. a}~LEED image of \vx{} after annealing to \SI{800}{\kelvin}. Some Gr and Ir first order reflections are encircled black and green, respectively. Intercalated S (red) gives rise to a $\sqrt{3}\times\sqrt{3}$ superstructure with respect to Ir. Some first order reflections of \vx{} (orange) and reflections of a $2 \times 2$ superstructure with respect to \vx{} (light blue) are encircled. \textbf{b}~Room temperature STM topograph of \SI{1.15}{\nm} island, with FFT as inset. A clear $2\times2$ superstructure is visible on the island. \textbf{c}~STM topograph of \vx{} sample, where $25\%$ additional V atoms were deposited on a sample grown at \SI{300}{\K} and annealed to \SI{500}{\K}. After room temperature V deposition, the sample was subsequently annealed to \SI{800}{\K} in UHV (no S pressure). \textbf{d}~STM topograph of \SI{1.15}{\nm} and \SI{1.45}{\nm} islands on sample without additional V atoms. A \ro{} superstructure is highlighted with blue circles on both islands, see text. Below the topograph, a height profile taken long the green line is shown. \textbf{e}~Side view of atomic structure model of \SI{1.15}{\nm} and \SI{1.45}{\nm} islands forming a continuous top layer.
			STM parameters: 
			STM parameters: \textbf{b}~\SI{300}{\K}, \SI{10 \times 10}{\nano \meter^2}, \SI{-1.0}{\V}, \SI{200}{\pico \ampere}; \textbf{c}~\SI{110}{\K}, \SI{80 \times 80}{\nano \meter^2}, \SI{1.5}{\V}, \SI{550}{\pico \ampere}; \textbf{d}~\SI{77}{\K}, \SI{16 \times 16}{\nano \meter^2}, \SI{-1.0}{\V}, \SI{200}{\pico \ampere}
			LEED taken with \SI{118}{\eV}. The contrast of the inner part of the LEED image has been enhanced for clarity.}
			
 \label{fig:Fig5}
\end{figure*}

In the LEED pattern of the multiheight sample annealed at \SI{800}{\K}, where the apparent minimum island height is \SI{1.15}{\nm}, we find multiple ordered structures, see Fig.~\ref{fig:Fig5}a. The Gr (black circle) and iridium (green circle) first order spots are visible, surrounded by hardly visible satellite spots from their moir\'e. Closer to the center, sharp $\sqrt{3} \times \sqrt{3}$ spots with respect to Ir [$(\sqrt{3}\times\sqrt{3})_\text{Ir}$] are present, which stem from S intercalation under Gr. The \vx{} spots (orange ellipses) are elongated, indicating a small epitaxial spread in orientation angles, which sets them apart from the perfectly oriented spots of the substrate. Besides the first order \vx{} spots, two other sets of spots with the same elongation are visible, which are identified as first and second order spots of a $2\times2$ superstructure with respect to \vx{}. The first order spots indicate a hexagonal symmetry with a lattice parameter of \SI{3.23 \pm 0.03}{\angstrom}. This is close to the lattice parameter of stoichiometric \va{} (\SI{3.21 \pm 0.02}{\angstrom})~\cite{Murphy1977, VanEfferen2021}. However, the strong $2\times2$ reflections are characteristic of \vai, a material of monoclinic symmetry which can be understood as a bulk crystal consisting of sheets of \va{} with a $2\times2$ layer of V atoms in between~\cite{Kawada1975}. 

An atomically resolved STM topograph, taken at room temperature, of a \SI{1.15}{\nm} high island after annealing to \SI{800}{\K}, depicted in Fig.~\ref{fig:Fig5}b, shows that the material has preserved its hexagonal symmetry on the surface - as expected from the LEED pattern. At room temperature and under favorable imaging conditions, along with the atomic lattice, a $2 \times 2$ superstructure is visible. The superstructure is slightly disordered. This is evident also in the fast Fourier transform (FFT) shown in the inset of Fig.~\ref{fig:Fig5}b, where the $2\times2$ spots (green circles) are broader than the atomic lattice spots (yellow circles). The presence of a $(2\times2)$ superstructure in STM is consistent with a $2\times2$ intercalation layer, as expected for \vai. The fact that it is hard to image and not present on all atomic resolution topographs obtained indicates that the $2\times2$ does not originate from the top layer, again as expected for an intercalation layer in a \vai-derived structure. 

To confirm that it is straightforward to self-intercalate V into existing layers, we evaporated $25\%$ additional V atoms on an already grown \vx{} sample annealed to \SI{500}{\K}. Prior to V deposition, the sample was comparable to the sample shown in Fig.~\ref{fig:Fig4}a, with uncovered single-layer \va{} and some islands of \SI{1.15}{\nm} apparent height. The evaporation was performed in UHV, with no additional S added. After deposition, the sample was annealed to \SI{800}{\K} in UHV - so again without the addition of any S. The result is shown in Fig.~\ref{fig:Fig5}c. Compared to Fig.~\ref{fig:Fig4}b, which was annealed to the same temperature, the ratio of \SI{1.45}{\nm} islands to \SI{1.15}{\nm} islands has increased and little to no thicker layers have formed. Crucially, no V atoms or clusters are present on top of the islands or the Gr. The V atoms have been incorporated in the islands, changing their stoichiometry. In the LEED of this sample, a clear $2\times2$ signal can be distinguished, consistent with our interpretation of a $2\times2$ intercalation structure, see Note~10 of the SI.

No structural differences exist between the surfaces of the \SI{1.15}{\nm} and \SI{1.45}{\nm} islands. On the contrary, a continuous transition from one island type to the other is possible, as evinced in Fig.~\ref{fig:Fig5}d. In the image, an island of \SI{1.45}{\nm} apparent height is seen to continuously transform into the island of \SI{1.15}{\nm} on the right hand side, while a step edge separates it from the \SI{1.15}{\nm} island in the upper left, as seen in the line profile below Fig.~\ref{fig:Fig5}d. Because of this continuity, we conclude that the structure of the islands is essentially the same. 

The thinnest possible form of \vai{} consists of two layers of \va{} intercalated by a single $2\times2$ sheet of V atoms and has a stoichiometry of V$_9$S$_{16}$, see Fig.~\ref{fig:Fig0}c. The second thinnest 2D \vai{}-derivative would then have a second intercalation layer under the lower S atoms. With these structures in mind, we give an atomic model that explains the continuity in Fig.~\ref{fig:Fig5}e. Since V atoms below the island would be highly reactive, we presume that they are passivated by S atoms. This additional layer of V and S accounts for the \SI{3}{\angstrom} apparent height difference between the islands, see in Fig.~\ref{fig:Fig5}e. The stoichiometry of the thicker island, with two V intercalation layers, is dubbed V$_5$S$_{8 - x}$, since the exact configuration of the bottom S atoms is not known. 2p$_{3/2}$ and S 2p$_{1/2}$

\begin{figure*}
	\centering
		\includegraphics[width=0.9\textwidth]{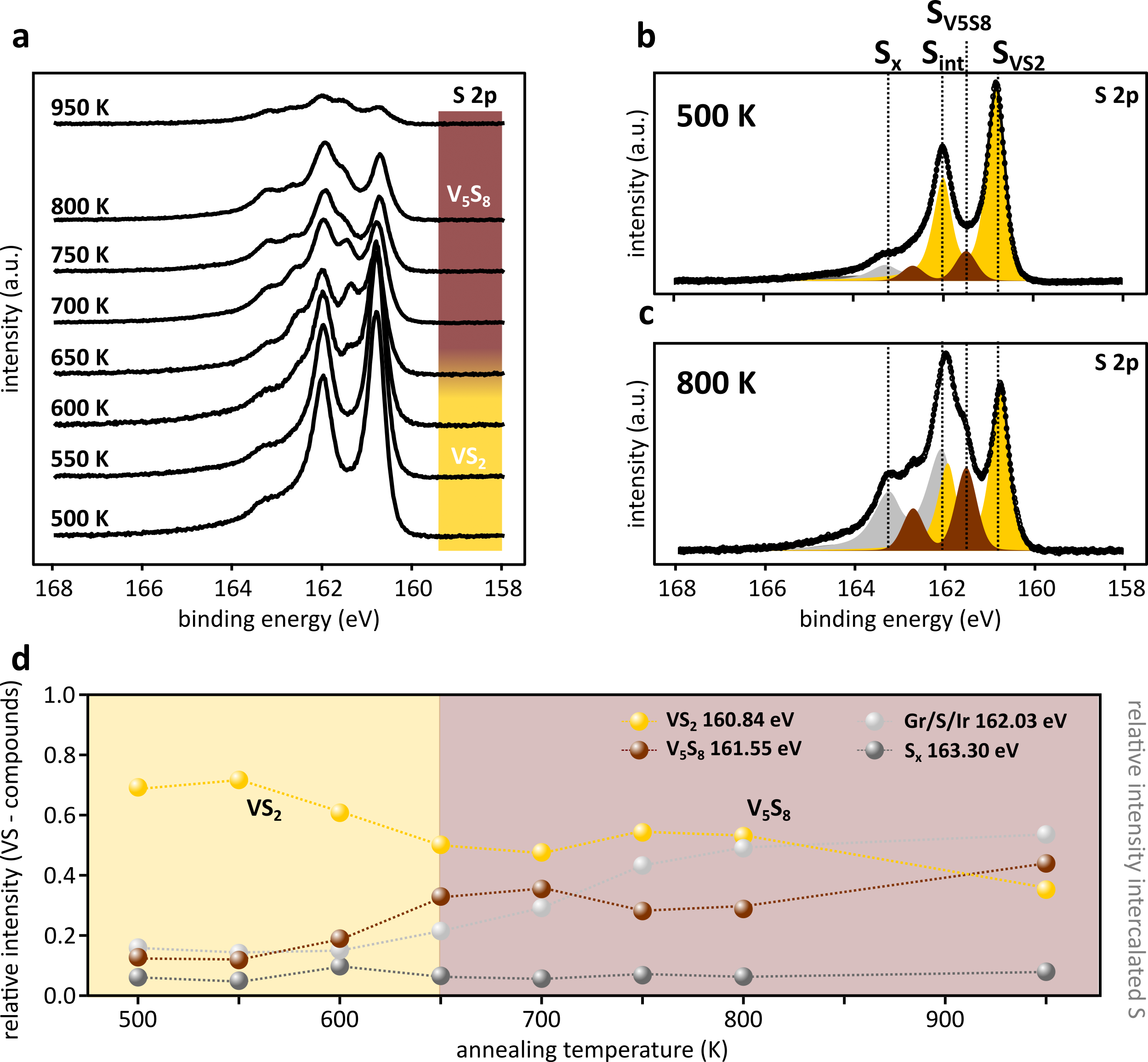}
			\caption{\textbf{XPS of multiheight VS$_x$. a,b,c}~Temperature resolved X-ray photoemission spectra (XPS) of the S 2p$_{3/2}$ and S 2p$_{1/2}$ core-levels of \va{} measured at a photon energy of $hv = \SI{260}{\eV}$. For each spectrum, the sample was annealed to the indicated temperature in a S background pressure of \SI{3\times 10^{-9}}{\milli\bar} and then measured at \SI{300}{\K}. The XPS spectra are fitted with 4 components. The fitted components are S$_\text{VS2}$ (yellow), S$_\text{V5S8}$ (brown), S$_\text{int}$ (silver) and S$_\text{x}$ (dark grey), see text for discussion. The data points are represented by black circles and the overall fit by a solid black line. \textbf{d}~Change of the relative intensities of the fit components as a function of temperature. Dashed lines are present to guide the eye. Note that the relative intensity of the intercalated S atoms S$_\text{int}$ has been plotted on the right y-axis versus the total S signal of the atoms in the vanadium compound.
			}
 \label{fig:Fig6}
\end{figure*}

Using XPS we investigated the S2p core level of \vx{} at different temperatures during an annealing experiment in a S-rich atmosphere corresponding to the STM sequence in Fig.~\ref{fig:Fig4}. The result is displayed in Fig.~\ref{fig:Fig6}a. After annealing to \SI{500}{\K}, the S$_\text{VS2}$ doublet at \SI{160.83(3)}{\eV} and \SI{162.02(3)}{\eV} is most prominent in the spectrum, indicating the dominant presence of stoichiometric \va, see Fig.~\ref{fig:Fig6}b. When the annealing temperature is increased, S$_\text{VS2}$ decreases in intensity and a new doublet S$_\text{V5S8}$ appears at the higher binding energy of \SI{161.55(3)}{\eV} (brown in Fig.~\ref{fig:Fig6}c). The total intensity of the signal is reduced significantly. However, unlike the annealing sequence in Fig.~\ref{fig:Fig3}, where a small coverage of \va{} was annealed without S pressure, the S$_\text{VS2}$ component in Fig.~\ref{fig:Fig6} remains dominant up to \SI{800}{\K}. When we fit the spectra, using the same fitting parameters for S$_\text{VS2}$ and S$_\text{int}$ as for the fits of Fig.~\ref{fig:Fig3}, we find that we obtain a reliable fit for a total of four peaks. Besides S$_\text{V5S8}$, we added a minor dark grey component S$_\text{x}$ located at \SI{163.30(10)}{\eV}, hardly visible in the spectra and of unkown origin.  Additional XP spectra of the sample with extra V atoms, shown in Fig.~\ref{fig:Fig5}c, can be found in Note~11 of the SI.

Tracking the relative intensities of all components in Fig.~\ref{fig:Fig6}d, it becomes apparent that S$_\text{V5S8}$ is already present at the lowest investigated temperature and grows at the expense of S$_\text{VS2}$, with an onset for the growth between \SIrange{550}{600}{\K}. S$_\text{x}$ remains of equal low intensity throughout the annealing process, while S$_\text{int}$ grows as the annealing temperature is increased. 

We interpret the two characteristic components in \vai-derived islands to stem from S atoms in two different chemical environments. S atoms bound only to 3 V neighbours as in \va{} give rise to the S$_\text{VS2}$ component, which undergoes a small shift on the order of $\approx \SI{100}{\meV}$ towards lower binding energies upon intercalation, while S atoms next to the V $2\times2$ intercalation layer, being bound to more than 3 V, give rise to the S$_\text{V5S8}$ component. Although the $2 \times 2$ superstructure in STM and LEED alone could stem from a lattice distortion, the strong shift ($\approx \SI{700}{\meV}$) in the binding energy of the S atoms from the S$_\text{VS2}$ to the S$_\text{V5S8}$ doublet signals a more significant change in the chemical environment of the atoms. A similar shift to higher binding energy was observed in self-intercalated bilayers of \ve{} in Ref.~\mbox{\citenum{Bonilla2020}}, where it was associated with a change in the electrostatic energy at the Se sites coordinated to more than 3 V atoms. Analogous to the XP spectra of \vai, again no significant changes are observed in the V 2p XP spectra, see Note~12 of the SI.

Islands thicker than \SI{1.45}{\nm} are most likely also \vai-derived. Consquently, the number of S atoms bordering $2\times2$-V planes increases upon annealing, causing the S$_\text{V5S8}$ component to rise, while the number of surface S atoms decreases, leading to the decrease of the S$_\text{VS2}$ component and the observed shift in relative ratio between the components. Since XPS is a surface sensitive technique, the S$_\text{VS2}$ component stemming from the top S atoms will generally outweigh XPS signals of S atoms deeper in the islands, which explains the dominant presence of S$_\text{VS2}$ even when most islands have one or more intercalation layers. We also cannot exclude the presence of (unintercalated/partially intercalated) vdWs gaps in the thicker islands. Their precise analysis lies beyond the scope of the manuscript.

Concerning the other components, the comparatively small increase in the relative intensity of S$_\text{int}$, compared to the increase of the same component in the XPS of single-layer \va, is probably due to the presence of S-intercalation already after annealing at \SI{500}{\K}, since all annealing steps were performed in a S-rich environment (note the high intensity of the S intercalation spots in the LEED of Fig.~\ref{fig:Fig5}a). The increase in the intensity of S$_\text{int}$ is thus a measure of the reduced surface area of the islands, exposing more of the Gr. The S$_\text{x}$ component has little weight and its origin cannot be uniquely determined. It could stem from adsorbed S, since it has a similarly high binding energy as the S$_\text{S}$ component in the monolayer.

To summarize at this point, we interpret the islands formed when annealing a large ($> 1$ ML) deposited amount of VS$_x$ in a S background pressure as being \vai-derived. This is indicated by the simultaneous presence of a $2\times2$ superstructure in LEED and STM and the new S$_\text{V5S8}$ component in the S 2p XPS spectra, with a binding energy \SI{700}{\meV} higher than S$_\text{VS2}$. We are able to create a more pure minimal-thickness \vai{} sample \textit{via} the evaporation of extra V atoms on a multiheight sample annealed to \SI{500}{\K}, and then annealing it to \SI{800}{\K}, as depicted in Fig.~\ref{fig:Fig5}c.

We note finally that in similar fashion V$_5$Se$_8$ was previously procured from seed layers of VSe$_2$ during thin film growth~\cite{Nakano2019}, or by increasing the substrate temperature during growth or annealing of single-layer VSe$_2$~\cite{Meng2022, Sumida2022}. In a similar vein, the chemical vapor deposition growth of \va{} nanosheets generally leads to the simultaneous presence of \va{} and \vai{}~\cite{Ji2017, Lee2022}. In each case, the small differences in formation energy between the self-intercalated material and the TMDC are emphasized. Here, we showed that providing extra V atoms after an initial growth step can help to favor one phase over the other and enables one to grow phase pure ultrathin \vai-derived material down to the minimal thickness of \SI{1.15}{\nm}. These minimal thickness islands with a stoichiometry of V$_9$S$_{16}$ can be considered as single layers of a new 2D material.

\subsection{\ro{} CDW in ultimately thin \vai-derived islands}
When we cool down the multiheight sample, we find that the islands of \SI{1.15}{\nm} and \SI{1.45}{\nm} apparent height undergo a structural phase transition. In both island types, a pronounced superstructure with a periodicity of $\SI{5.5 \pm 0.1}{\angstrom} = \sqrt{3}a_\text{VS2}$ is observed in topography at temperatures $\le\SI{77}{\K}$. This superstructure can already be seen clearly in Fig.~\ref{fig:Fig5}d, where the \ro{} superstructure maxima are encircled in blue. It coexists with the CDW in the monolayer, see Note~13 of the SI.

\begin{figure*}
	\centering
		\includegraphics[width=0.9\textwidth]{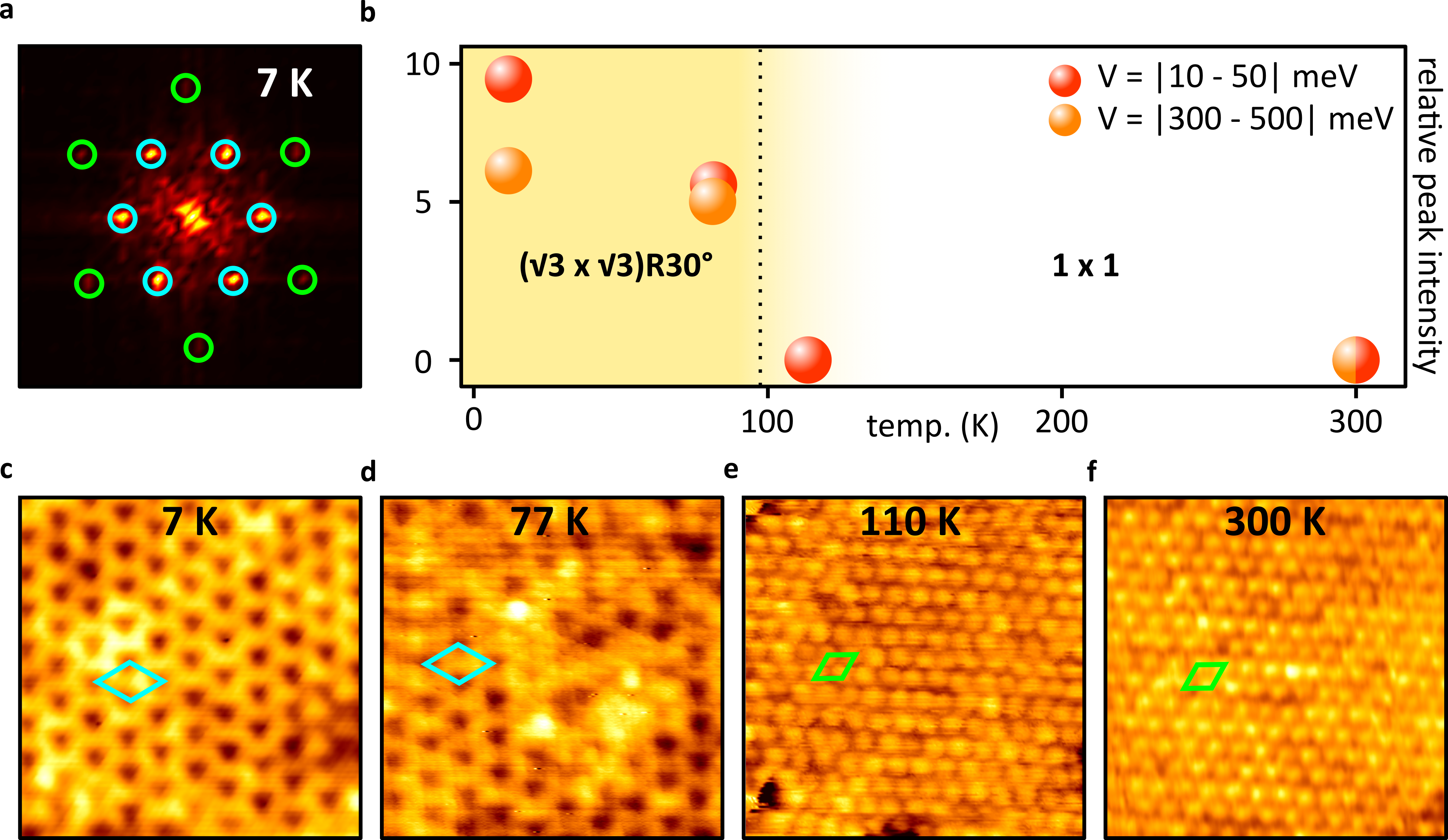}
			\caption{\textbf{\ro{} CDW transition in thin \vai-derived islands. a}~FFT of \SI{1.15}{\nm} V$_9$S$_{16}$ island taken at \SI{7}{\K}, with the superstructure spots stemming from the \ro{} and atomic lattice encircled in cyan and green, respectively. \textbf{b}~Temperature dependence of \ro{} CDW in V$_9$S$_{16}$. Thermal evolution of intensities of CDW peaks normalized to intensities of Bragg peaks, from FFTs of \SI{5 \times 5}{\nm^2} STM topographs of \SI{1.15}{\nm} (V$_9$S$_{16}$) and \SI{1.45}{\nm} high (V$_5$S$_{8+x}$) islands. \textbf{c-f}~Selected topographs at each temperature used in \textbf{a}, with the unit cells of the \ro{} (cyan) and the atomic lattice (green) indicated. The STM topographies used for the analysis were taken both close to the Fermi level (\SI{|10-50|}{\meV}) and far from it (\SI{|300-500|}{\meV}), to make certain that the disappearance is not an artefact stemming from different tunneling conditions.
			STM/STS parameters: \textbf{c-f}~All topographs \SI{5 \times 5}{\nano \meter^2}; \textbf{c}~\SI{-0.5}{V}, \SI{50}{\pico \ampere}; \textbf{d}~\SI{-0.5}{V}, \SI{200}{\pico \ampere}; \textbf{e}~\SI{0.014}{V}, \SI{500}{\pico \ampere}; \textbf{f}~\SI{0.010}{V}, \SI{3}{\nano \ampere}.
			}
 \label{fig:Fig7}
\end{figure*}

Upon heating the sample from \SI{7}{\K} back to room temperature, the \ro{} superstructure vanishes and the atomic lattice becomes visible. As a measure of the strength of the superstructure, we take the relative peak intensity of the $1\times1$ Bragg peaks with respect to that of the \ro{} peaks~\cite{Ugeda2016}. In the \SI{7}{\K} FFT shown in Fig~\ref{fig:Fig7}a, these are encircled in green and cyan, respectively. Plotting the relative intensities for FFTs obtained at different temperatures in Fig~\ref{fig:Fig7}b, the \ro{} can be seen to disappear between \SI{77}{\K} and \SI{110}{\K} (see Note~14 of the SI for all topographies and FFTs). Some representative topographies are shown Fig~\ref{fig:Fig7}c-f, which make clear that the atomic lattice is recovered at \SI{110}{\K}. From the strong temperature dependence of the \ro{} superstructure, we conclude that it is likely a charge density wave. This CDW is not to be confused with the one that develops in single-layer \va, which is unidirectional and has a much higher transition temperature~\cite{VanEfferen2021}, see Table~\ref{Table1} for the differences.

\begin{figure*}
	\centering
		\includegraphics[width=0.9\textwidth]{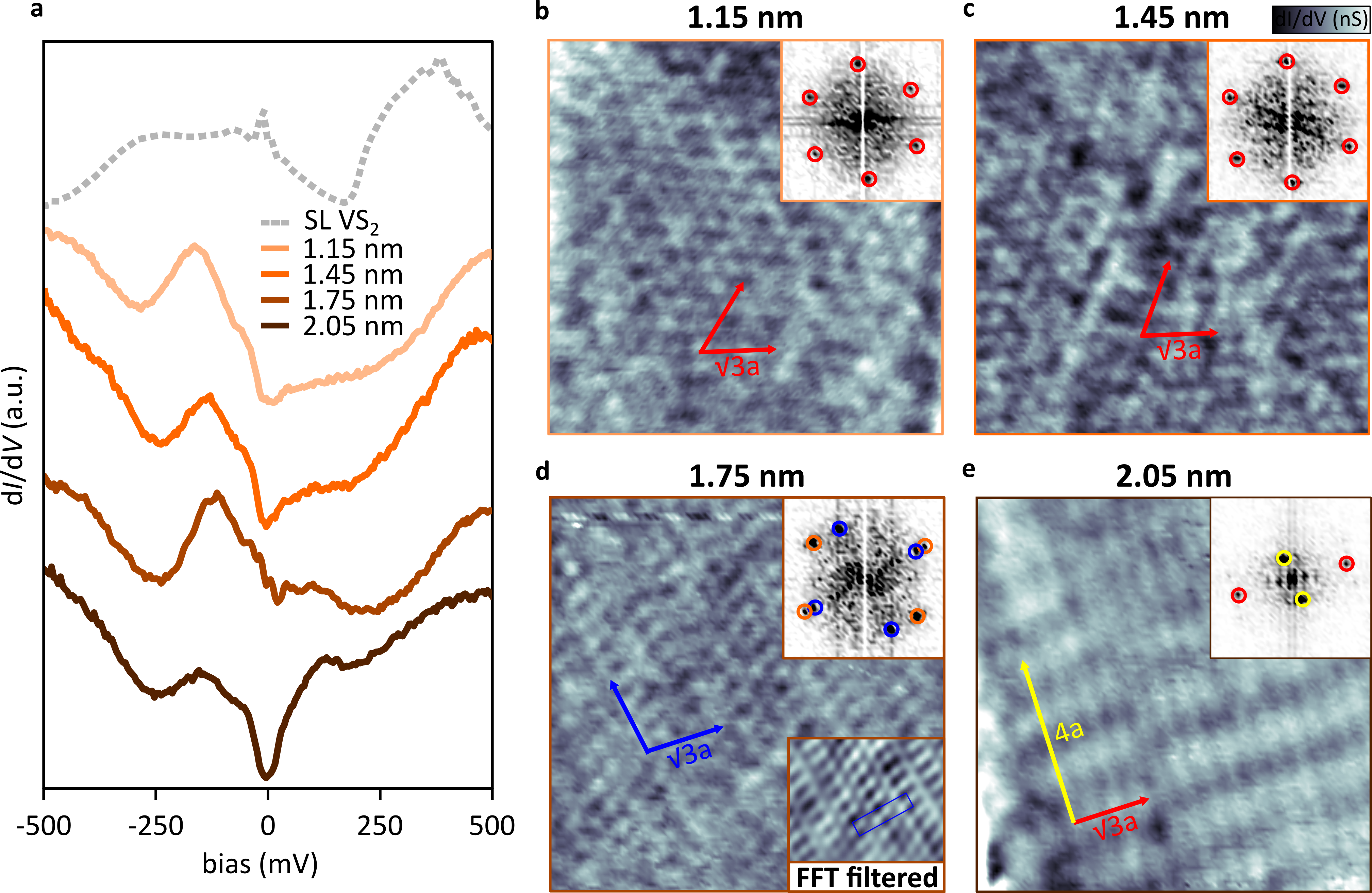}
			\caption{\textbf{Multiple CDW orders in \vai-derived islands. a}~d$I$/d$V$ spectra of islands of different apparent heights. Spectrum of single layer (SL) \va{} (dashed line) added for comparison. \textbf{b-e}~Constant-current d$I$/d$V$ maps of different islands, all taken with V$_\text{set} =\SI{-0.5}{\V}$. The insets show FFTs of the images. Note that the FFTs only display the superstructure, no atomic spots are present. The superstructure spots are encircled for emphasis. In the images, arrows are drawn to indicate the direction of particular periodicities. Each arrow has the size of three wavelengths of the corresponding structure.
			STM/STS parameters:  \textbf{a}~$V_\text{stab} = \SI{-500}{\milli \V}$, $I_\text{stab} = \SI{250}{\pico \ampere}$; \textbf{b-e}~\SI{9 \times 9}{\nano \meter^2}, \SI{-0.5}{\V}, \SI{100}{\pico \ampere}. STS taken with $f_\text{mod} = \SI{871}{\Hz}$ and $V_\text{mod} = \SI{5}{\milli \V}$. Maps obtained at \SI{7}{\K}.
			}
 \label{fig:Fig8}
\end{figure*}

Finally, turning towards the electronic structure of the multiheight islands, we see in  Fig.~\ref{fig:Fig8}a that, in contrast to ML \va, which hosts a CDW with a full gap above the Fermi level $E_\text{F}$~\cite{VanEfferen2021}, d$I$/d$V$ spectra taken on the thinnest \vai-derived islands (\SI{1.15}{\nm} and \SI{1.45}{\nm}) show a partial gap around $E_\text{F}$, while a large density of states (DOS) peak is observed below $E_\text{F}$. Comparing d$I$/d$V$ spectra of islands of different apparent heights, taken with the same tip, we see essentially no differences between islands of \SI{1.15}{\nm} and \SI{1.45}{\nm} apparent height, which is in line with the fact that both have the same \ro{} CDW. 

Islands of \SI{1.75}{\nm}, however, have a spectrum with no clear gap around $E_\text{F}$ and instead a depression around \SI{250}{\meV}. At \SI{2.05}{\nm} apparent height, yet another spectrum is obtained, which lacks the large DOS below $E_\text{F}$ that characterizes the other spectra and instead displays a more symmetric distribution of states on either side of $E_\text{F}$. While it is tempting to correlate the presence of a gap at the Fermi level with the CDW, a more thorough investigation, backed up by DFT calculations is necessary to fully resolve this issue. Gaps at the Fermi level have also been argued to arise due to \textit{e.g.} inelastic tunnelling effects~\mbox{\cite{Hou2020, Wen2020}}, whereas CDWs do not always lead to Fermi-level gaps which can be resolved with STM~\mbox{\cite{VanEfferen2021}}. In this case, temperature-resolved ARPES data could give more insight into the mechanism of CDW formation.

d$I$/d$V$ maps taken on the islands at \SI{7}{\K}, as shown in Fig.~\ref{fig:Fig8}b-e, reveal that these differences in electronic structure are accompanied by the presence of different CDWs. Note that all maps shown are taken at the same bias voltage $V_\text{set} = \SI{-0.5}{\meV}$, with the same tip, see Note~15 of the SI for an overview STM image showing all investigated islands. While the two lower islands both show the aforementioned \ro{}, the lattice symmetry is broken in the higher islands, which exhibit striped superstructures. Upon closer inspection, the \SI{1.75}{\nm} superstructure can be understood as a \ro{} structure, together with dimerization along one of the high-symmetry directions of the lattice, which is highlighted in the inset Figs.~\ref{fig:Fig8}d. The dimerization breaks the symmetry and leads to a distortion of the $\sqrt{3}a$ FFT spots. The superstructure of the \SI{2.05}{\nm} island shown in Fig.~\ref{fig:Fig8}e, on the other hand, can be readily decomposed into a $\sqrt{3}a$ and a $4a$ component, which are orthogonal to one another. 

These differences illustrate the complexity of the system and would require a further and more detailed investigation. However, regardless of the actual composition, this variety of superstructures of slightly differing thickness and stoichiometry is striking. It underscores both the ubiquity of CDWs in low-dimensional vanadium chalcogenides and the metastability of any particular periodicity. It can be seen as an analogue to the case of \ve, which hosts a plentitude of CDW phases, depending on its substrate and the number of layers~\cite{Zhang2017, Duvjir2021, Wang2021, Chua2022, Fumega2023}.

With respect to \vai{} in particular, the fact that it hosts a CDW when thinned down to its 2D limit could shed new light on the layer-dependence of its magnetism, which changes from anti-ferromagnetic to ferromagnetic when thinned down to \SI{3.2}{\nm}~\cite{Hardy2016, Niu2017, Zhang2020}. Since bulk \vai{} is not known to have a CDW, the interplay between the formation of the \ro{} CDW and the magnetic structure would be a compelling subject for further study.

\section{Conclusion}
In conclusion, we have grown and investigated two novel 2D vanadium-sulphur compounds: \vas{} and V$_9$S$_{16}$, the latter being the thinnest possible \vai-derived material.

\vas{} is created from ML \va{} by the formation of S-defects through annealing without additional sulphur vapor. At an annealing temperature of \SI{800}{\K}, the S defects have spontaneously ordered into a homogeneous array of 1D trenches. We have shown that this 1D pattern templates S adsorption and speculate that also other adsorbates can be templated through the vacancy row pattern. Based on size and symmetry of the experimental unit cell, \textit{ab-initio} DFT calculations were able to provide a stable structure model for \vas{}, which is fully consistent with the STM data.

\vai{}-derived islands are obtained automatically when a larger coverage ($> 0.5$ ML) of VS$_x$ is grown. Second layer nucleation on top of single-layer \va{} does not occur as bilayer \va{}; instead, V atoms intercalate between \va{} layers in a $2 \times 2$ pattern and form V$_9$S$_{16}$. Annealing a mixture of single-layer \va{} and V$_9$S$_{16}$ in a S-rich atmosphere to \SI{800}{\K} leads to further self-intercalation of V atoms and the creation of a variety of structures of different apparent height. However, when additional V atoms are supplied after growth, phase-pure minimal thickness \vai-derived material is obtained as a mixture of V$_9$S$_{16}$ of \SI{1.15}{\nm} thickness and V$_5$S$_{8+x}$ with \SI{1.45}{\nm} thickness. Using low-temperature STM, these 2D \vai{}-derived structures were shown to undergo a CDW transition to a \ro{} phase, which develops below \SI{110}{\K}, whereas the thicker islands were found to exhibit different superstructures, presumably all of CDW origin. 

Our findings provide a recipe for the growth of two new single-layer vanadium compounds. Given the thickness dependent magnetic and electronic properties of \vai, having access to its 2D form is of particular interest for the further development of 2D magnetic materials.  

\section{Methods}
The experiments were conducted in five ultra high vacuum systems (base pressure in the low \SI{10^{-10}}{\milli\bar} range). Three systems were equipped with sample preparation facilities, scanning tunneling microscopy (STM) - operating at base temperatures given in the figure captions - and low energy electron diffraction (LEED). The fourth system was the FlexPES beamline endstation preparation chamber at MAX IV, Laboratory Lund, Sweden. There the samples were grown following the same recipes as in the STM system and sample quality was checked by LEED. The fifth system was an STM-only system, to which samples from FlexPES were transferred with a vacuum suitcase and then measured by STM at room temperature and \SI{110}{\K}.

Ir(111) was cleaned by cycles of noble gas sputtering (\SI{4.5}{\kilo\eV} Xe$^+$ at $75\degree$ with respect
to the surface normal or with \SI{1}{\kilo\eV} Ar$^+$ at normal incidence) and annealing to \SI{1500}{\K}. A closed monolayer of single-crystalline Gr on Ir(111) is grown by room temperature exposure of Ir(111) to ethylene until saturation, subsequent annealing to \SI{1500}{\K} and followed by exposure to \SI{200}{\L} ethylene at \SI{1200}{\K}~\cite{VanGastel2009}.

The synthesis of vanadium sulphides on Gr/Ir(111) is based on a two-step molecular beam epitaxy approach introduced in detail in Ref.~\cite{Hall2018} for MoS$_2$. In the first step, the sample is held at room temperature and V is evaporated at a rate of $F_\text{V} = 2.5 \times 10^{16}$ atoms $ \text{m}^{-2} s^{-1}$ into a sulphur background pressure of $P_\text{S}^\text{g} = \SI[parse-numbers=false]{5 \times 10^{-9}}{\milli\bar}$ built up by thermal decomposition of pyrite inside a Knudsen cell. In a second step, the sample is annealed to \SI{800}{\K} with or without S pressure. 

The XPS experiments were conducted at the FlexPES beamline at MAX IV Laboratory, Lund, Sweden~\cite{Preobrajenski2023}. The S 2p core levels were measured with a photon energy of \SI{260}{\eV}. The measurements were conducted at room temperature with a spot size of \SI{50 \times 50}{\um}. The spectra for each sample were fitted simultaneously for all temperatures with pseudo-Voight functions. The width, skew and ratio of Gaussian to Lorentzian contributions were fixed for each component, meaning that they were not allowed to vary between spectra taken at different annealing temperatures. The center energy of each component was granted a maximum deviation of $\pm \SI{100}{\meV}$ to account for small shifts in the chemical environment between spectra while the intensities (total area) of the components were unconstrained. The fitting was performed with the lmfit module of python, using a basinhopping algorithm.

Our first-principles spin-polarized calculations were performed by using density functional theory (DFT)~\mbox{\cite{Hohenberg1964}} and the projector augmented plane wave method~\mbox{\cite{Blochl1994}} as implemented in the VASP code~\mbox{\cite{Kresse1993, Kresse1996}}. For the plane wave expansion of the Kohn–Sham wave functions~\mbox{\cite{Kohn1965}} we used a cutoff energy of \mbox{\SI{500}{\eV}}. We performed the structural relaxation using the vdW-DF2~\mbox{\cite{Lee2010a}} with a revised Becke (B86b) exchange~\mbox{\cite{Becke1986, Hamada2014, Huttmann2015}} functional to properly account for the nonlocal correlation effects like van der Waals interactions~\mbox{\cite{Huttmann2015}}. The analysis of the electronic structures was done by using the PBE exchange-correlation energy functional~\mbox{\cite{Perdew1996}}.

\section{Acknowledgements}
This work was funded by the Deutsche Forschungsgemeinschaft (DFG, German Research foundation) through CRC 1238 (project no. 277146847, subprojects A01, B06 and C01). 
V.B. and J.K. acknowledge financial support from the Swedish Research Council (grant number 2017-04840).
We acknowledge MAX IV Laboratory for time on Beamline FlexPES under Proposal 20210859. Research conducted at MAX IV, a Swedish national user facility, is supported by the Swedish Research council under contract 2018-07152, the Swedish Governmental Agency for Innovation Systems under contract 2018-04969, and Formas under contract 2019-02496.
The authors gratefully acknowledge the computing time granted by the JARA Vergabegremium and provided on the JARA Partition part of the supercomputer JURECA at Forschungszentrum Jülich.

	\bibliography{./library}

\providecommand{\latin}[1]{#1}
\makeatletter
\providecommand{\doi}
  {\begingroup\let\do\@makeother\dospecials
  \catcode`\{=1 \catcode`\}=2 \doi@aux}
\providecommand{\doi@aux}[1]{\endgroup\texttt{#1}}
\makeatother
\providecommand*\mcitethebibliography{\thebibliography}
\csname @ifundefined\endcsname{endmcitethebibliography}
  {\let\endmcitethebibliography\endthebibliography}{}
\begin{mcitethebibliography}{80}
\providecommand*\natexlab[1]{#1}
\providecommand*\mciteSetBstSublistMode[1]{}
\providecommand*\mciteSetBstMaxWidthForm[2]{}
\providecommand*\mciteBstWouldAddEndPuncttrue
  {\def\EndOfBibitem{\unskip.}}
\providecommand*\mciteBstWouldAddEndPunctfalse
  {\let\EndOfBibitem\relax}
\providecommand*\mciteSetBstMidEndSepPunct[3]{}
\providecommand*\mciteSetBstSublistLabelBeginEnd[3]{}
\providecommand*\EndOfBibitem{}
\mciteSetBstSublistMode{f}
\mciteSetBstMaxWidthForm{subitem}{(\alph{mcitesubitemcount})}
\mciteSetBstSublistLabelBeginEnd
  {\mcitemaxwidthsubitemform\space}
  {\relax}
  {\relax}

\bibitem[Ma \latin{et~al.}(2014)Ma, Isarraraz, Wang, Preciado, Klee, Bobek,
  Yamaguchi, Li, Odenthal, and Nguyen]{Ma2014}
Ma,~Q.; Isarraraz,~M.; Wang,~C.~S.; Preciado,~E.; Klee,~V.; Bobek,~S.;
  Yamaguchi,~K.; Li,~E.; Odenthal,~P.~M.; Nguyen,~A. {Postgrowth tuning of the
  bandgap of single-layer molybdenum disulfide films by sulfur/selenium
  exchange}. \emph{ACS Nano} \textbf{2014}, \emph{8}, 4672--4677\relax
\mciteBstWouldAddEndPuncttrue
\mciteSetBstMidEndSepPunct{\mcitedefaultmidpunct}
{\mcitedefaultendpunct}{\mcitedefaultseppunct}\relax
\EndOfBibitem
\bibitem[Lin \latin{et~al.}(2015)Lin, Bj\"orkman, Komsa, Teng, Yeh, Huang, Lin,
  Jadczak, Huang, Chiu, Krasheninnikov, and Suenaga]{Lin2015}
Lin,~Y.-C.; Bj\"orkman,~T.; Komsa,~H.-P.; Teng,~P.-Y.; Yeh,~C.-H.;
  Huang,~F.-S.; Lin,~K.-H.; Jadczak,~J.; Huang,~Y.-S.; Chiu,~P.-W.;
  Krasheninnikov,~A.~V.; Suenaga,~K. {Three-Fold Rotational Defects in
  Two-Dimensional Transition Metal Dichalcogenides}. \emph{Nat. Commun.}
  \textbf{2015}, \emph{6}, 6736\relax
\mciteBstWouldAddEndPuncttrue
\mciteSetBstMidEndSepPunct{\mcitedefaultmidpunct}
{\mcitedefaultendpunct}{\mcitedefaultseppunct}\relax
\EndOfBibitem
\bibitem[Coelho \latin{et~al.}(2018)Coelho, Komsa, {Coy Diaz}, Ma,
  Krasheninnikov, and Batzill]{Coelho2018}
Coelho,~P.~M.; Komsa,~H.-P.; {Coy Diaz},~H.; Ma,~Y.; Krasheninnikov,~A.~V.;
  Batzill,~M. {Post-Synthesis Modifications of Two-Dimensional MoSe$_2$ or
  MoTe$_2$ by Incorporation of Excess Metal Atoms into the Crystal Structure}.
  \emph{ACS Nano} \textbf{2018}, \emph{12}, 3975--3984\relax
\mciteBstWouldAddEndPuncttrue
\mciteSetBstMidEndSepPunct{\mcitedefaultmidpunct}
{\mcitedefaultendpunct}{\mcitedefaultseppunct}\relax
\EndOfBibitem
\bibitem[Huang \latin{et~al.}(2014)Huang, Wu, Sanchez, Peters, Beanland, Ross,
  Rivera, Yao, Cobden, and Xu]{Huang2014}
Huang,~C.; Wu,~S.; Sanchez,~A.~M.; Peters,~J. J.~P.; Beanland,~R.; Ross,~J.~S.;
  Rivera,~P.; Yao,~W.; Cobden,~D.~H.; Xu,~X. {Lateral heterojunctions within
  monolayer MoSe$_2$–WSe$_2$ semiconductors}. \emph{Nat. Mater.}
  \textbf{2014}, \emph{13}, 1096--1101\relax
\mciteBstWouldAddEndPuncttrue
\mciteSetBstMidEndSepPunct{\mcitedefaultmidpunct}
{\mcitedefaultendpunct}{\mcitedefaultseppunct}\relax
\EndOfBibitem
\bibitem[Vaňo \latin{et~al.}(2021)Vaňo, Amini, Ganguli, Chen, Lado,
  Kezilebieke, and Liljeroth]{Vano2021}
Vaňo,~V.; Amini,~M.; Ganguli,~S.~C.; Chen,~G.; Lado,~J.~L.; Kezilebieke,~S.;
  Liljeroth,~P. {Artificial heavy fermions in a van der Waals heterostructure}.
  \emph{Nature} \textbf{2021}, \emph{599}, 582--586\relax
\mciteBstWouldAddEndPuncttrue
\mciteSetBstMidEndSepPunct{\mcitedefaultmidpunct}
{\mcitedefaultendpunct}{\mcitedefaultseppunct}\relax
\EndOfBibitem
\bibitem[Lin \latin{et~al.}(2014)Lin, Dumcenco, Huang, and Suenaga]{Lin2014a}
Lin,~Y.-C.; Dumcenco,~D.~O.; Huang,~Y.-S.; Suenaga,~K. {Atomic mechanism of the
  semiconducting-to-metallic phase transition in single-layered MoS$_2$}.
  \emph{Nat. Nanotechnol.} \textbf{2014}, \emph{9}, 391--396\relax
\mciteBstWouldAddEndPuncttrue
\mciteSetBstMidEndSepPunct{\mcitedefaultmidpunct}
{\mcitedefaultendpunct}{\mcitedefaultseppunct}\relax
\EndOfBibitem
\bibitem[Sutter \latin{et~al.}(2016)Sutter, Huang, Komsa, Ghorbani-Asl,
  Krasheninnikov, and Sutter]{Sutter2016}
Sutter,~E.; Huang,~Y.; Komsa,~H.-P.; Ghorbani-Asl,~M.; Krasheninnikov,~A.~V.;
  Sutter,~P. {Electron-beam induced transformations of layered tin
  dichalcogenides}. \emph{Nano Lett.} \textbf{2016}, \emph{16},
  4410--4416\relax
\mciteBstWouldAddEndPuncttrue
\mciteSetBstMidEndSepPunct{\mcitedefaultmidpunct}
{\mcitedefaultendpunct}{\mcitedefaultseppunct}\relax
\EndOfBibitem
\bibitem[Wang \latin{et~al.}(2017)Wang, Xiao, Zhu, Li, Alsaid, Fong, Zhou,
  Wang, Shi, and Wang]{Wang2017}
Wang,~Y.; Xiao,~J.; Zhu,~H.; Li,~Y.; Alsaid,~Y.; Fong,~K.~Y.; Zhou,~Y.;
  Wang,~S.; Shi,~W.; Wang,~Y. {Structural phase transition in monolayer
  MoTe$_2$ driven by electrostatic doping}. \emph{Nature} \textbf{2017},
  \emph{550}, 487--491\relax
\mciteBstWouldAddEndPuncttrue
\mciteSetBstMidEndSepPunct{\mcitedefaultmidpunct}
{\mcitedefaultendpunct}{\mcitedefaultseppunct}\relax
\EndOfBibitem
\bibitem[van Efferen \latin{et~al.}(2022)van Efferen, Murray, Fischer, Busse,
  Komsa, Michely, and Jolie]{VanEfferen2022}
van Efferen,~C.; Murray,~C.; Fischer,~J.; Busse,~C.; Komsa,~H.-P.; Michely,~T.;
  Jolie,~W. {Metal-insulator transition in monolayer MoS$_2$ via contactless
  chemical doping}. \emph{2D Mater.} \textbf{2022}, \emph{9}, 25026\relax
\mciteBstWouldAddEndPuncttrue
\mciteSetBstMidEndSepPunct{\mcitedefaultmidpunct}
{\mcitedefaultendpunct}{\mcitedefaultseppunct}\relax
\EndOfBibitem
\bibitem[Friend and Yoffe(1987)Friend, and Yoffe]{Friend1987}
Friend,~R.~H.; Yoffe,~A.~D. {Electronic properties of intercalation complexes
  of the transition metal dichalcogenides}. \emph{Adv. Phys.} \textbf{1987},
  \emph{36}, 1--94\relax
\mciteBstWouldAddEndPuncttrue
\mciteSetBstMidEndSepPunct{\mcitedefaultmidpunct}
{\mcitedefaultendpunct}{\mcitedefaultseppunct}\relax
\EndOfBibitem
\bibitem[Wan \latin{et~al.}(2016)Wan, Lacey, Dai, Bao, Fuhrer, and Hu]{Wan2016}
Wan,~J.; Lacey,~S.~D.; Dai,~J.; Bao,~W.; Fuhrer,~M.~S.; Hu,~L. {Tuning
  two-dimensional nanomaterials by intercalation: materials, properties and
  applications}. \emph{Chem. Soc. Rev.} \textbf{2016}, \emph{45},
  6742--6765\relax
\mciteBstWouldAddEndPuncttrue
\mciteSetBstMidEndSepPunct{\mcitedefaultmidpunct}
{\mcitedefaultendpunct}{\mcitedefaultseppunct}\relax
\EndOfBibitem
\bibitem[Hu \latin{et~al.}(2018)Hu, Wu, Han, He, Ni, and Chen]{Hu2018}
Hu,~Z.; Wu,~Z.; Han,~C.; He,~J.; Ni,~Z.; Chen,~W. {Two-dimensional transition
  metal dichalcogenides: interface and defect engineering}. \emph{Chem. Soc.
  Rev.} \textbf{2018}, \emph{47}, 3100--3128\relax
\mciteBstWouldAddEndPuncttrue
\mciteSetBstMidEndSepPunct{\mcitedefaultmidpunct}
{\mcitedefaultendpunct}{\mcitedefaultseppunct}\relax
\EndOfBibitem
\bibitem[Komsa \latin{et~al.}(2012)Komsa, Kotakoski, Kurasch, Lehtinen, Kaiser,
  and Krasheninnikov]{Komsa2012a}
Komsa,~H.-P.; Kotakoski,~J.; Kurasch,~S.; Lehtinen,~O.; Kaiser,~U.;
  Krasheninnikov,~A.~V. {Two-Dimensional Transition Metal Dichalcogenides under
  Electron Irradiation: Defect Production and Doping}. \emph{Phys. Rev. Lett.}
  \textbf{2012}, \emph{109}, 35503\relax
\mciteBstWouldAddEndPuncttrue
\mciteSetBstMidEndSepPunct{\mcitedefaultmidpunct}
{\mcitedefaultendpunct}{\mcitedefaultseppunct}\relax
\EndOfBibitem
\bibitem[Yu \latin{et~al.}(2019)Yu, Li, Herng, Wang, Zhao, Chi, Fu, Abdelwahab,
  Zhou, Dan, Chen, Chen, Li, Lu, Pennycook, Feng, Ding, and Loh]{Yu2019}
Yu,~W. \latin{et~al.}  {Chemically Exfoliated VSe$_2$ Monolayers with
  Room-Temperature Ferromagnetism}. \emph{Adv. Mater.} \textbf{2019},
  \emph{31}, 1903779\relax
\mciteBstWouldAddEndPuncttrue
\mciteSetBstMidEndSepPunct{\mcitedefaultmidpunct}
{\mcitedefaultendpunct}{\mcitedefaultseppunct}\relax
\EndOfBibitem
\bibitem[Chua \latin{et~al.}(2020)Chua, Yang, He, Yu, Yu, Bussolotti, Wong,
  Loh, Breese, Goh, Huang, and Wee]{Chua2020}
Chua,~R.; Yang,~J.; He,~X.; Yu,~X.; Yu,~W.; Bussolotti,~F.; Wong,~P. K.~J.;
  Loh,~K.~P.; Breese,~M. B.~H.; Goh,~K. E.~J.; Huang,~Y.~L.; Wee,~A. T.~S. {Can
  Reconstructed Se-Deficient Line Defects in Monolayer VSe$_2$ Induce
  Magnetism?} \emph{Adv. Mater.} \textbf{2020}, \emph{32}, 2000693\relax
\mciteBstWouldAddEndPuncttrue
\mciteSetBstMidEndSepPunct{\mcitedefaultmidpunct}
{\mcitedefaultendpunct}{\mcitedefaultseppunct}\relax
\EndOfBibitem
\bibitem[Liu \latin{et~al.}(2019)Liu, Lei, Zhu, Tao, Qi, Bao, Wu, Huang, Zhang,
  Lin, Wang, Du, Pantelides, and Gao]{Liu2019}
Liu,~Z.-L.; Lei,~B.; Zhu,~Z.-L.; Tao,~L.; Qi,~J.; Bao,~D.-L.; Wu,~X.;
  Huang,~L.; Zhang,~Y.-Y.; Lin,~X.; Wang,~Y.-L.; Du,~S.; Pantelides,~S.~T.;
  Gao,~H.-J. {Spontaneous Formation of 1D Pattern in Monolayer VSe$_2$ with
  Dispersive Adsorption of Pt Atoms for HER Catalysis}. \emph{Nano Lett.}
  \textbf{2019}, \emph{19}, 4897--4903\relax
\mciteBstWouldAddEndPuncttrue
\mciteSetBstMidEndSepPunct{\mcitedefaultmidpunct}
{\mcitedefaultendpunct}{\mcitedefaultseppunct}\relax
\EndOfBibitem
\bibitem[Wang \latin{et~al.}(2020)Wang, Zhang, Si, Zhang, Wu, Gao, Wei, Sun,
  Liao, and Zhang]{Wang2020}
Wang,~X.; Zhang,~Y.; Si,~H.; Zhang,~Q.; Wu,~J.; Gao,~L.; Wei,~X.; Sun,~Y.;
  Liao,~Q.; Zhang,~Z. {Single-atom vacancy defect to trigger high-efficiency
  hydrogen evolution of MoS$_2$}. \emph{JACS} \textbf{2020}, \emph{142},
  4298--4308\relax
\mciteBstWouldAddEndPuncttrue
\mciteSetBstMidEndSepPunct{\mcitedefaultmidpunct}
{\mcitedefaultendpunct}{\mcitedefaultseppunct}\relax
\EndOfBibitem
\bibitem[Lin \latin{et~al.}(2015)Lin, Pantelides, and Zhou]{Lin2015a}
Lin,~J.; Pantelides,~S.~T.; Zhou,~W. {Vacancy-Induced Formation and Growth of
  Inversion Domains in Transition-Metal Dichalcogenide Monolayer}. \emph{ACS
  Nano} \textbf{2015}, \emph{9}, 5189--5197\relax
\mciteBstWouldAddEndPuncttrue
\mciteSetBstMidEndSepPunct{\mcitedefaultmidpunct}
{\mcitedefaultendpunct}{\mcitedefaultseppunct}\relax
\EndOfBibitem
\bibitem[Liu \latin{et~al.}(2023)Liu, Zhao, Sun, Deng, Zou, Hu, Wang, Chu, Li,
  Wu, Ke, and Ajayan]{Liu2023}
Liu,~X.-C.; Zhao,~S.; Sun,~X.; Deng,~L.; Zou,~X.; Hu,~Y.; Wang,~Y.-X.;
  Chu,~C.-W.; Li,~J.; Wu,~J.; Ke,~F.-S.; Ajayan,~P.~M. {Spontaneous
  self-intercalation of copper atoms into transition metal dichalcogenides}.
  \emph{Sci. Adv.} \textbf{2023}, \emph{6}, eaay4092\relax
\mciteBstWouldAddEndPuncttrue
\mciteSetBstMidEndSepPunct{\mcitedefaultmidpunct}
{\mcitedefaultendpunct}{\mcitedefaultseppunct}\relax
\EndOfBibitem
\bibitem[Zhao \latin{et~al.}(2020)Zhao, Song, Wang, Riis-Jensen, Fu, Deng, Wan,
  Kang, Ning, Dan, Venkatesan, Liu, Zhou, Thygesen, Luo, Pennycook, and
  Loh]{Zhao2020}
Zhao,~X. \latin{et~al.}  {Engineering covalently bonded 2D layered materials by
  self-intercalation}. \emph{Nature} \textbf{2020}, \emph{581}, 171--177\relax
\mciteBstWouldAddEndPuncttrue
\mciteSetBstMidEndSepPunct{\mcitedefaultmidpunct}
{\mcitedefaultendpunct}{\mcitedefaultseppunct}\relax
\EndOfBibitem
\bibitem[Wang \latin{et~al.}(2014)Wang, Shen, Wang, Yu, and Chen]{Wang2014}
Wang,~X.; Shen,~X.; Wang,~Z.; Yu,~R.; Chen,~L. {Atomic-scale clarification of
  structural transition of MoS$_2$ upon sodium intercalation}. \emph{ACS Nano}
  \textbf{2014}, \emph{8}, 11394--11400\relax
\mciteBstWouldAddEndPuncttrue
\mciteSetBstMidEndSepPunct{\mcitedefaultmidpunct}
{\mcitedefaultendpunct}{\mcitedefaultseppunct}\relax
\EndOfBibitem
\bibitem[Tan \latin{et~al.}(2017)Tan, Abdelwahab, Ding, Zhao, Yang, Loke, Lin,
  Verzhbitskiy, Poh, and Xu]{Tan2017}
Tan,~S. J.~R.; Abdelwahab,~I.; Ding,~Z.; Zhao,~X.; Yang,~T.; Loke,~G. Z.~J.;
  Lin,~H.; Verzhbitskiy,~I.; Poh,~S.~M.; Xu,~H. {Chemical stabilization of 1T'
  phase transition metal dichalcogenides with giant optical Kerr nonlinearity}.
  \emph{JACS} \textbf{2017}, \emph{139}, 2504--2511\relax
\mciteBstWouldAddEndPuncttrue
\mciteSetBstMidEndSepPunct{\mcitedefaultmidpunct}
{\mcitedefaultendpunct}{\mcitedefaultseppunct}\relax
\EndOfBibitem
\bibitem[Zhang \latin{et~al.}(2022)Zhang, Rousuli, Zhang, Luo, Guo, Cong, Lin,
  Bao, Zhang, Xu, Feng, Shen, Zhao, Yao, Wu, Ji, Chen, Tan, Xue, Xu, Duan, Yu,
  and Zhou]{Zhang2022}
Zhang,~H. \latin{et~al.}  {Tailored Ising superconductivity in intercalated
  bulk NbSe$_2$}. \emph{Nat. Phys.} \textbf{2022}, \emph{18}, 1425--1430\relax
\mciteBstWouldAddEndPuncttrue
\mciteSetBstMidEndSepPunct{\mcitedefaultmidpunct}
{\mcitedefaultendpunct}{\mcitedefaultseppunct}\relax
\EndOfBibitem
\bibitem[Kanetani \latin{et~al.}(2012)Kanetani, Sugawara, Sato, Shimizu, Iwaya,
  Hitosugi, and Takahashi]{Kanetani2012}
Kanetani,~K.; Sugawara,~K.; Sato,~T.; Shimizu,~R.; Iwaya,~K.; Hitosugi,~T.;
  Takahashi,~T. {Ca intercalated bilayer graphene as a thinnest limit of
  superconducting C$_6$Ca}. \emph{PNAS} \textbf{2012}, \emph{109},
  19610--19613\relax
\mciteBstWouldAddEndPuncttrue
\mciteSetBstMidEndSepPunct{\mcitedefaultmidpunct}
{\mcitedefaultendpunct}{\mcitedefaultseppunct}\relax
\EndOfBibitem
\bibitem[Lasek \latin{et~al.}(2020)Lasek, Coelho, Zberecki, Xin, Kolekar, Li,
  and Batzill]{Lasek2020}
Lasek,~K.; Coelho,~P.~M.; Zberecki,~K.; Xin,~Y.; Kolekar,~S.~K.; Li,~J.;
  Batzill,~M. {Molecular Beam Epitaxy of Transition Metal (Ti-, V-, and Cr-)
  Tellurides: From Monolayer Ditellurides to Multilayer Self-Intercalation
  Compounds}. \emph{ACS Nano} \textbf{2020}, \emph{14}, 8473--8484\relax
\mciteBstWouldAddEndPuncttrue
\mciteSetBstMidEndSepPunct{\mcitedefaultmidpunct}
{\mcitedefaultendpunct}{\mcitedefaultseppunct}\relax
\EndOfBibitem
\bibitem[Ma \latin{et~al.}(2012)Ma, Dai, Guo, Niu, Zhu, and Huang]{Ma2012}
Ma,~Y.; Dai,~Y.; Guo,~M.; Niu,~C.; Zhu,~Y.; Huang,~B. {Evidence of the
  Existence of Magnetism in Pristine $\mathrm{VX}_{2}$ Monolayers (X = S, Se)
  and Their Strain-Induced Tunable Magnetic Properties}. \emph{ACS Nano}
  \textbf{2012}, \emph{6}, 1695--1701\relax
\mciteBstWouldAddEndPuncttrue
\mciteSetBstMidEndSepPunct{\mcitedefaultmidpunct}
{\mcitedefaultendpunct}{\mcitedefaultseppunct}\relax
\EndOfBibitem
\bibitem[Zhang \latin{et~al.}(2013)Zhang, Liu, and Lau]{Zhang2013}
Zhang,~H.; Liu,~L.-M.; Lau,~W.-M. {Dimension-dependent phase transition and
  magnetic properties of $\mathrm{VS}_{2}$}. \emph{J. Mater. Chem. A}
  \textbf{2013}, \emph{1}, 10821--10828\relax
\mciteBstWouldAddEndPuncttrue
\mciteSetBstMidEndSepPunct{\mcitedefaultmidpunct}
{\mcitedefaultendpunct}{\mcitedefaultseppunct}\relax
\EndOfBibitem
\bibitem[Isaacs and Marianetti(2016)Isaacs, and Marianetti]{Isaacs2016}
Isaacs,~E.~B.; Marianetti,~C.~A. {Electronic Correlations in Monolayer
  $\mathrm{VS}_{2}$}. \emph{Phys. Rev. B} \textbf{2016}, \emph{94}, 35120\relax
\mciteBstWouldAddEndPuncttrue
\mciteSetBstMidEndSepPunct{\mcitedefaultmidpunct}
{\mcitedefaultendpunct}{\mcitedefaultseppunct}\relax
\EndOfBibitem
\bibitem[Zhuang and Hennig(2016)Zhuang, and Hennig]{Zhuang2016}
Zhuang,~H.~L.; Hennig,~R.~G. {Stability and magnetism of strongly correlated
  single-layer $\mathrm{VS}_{2}$}. \emph{Phys. Rev. B} \textbf{2016},
  \emph{93}, 54429\relax
\mciteBstWouldAddEndPuncttrue
\mciteSetBstMidEndSepPunct{\mcitedefaultmidpunct}
{\mcitedefaultendpunct}{\mcitedefaultseppunct}\relax
\EndOfBibitem
\bibitem[Mulazzi \latin{et~al.}(2010)Mulazzi, Chainani, Katayama, Eguchi,
  Matsunami, Ohashi, Senba, Nohara, Uchida, Takagi, and Shin]{Mulazzi2010}
Mulazzi,~M.; Chainani,~A.; Katayama,~N.; Eguchi,~R.; Matsunami,~M.; Ohashi,~H.;
  Senba,~Y.; Nohara,~M.; Uchida,~M.; Takagi,~H.; Shin,~S. {Absence of nesting
  in the charge-density-wave system 1T-$\mathrm{VS}_{2}$ as seen by
  photoelectron spectroscopy}. \emph{Phys. Rev. B} \textbf{2010}, \emph{82},
  75130\relax
\mciteBstWouldAddEndPuncttrue
\mciteSetBstMidEndSepPunct{\mcitedefaultmidpunct}
{\mcitedefaultendpunct}{\mcitedefaultseppunct}\relax
\EndOfBibitem
\bibitem[Gauzzi \latin{et~al.}(2014)Gauzzi, Sellam, Rousse, Klein, Taverna,
  Giura, Calandra, Loupias, Gozzo, Gilioli, Bolzoni, Allodi, {De Renzi},
  Calestani, and Roy]{Gauzzi2014}
Gauzzi,~A.; Sellam,~A.; Rousse,~G.; Klein,~Y.; Taverna,~D.; Giura,~P.;
  Calandra,~M.; Loupias,~G.; Gozzo,~F.; Gilioli,~E.; Bolzoni,~F.; Allodi,~G.;
  {De Renzi},~R.; Calestani,~G.~L.; Roy,~P. {Possible Phase Separation and Weak
  Localization in the Absence of a Charge-Density Wave in Single-Phase
  1T-$\mathrm{VS}_{2}$}. \emph{Phys. Rev. B} \textbf{2014}, \emph{89},
  235125\relax
\mciteBstWouldAddEndPuncttrue
\mciteSetBstMidEndSepPunct{\mcitedefaultmidpunct}
{\mcitedefaultendpunct}{\mcitedefaultseppunct}\relax
\EndOfBibitem
\bibitem[Arnold \latin{et~al.}(2018)Arnold, Stan, Mahatha, Lund, Curcio,
  Dendzik, Bana, Travaglia, Bignardi, Lacovig, Lizzit, Li, Bianchi, Miwa,
  Bremholm, Lizzit, Hofmann, and Sanders]{Arnold2018}
Arnold,~F. \latin{et~al.}  {Novel Single-Layer Vanadium Sulphide Phases}.
  \emph{2D Mater.} \textbf{2018}, \emph{5}, 045009\relax
\mciteBstWouldAddEndPuncttrue
\mciteSetBstMidEndSepPunct{\mcitedefaultmidpunct}
{\mcitedefaultendpunct}{\mcitedefaultseppunct}\relax
\EndOfBibitem
\bibitem[van Efferen \latin{et~al.}(2021)van Efferen, Berges, Hall, van Loon,
  Kraus, Schobert, Wekking, Huttmann, Plaar, Rothenbach, Ollefs, Arruda,
  Brookes, Sch{\"{o}}nhoff, Kummer, Wende, Wehling, and
  Michely]{VanEfferen2021}
van Efferen,~C. \latin{et~al.}  {A full gap above the Fermi level: the charge
  density wave of monolayer VS$_2$}. \emph{Nat. Comm.} \textbf{2021},
  \emph{12}, 6837\relax
\mciteBstWouldAddEndPuncttrue
\mciteSetBstMidEndSepPunct{\mcitedefaultmidpunct}
{\mcitedefaultendpunct}{\mcitedefaultseppunct}\relax
\EndOfBibitem
\bibitem[Coelho \latin{et~al.}(2019)Coelho, {Nguyen Cong}, Bonilla, Kolekar,
  Phan, Avila, Asensio, Oleynik, and Batzill]{Coelho2019}
Coelho,~P.~M.; {Nguyen Cong},~K.; Bonilla,~M.; Kolekar,~S.; Phan,~M.-H.;
  Avila,~J.; Asensio,~M.~C.; Oleynik,~I.~I.; Batzill,~M. {Charge Density Wave
  State Suppresses Ferromagnetic Ordering in $\mathrm{VSe}_{2}$ Monolayers}.
  \emph{J. Phys. Chem. C} \textbf{2019}, \emph{123}, 14089--14096\relax
\mciteBstWouldAddEndPuncttrue
\mciteSetBstMidEndSepPunct{\mcitedefaultmidpunct}
{\mcitedefaultendpunct}{\mcitedefaultseppunct}\relax
\EndOfBibitem
\bibitem[Hardy \latin{et~al.}(2016)Hardy, Yuan, Guo, Zhou, Lou, and
  Natelson]{Hardy2016}
Hardy,~W.~J.; Yuan,~J.; Guo,~H.; Zhou,~P.; Lou,~J.; Natelson,~D.
  {Thickness-Dependent and Magnetic-Field-Driven Suppression of
  Antiferromagnetic Order in Thin V$_5$S$_8$ Single Crystals}. \emph{ACS Nano}
  \textbf{2016}, \emph{10}, 5941--5946\relax
\mciteBstWouldAddEndPuncttrue
\mciteSetBstMidEndSepPunct{\mcitedefaultmidpunct}
{\mcitedefaultendpunct}{\mcitedefaultseppunct}\relax
\EndOfBibitem
\bibitem[Niu \latin{et~al.}(2017)Niu, Yan, Ji, Liu, Li, Gao, Zhang, Yu, and
  Wu]{Niu2017}
Niu,~J.; Yan,~B.; Ji,~Q.; Liu,~Z.; Li,~M.; Gao,~P.; Zhang,~Y.; Yu,~D.; Wu,~X.
  {Anomalous Hall effect and magnetic orderings in nanothick V$_5$S$_8$}.
  \emph{Phys. Rev. B} \textbf{2017}, \emph{96}, 75402\relax
\mciteBstWouldAddEndPuncttrue
\mciteSetBstMidEndSepPunct{\mcitedefaultmidpunct}
{\mcitedefaultendpunct}{\mcitedefaultseppunct}\relax
\EndOfBibitem
\bibitem[Zhang \latin{et~al.}(2020)Zhang, Zhang, and Du]{Zhang2020}
Zhang,~R.-Z.; Zhang,~Y.-Y.; Du,~S.-X. {Thickness-dependent magnetic order and
  phase transition in V$_5$S$_8$}. \emph{Chin. Phys. B} \textbf{2020},
  \emph{29}, 77504\relax
\mciteBstWouldAddEndPuncttrue
\mciteSetBstMidEndSepPunct{\mcitedefaultmidpunct}
{\mcitedefaultendpunct}{\mcitedefaultseppunct}\relax
\EndOfBibitem
\bibitem[Nozaki \latin{et~al.}(1975)Nozaki, Ishizawa, Saeki, and
  Nakahira]{Nozaki1975}
Nozaki,~H.; Ishizawa,~Y.; Saeki,~M.; Nakahira,~M. {Electrical properties of
  V$_5$S$_8$ single crystals}. \emph{Phys. Lett. A} \textbf{1975}, \emph{54},
  29--30\relax
\mciteBstWouldAddEndPuncttrue
\mciteSetBstMidEndSepPunct{\mcitedefaultmidpunct}
{\mcitedefaultendpunct}{\mcitedefaultseppunct}\relax
\EndOfBibitem
\bibitem[Moutaabbid \latin{et~al.}(2016)Moutaabbid, {Le Godec}, Taverna,
  Baptiste, Klein, Loupias, and Gauzzi]{Moutaabbid2016}
Moutaabbid,~H.; {Le Godec},~Y.; Taverna,~D.; Baptiste,~B.; Klein,~Y.;
  Loupias,~G.; Gauzzi,~A. {High-Pressure Control of Vanadium Self-Intercalation
  and Enhanced Metallic Properties in 1T-$\mathrm{V}_{1+x}\mathrm{S}_{2}$
  Single Crystals}. \emph{Inorg. Chem.} \textbf{2016}, \emph{55},
  6481--6486\relax
\mciteBstWouldAddEndPuncttrue
\mciteSetBstMidEndSepPunct{\mcitedefaultmidpunct}
{\mcitedefaultendpunct}{\mcitedefaultseppunct}\relax
\EndOfBibitem
\bibitem[Bensch and Koy(1993)Bensch, and Koy]{Bensch1993}
Bensch,~W.; Koy,~J. {The single crystal structure of V$_5$S$_8$ determined at
  two different temperatures: anisotropic changes of the metal atom network}.
  \emph{Inorganica Chim. Acta} \textbf{1993}, \emph{206}, 221--223\relax
\mciteBstWouldAddEndPuncttrue
\mciteSetBstMidEndSepPunct{\mcitedefaultmidpunct}
{\mcitedefaultendpunct}{\mcitedefaultseppunct}\relax
\EndOfBibitem
\bibitem[Niu \latin{et~al.}(2020)Niu, Zhang, Li, Yang, Yan, Chen, Zhang, Zhang,
  Ren, and Gao]{Niu2020}
Niu,~J.; Zhang,~W.; Li,~Z.; Yang,~S.; Yan,~D.; Chen,~S.; Zhang,~Z.; Zhang,~Y.;
  Ren,~X.; Gao,~P. {Intercalation of van der Waals layered materials: A route
  towards engineering of electron correlation}. \emph{Chin. Phys. B}
  \textbf{2020}, \emph{29}, 97104\relax
\mciteBstWouldAddEndPuncttrue
\mciteSetBstMidEndSepPunct{\mcitedefaultmidpunct}
{\mcitedefaultendpunct}{\mcitedefaultseppunct}\relax
\EndOfBibitem
\bibitem[Zhou \latin{et~al.}(2022)Zhou, Zhao, Wu, Liu, Chen, Xi, Wang, Liu,
  Zhou, and Pennycook]{Zhou2022}
Zhou,~Z.; Zhao,~X.; Wu,~L.; Liu,~H.; Chen,~J.; Xi,~C.; Wang,~Z.; Liu,~E.;
  Zhou,~W.; Pennycook,~S.~J. {Dimensional crossover in self-intercalated
  antiferromagnetic V$_5$S$_8$ nanoflakes}. \emph{Phys. Rev. B} \textbf{2022},
  \emph{105}, 235433\relax
\mciteBstWouldAddEndPuncttrue
\mciteSetBstMidEndSepPunct{\mcitedefaultmidpunct}
{\mcitedefaultendpunct}{\mcitedefaultseppunct}\relax
\EndOfBibitem
\bibitem[Hall \latin{et~al.}(2018)Hall, Pieli{\'{c}}, Murray, Jolie, Wekking,
  Busse, Kralj, and Michely]{Hall2018}
Hall,~J.; Pieli{\'{c}},~B.; Murray,~C.; Jolie,~W.; Wekking,~T.; Busse,~C.;
  Kralj,~M.; Michely,~T. {Molecular Beam Epitaxy of Quasi-Freestanding
  Transition Metal Disulphide Monolayers on van der Waals Substrates: A Growth
  Study}. \emph{2D Mater.} \textbf{2018}, \emph{5}, 025005\relax
\mciteBstWouldAddEndPuncttrue
\mciteSetBstMidEndSepPunct{\mcitedefaultmidpunct}
{\mcitedefaultendpunct}{\mcitedefaultseppunct}\relax
\EndOfBibitem
\bibitem[Murray \latin{et~al.}(2019)Murray, Jolie, Fischer, Hall, van Efferen,
  Ehlen, Gr\"uneis, Busse, and Michely]{Murray2019}
Murray,~C.; Jolie,~W.; Fischer,~J.~A.; Hall,~J.; van Efferen,~C.; Ehlen,~N.;
  Gr\"uneis,~A.; Busse,~C.; Michely,~T. {Comprehensive Tunneling Spectroscopy
  of Quasifreestanding ${\mathrm{MoS}}_{2}$ on Graphene on Ir(111)}.
  \emph{Phys. Rev. B} \textbf{2019}, \emph{99}, 115434\relax
\mciteBstWouldAddEndPuncttrue
\mciteSetBstMidEndSepPunct{\mcitedefaultmidpunct}
{\mcitedefaultendpunct}{\mcitedefaultseppunct}\relax
\EndOfBibitem
\bibitem[Kawada \latin{et~al.}(1975)Kawada, Nakano-Onoda, Ishii, Saeki, and
  Nakahira]{Kawada1975}
Kawada,~I.; Nakano-Onoda,~M.; Ishii,~M.; Saeki,~M.; Nakahira,~M. {Crystal
  structures of V$_3$S$_4$ and V$_5$S$_8$}. \emph{J. Solid State Chem.}
  \textbf{1975}, \emph{15}, 246--252\relax
\mciteBstWouldAddEndPuncttrue
\mciteSetBstMidEndSepPunct{\mcitedefaultmidpunct}
{\mcitedefaultendpunct}{\mcitedefaultseppunct}\relax
\EndOfBibitem
\bibitem[Pieli{\'{c}} \latin{et~al.}(2020)Pieli{\'{c}}, Hall, Despoja,
  Raki{\'{c}}, Petrovi{\'{c}}, Sohani, Busse, Michely, and Kralj]{Pielic2020}
Pieli{\'{c}},~B.; Hall,~J.; Despoja,~V.; Raki{\'{c}},~I.~{\v{S}}.;
  Petrovi{\'{c}},~M.; Sohani,~A.; Busse,~C.; Michely,~T.; Kralj,~M. {Sulfur
  Structures on Bare and Graphene-Covered Ir(111)}. \emph{J. Phys. Chem. C.}
  \textbf{2020}, \emph{124}, 6659--6668\relax
\mciteBstWouldAddEndPuncttrue
\mciteSetBstMidEndSepPunct{\mcitedefaultmidpunct}
{\mcitedefaultendpunct}{\mcitedefaultseppunct}\relax
\EndOfBibitem
\bibitem[Kamber \latin{et~al.}(2021)Kamber, Pakdel, Stan, Kamlapure, Kiraly,
  Arnold, Eich, Ngankeu, Bianchi, and Miwa]{Kamber2021}
Kamber,~U.; Pakdel,~S.; Stan,~R.-M.; Kamlapure,~A.; Kiraly,~B.; Arnold,~F.;
  Eich,~A.; Ngankeu,~A.~S.; Bianchi,~M.; Miwa,~J.~A. {Moir{\'{e}}-induced
  electronic structure modifications in monolayer V$_2$S$_3$ on Au (111)}.
  \emph{Phys. Rev. B} \textbf{2021}, \emph{103}, 115414\relax
\mciteBstWouldAddEndPuncttrue
\mciteSetBstMidEndSepPunct{\mcitedefaultmidpunct}
{\mcitedefaultendpunct}{\mcitedefaultseppunct}\relax
\EndOfBibitem
\bibitem[Komsa \latin{et~al.}(2013)Komsa, Kurasch, Lehtinen, Kaiser, and
  Krasheninnikov]{Komsa2013}
Komsa,~H.-P.; Kurasch,~S.; Lehtinen,~O.; Kaiser,~U.; Krasheninnikov,~A.~V.
  {From Point to Extended Defects in Two-Dimensional $\mathrm{MoS}_{2}$:
  Evolution of Atomic Structure Under Electron Irradiation}. \emph{Phys. Rev.
  B} \textbf{2013}, \emph{88}, 035301\relax
\mciteBstWouldAddEndPuncttrue
\mciteSetBstMidEndSepPunct{\mcitedefaultmidpunct}
{\mcitedefaultendpunct}{\mcitedefaultseppunct}\relax
\EndOfBibitem
\bibitem[Lu \latin{et~al.}(2015)Lu, Carvalho, Chan, Liu, Liu, Tok, Loh, {Castro
  Neto}, and Sow]{Lu2015a}
Lu,~J.; Carvalho,~A.; Chan,~X.~K.; Liu,~H.; Liu,~B.; Tok,~E.~S.; Loh,~K.~P.;
  {Castro Neto},~A.~H.; Sow,~C.~H. {Atomic Healing of Defects in Transition
  Metal Dichalcogenides}. \emph{Nano Lett.} \textbf{2015}, \emph{15},
  3524--3532\relax
\mciteBstWouldAddEndPuncttrue
\mciteSetBstMidEndSepPunct{\mcitedefaultmidpunct}
{\mcitedefaultendpunct}{\mcitedefaultseppunct}\relax
\EndOfBibitem
\bibitem[Elibol \latin{et~al.}(2018)Elibol, Susi, Argentero, {Reza Ahmadpour
  Monazam}, Pennycook, Meyer, and Kotakoski]{Elibol2018}
Elibol,~K.; Susi,~T.; Argentero,~G.; {Reza Ahmadpour Monazam},~M.;
  Pennycook,~T.~J.; Meyer,~J.~C.; Kotakoski,~J. {Atomic Structure of Intrinsic
  and Electron-Irradiation-Induced Defects in MoTe$_2$}. \emph{Chem. Mater.}
  \textbf{2018}, \emph{30}, 1230--1238\relax
\mciteBstWouldAddEndPuncttrue
\mciteSetBstMidEndSepPunct{\mcitedefaultmidpunct}
{\mcitedefaultendpunct}{\mcitedefaultseppunct}\relax
\EndOfBibitem
\bibitem[Zhao \latin{et~al.}(2019)Zhao, Ji, Chen, Fu, Dan, Liu, Pennycook,
  Zhou, and Loh]{Zhao2019}
Zhao,~X.; Ji,~Y.; Chen,~J.; Fu,~W.; Dan,~J.; Liu,~Y.; Pennycook,~S.~J.;
  Zhou,~W.; Loh,~K.~P. {Healing of Planar Defects in 2D Materials via Grain
  Boundary Sliding}. \emph{Adv. Mater.} \textbf{2019}, \emph{31}, 1900237\relax
\mciteBstWouldAddEndPuncttrue
\mciteSetBstMidEndSepPunct{\mcitedefaultmidpunct}
{\mcitedefaultendpunct}{\mcitedefaultseppunct}\relax
\EndOfBibitem
\bibitem[Wong \latin{et~al.}(2019)Wong, Zhang, Bussolotti, Yin, Herng, Zhang,
  Huang, Vinai, Krishnamurthi, Bukhvalov, Zheng, Chua, N'Diaye, Morton, Yang,
  {Ou Yang}, Torelli, Chen, Goh, Ding, Lin, Brocks, de~Jong, {Castro Neto}, and
  Wee]{Wong2019}
Wong,~P. K.~J. \latin{et~al.}  {Evidence of Spin Frustration in a Vanadium
  Diselenide Monolayer Magnet}. \emph{Adv. Mater.} \textbf{2019}, \emph{31},
  1901185\relax
\mciteBstWouldAddEndPuncttrue
\mciteSetBstMidEndSepPunct{\mcitedefaultmidpunct}
{\mcitedefaultendpunct}{\mcitedefaultseppunct}\relax
\EndOfBibitem
\bibitem[Murphy \latin{et~al.}(1977)Murphy, Cros, {Di Salvo}, and
  Waszczak]{Murphy1977}
Murphy,~D.~W.; Cros,~C.; {Di Salvo},~F.~J.; Waszczak,~J.~V. {Preparation and
  Properties of $\mathrm{Li}_{x}\mathrm{VS}_{2}$}. \emph{Inorg. Chem.}
  \textbf{1977}, \emph{16}, 3027--3031\relax
\mciteBstWouldAddEndPuncttrue
\mciteSetBstMidEndSepPunct{\mcitedefaultmidpunct}
{\mcitedefaultendpunct}{\mcitedefaultseppunct}\relax
\EndOfBibitem
\bibitem[Bonilla \latin{et~al.}(2020)Bonilla, Kolekar, Li, Xin, Coelho, Lasek,
  Zberecki, Lizzit, Tosi, Lacovig, Lizzit, and Batzill]{Bonilla2020}
Bonilla,~M.; Kolekar,~S.; Li,~J.; Xin,~Y.; Coelho,~P.~M.; Lasek,~K.;
  Zberecki,~K.; Lizzit,~D.; Tosi,~E.; Lacovig,~P.; Lizzit,~S.; Batzill,~M.
  {Compositional Phase Change of Early Transition Metal Diselenide (VSe$_2$ and
  TiSe$_2$) Ultrathin Films by Postgrowth Annealing}. \emph{Adv. Mater.
  Interfaces} \textbf{2020}, \emph{7}, 2000497\relax
\mciteBstWouldAddEndPuncttrue
\mciteSetBstMidEndSepPunct{\mcitedefaultmidpunct}
{\mcitedefaultendpunct}{\mcitedefaultseppunct}\relax
\EndOfBibitem
\bibitem[Nakano \latin{et~al.}(2019)Nakano, Wang, Yoshida, Matsuoka, Majima,
  Ikeda, Hirata, Takeda, Wadati, Kohama, Ohigashi, Sakano, Ishizaka, and
  Iwasa]{Nakano2019}
Nakano,~M.; Wang,~Y.; Yoshida,~S.; Matsuoka,~H.; Majima,~Y.; Ikeda,~K.;
  Hirata,~Y.; Takeda,~Y.; Wadati,~H.; Kohama,~Y.; Ohigashi,~Y.; Sakano,~M.;
  Ishizaka,~K.; Iwasa,~Y. {Intrinsic 2D Ferromagnetism in V$_5$Se$_8$ Epitaxial
  Thin Films}. \emph{Nano Lett.} \textbf{2019}, \emph{19}, 8806--8810\relax
\mciteBstWouldAddEndPuncttrue
\mciteSetBstMidEndSepPunct{\mcitedefaultmidpunct}
{\mcitedefaultendpunct}{\mcitedefaultseppunct}\relax
\EndOfBibitem
\bibitem[Meng \latin{et~al.}(2022)Meng, Zong, Tian, Chen, Xie, Yu, Qiu, Wang,
  Zhang, Wang, Li, Wang, and Zhang]{Meng2022}
Meng,~Q.; Zong,~J.; Tian,~Q.; Chen,~W.; Xie,~X.; Yu,~F.; Qiu,~X.; Wang,~K.;
  Zhang,~Y.; Wang,~P.; Li,~F.-S.; Wang,~C.; Zhang,~Y. {Selectable Growth and
  Electronic Structures of Monolayer 1T-VSe$_2$ and V$_5$Se$_8$ Films on
  Bilayer Graphene}. \emph{Phys. Status Solidi RRL} \textbf{2022}, \emph{16},
  2100601\relax
\mciteBstWouldAddEndPuncttrue
\mciteSetBstMidEndSepPunct{\mcitedefaultmidpunct}
{\mcitedefaultendpunct}{\mcitedefaultseppunct}\relax
\EndOfBibitem
\bibitem[Sumida \latin{et~al.}(2022)Sumida, Kusaka, Takeda, Kobayashi, and
  Hirahara]{Sumida2022}
Sumida,~K.; Kusaka,~S.; Takeda,~Y.; Kobayashi,~K.; Hirahara,~T. {Formation of
  monolayer V$_5$Se$_8$ from multilayer VSe$_2$ films via V- and
  Se-desorption}. \emph{Phys. Rev. B} \textbf{2022}, \emph{106}, 195421\relax
\mciteBstWouldAddEndPuncttrue
\mciteSetBstMidEndSepPunct{\mcitedefaultmidpunct}
{\mcitedefaultendpunct}{\mcitedefaultseppunct}\relax
\EndOfBibitem
\bibitem[Ji \latin{et~al.}(2017)Ji, Li, Wang, Niu, Gong, Zhang, Fang, Zhang,
  Shi, Liao, Wu, Gu, Liu, and Zhang]{Ji2017}
Ji,~Q.; Li,~C.; Wang,~J.; Niu,~J.; Gong,~Y.; Zhang,~Z.; Fang,~Q.; Zhang,~Y.;
  Shi,~J.; Liao,~L.; Wu,~X.; Gu,~L.; Liu,~Z.; Zhang,~Y. {Metallic Vanadium
  Disulfide Nanosheets as a Platform Material for Multifunctional Electrode
  Applications}. \emph{Nano Lett.} \textbf{2017}, \emph{17}, 4908--4916\relax
\mciteBstWouldAddEndPuncttrue
\mciteSetBstMidEndSepPunct{\mcitedefaultmidpunct}
{\mcitedefaultendpunct}{\mcitedefaultseppunct}\relax
\EndOfBibitem
\bibitem[Lee \latin{et~al.}(2022)Lee, Park, Chae, Kim, Kim, Choi, Lee, Chang,
  Chun, and Jung]{Lee2022}
Lee,~S.-H.; Park,~Y.~C.; Chae,~J.; Kim,~G.; Kim,~H.~J.; Choi,~B.~K.;
  Lee,~I.~H.; Chang,~Y.~J.; Chun,~S.-H.; Jung,~M. {Strong electron–phonon
  coupling driven charge density wave states in stoichiometric 1T-VS$_2$
  crystals}. \emph{J. Mater. Chem. C} \textbf{2022}, \emph{10},
  16657--16665\relax
\mciteBstWouldAddEndPuncttrue
\mciteSetBstMidEndSepPunct{\mcitedefaultmidpunct}
{\mcitedefaultendpunct}{\mcitedefaultseppunct}\relax
\EndOfBibitem
\bibitem[Ugeda \latin{et~al.}(2016)Ugeda, Bradley, Zhang, Onishi, Chen, Ruan,
  Ojeda-Aristizabal, Ryu, Edmonds, Tsai, Riss, Mo, Lee, Zettl, Hussain, Shen,
  and Crommie]{Ugeda2016}
Ugeda,~M.~M. \latin{et~al.}  {Characterization of collective ground states in
  single-layer NbSe$_2$}. \emph{Nat. Phys.} \textbf{2016}, \emph{12},
  92--97\relax
\mciteBstWouldAddEndPuncttrue
\mciteSetBstMidEndSepPunct{\mcitedefaultmidpunct}
{\mcitedefaultendpunct}{\mcitedefaultseppunct}\relax
\EndOfBibitem
\bibitem[Hou \latin{et~al.}(2020)Hou, Zhang, Tu, Gu, Zhang, Gong, Tu, Wang, Lv,
  Weng, Ren, Chen, Zhu, Hao, and Shan]{Hou2020}
Hou,~X.-Y.; Zhang,~F.; Tu,~X.-H.; Gu,~Y.-D.; Zhang,~M.-D.; Gong,~J.; Tu,~Y.-B.;
  Wang,~B.-T.; Lv,~W.-G.; Weng,~H.-M.; Ren,~Z.-A.; Chen,~G.-F.; Zhu,~X.-D.;
  Hao,~N.; Shan,~L. {Inelastic Electron Tunneling in 2H-Ta$_x$Nb$_{1-x}$Se$_2$
  Evidenced by Scanning Tunneling Spectroscopy}. \emph{Phys. Rev. Lett.}
  \textbf{2020}, \emph{124}, 106403\relax
\mciteBstWouldAddEndPuncttrue
\mciteSetBstMidEndSepPunct{\mcitedefaultmidpunct}
{\mcitedefaultendpunct}{\mcitedefaultseppunct}\relax
\EndOfBibitem
\bibitem[Wen \latin{et~al.}(2020)Wen, Xie, Wu, Shen, Kong, Lian, Li, Xing, and
  Yan]{Wen2020}
Wen,~C.; Xie,~Y.; Wu,~Y.; Shen,~S.; Kong,~P.; Lian,~H.; Li,~J.; Xing,~H.;
  Yan,~S. {Impurity-pinned incommensurate charge density wave and local phonon
  excitations in 2H-NbS$_2$}. \emph{Phys. Rev. B} \textbf{2020}, \emph{101},
  241404\relax
\mciteBstWouldAddEndPuncttrue
\mciteSetBstMidEndSepPunct{\mcitedefaultmidpunct}
{\mcitedefaultendpunct}{\mcitedefaultseppunct}\relax
\EndOfBibitem
\bibitem[Zhang \latin{et~al.}(2017)Zhang, Gong, Nie, Min, Liang, Oh, Zhang,
  Wang, Hong, Colombo, Wallace, and Cho]{Zhang2017}
Zhang,~C.; Gong,~C.; Nie,~Y.; Min,~K.~A.; Liang,~C.; Oh,~Y.~J.; Zhang,~H.;
  Wang,~W.; Hong,~S.; Colombo,~L.; Wallace,~R.~M.; Cho,~K. {Systematic study of
  electronic structure and band alignment of monolayer transition metal
  dichalcogenides in Van der Waals heterostructures}. \emph{2D Mater.}
  \textbf{2017}, \emph{4}, 015026\relax
\mciteBstWouldAddEndPuncttrue
\mciteSetBstMidEndSepPunct{\mcitedefaultmidpunct}
{\mcitedefaultendpunct}{\mcitedefaultseppunct}\relax
\EndOfBibitem
\bibitem[Duvjir \latin{et~al.}(2021)Duvjir, Choi, {Thi Ly}, Lam, Jang, Dung,
  Chang, and Kim]{Duvjir2021}
Duvjir,~G.; Choi,~B.~K.; {Thi Ly},~T.; Lam,~N.~H.; Jang,~K.; Dung,~D.~D.;
  Chang,~Y.~J.; Kim,~J. {Multiple charge density wave phases of monolayer
  VSe$_2$ manifested by graphene substrates}. \emph{Nanotechnology}
  \textbf{2021}, \emph{32}, 364002\relax
\mciteBstWouldAddEndPuncttrue
\mciteSetBstMidEndSepPunct{\mcitedefaultmidpunct}
{\mcitedefaultendpunct}{\mcitedefaultseppunct}\relax
\EndOfBibitem
\bibitem[Wang \latin{et~al.}(2021)Wang, Zhou, Loh, and Feng]{Wang2021}
Wang,~Z.; Zhou,~J.; Loh,~K.~P.; Feng,~Y.~P. {Controllable phase transitions
  between multiple charge density waves in monolayer 1T-VSe$_2$ via charge
  doping}. \emph{Appl. Phys. Lett.} \textbf{2021}, \emph{119}, 163101\relax
\mciteBstWouldAddEndPuncttrue
\mciteSetBstMidEndSepPunct{\mcitedefaultmidpunct}
{\mcitedefaultendpunct}{\mcitedefaultseppunct}\relax
\EndOfBibitem
\bibitem[Chua \latin{et~al.}(2022)Chua, Henke, Saha, Huang, Gou, He, Das, van
  Wezel, Soumyanarayanan, and Wee]{Chua2022}
Chua,~R.; Henke,~J.; Saha,~S.; Huang,~Y.; Gou,~J.; He,~X.; Das,~T.; van
  Wezel,~J.; Soumyanarayanan,~A.; Wee,~A. T.~S. {Coexisting Charge-Ordered
  States with Distinct Driving Mechanisms in Monolayer VSe$_2$}. \emph{ACS
  Nano} \textbf{2022}, \emph{16}, 783--791\relax
\mciteBstWouldAddEndPuncttrue
\mciteSetBstMidEndSepPunct{\mcitedefaultmidpunct}
{\mcitedefaultendpunct}{\mcitedefaultseppunct}\relax
\EndOfBibitem
\bibitem[Fumega \latin{et~al.}(2023)Fumega, Diego, Pardo, Blanco-Canosa, and
  Errea]{Fumega2023}
Fumega,~A.~O.; Diego,~J.; Pardo,~V.; Blanco-Canosa,~S.; Errea,~I.
  {Anharmonicity Reveals the Tunability of the Charge Density Wave Orders in
  Monolayer VSe$_2$}. \emph{Nano Lett.} \textbf{2023}, \emph{23},
  1794--1800\relax
\mciteBstWouldAddEndPuncttrue
\mciteSetBstMidEndSepPunct{\mcitedefaultmidpunct}
{\mcitedefaultendpunct}{\mcitedefaultseppunct}\relax
\EndOfBibitem
\bibitem[van Gastel \latin{et~al.}(2009)van Gastel, N'Diaye, Wall, Coraux,
  Busse, Buckanie, zu~Heringdorf, von Hoegen, Michely, and
  Poelsema]{VanGastel2009}
van Gastel,~R.; N'Diaye,~A.~T.; Wall,~D.; Coraux,~J.; Busse,~C.;
  Buckanie,~N.~M.; zu~Heringdorf,~F.-J.; von Hoegen,~M.; Michely,~T.;
  Poelsema,~B. {Selecting a Single Orientation for Millimeter Sized Graphene
  Sheets}. \emph{Appl. Phys. Lett.} \textbf{2009}, \emph{95}, 121901\relax
\mciteBstWouldAddEndPuncttrue
\mciteSetBstMidEndSepPunct{\mcitedefaultmidpunct}
{\mcitedefaultendpunct}{\mcitedefaultseppunct}\relax
\EndOfBibitem
\bibitem[Preobrajenski \latin{et~al.}(2023)Preobrajenski, Generalov,
  {\"{O}}hrwall, Tchaplyguine, Tarawneh, Appelfeller, Frampton, and
  Walsh]{Preobrajenski2023}
Preobrajenski,~A.; Generalov,~A.; {\"{O}}hrwall,~G.; Tchaplyguine,~M.;
  Tarawneh,~H.; Appelfeller,~S.; Frampton,~E.; Walsh,~N. {FlexPES: a versatile
  soft X-ray beamline at MAX IV Laboratory}. \emph{J. Synchrotron Radiat.}
  \textbf{2023}, \emph{30}\relax
\mciteBstWouldAddEndPuncttrue
\mciteSetBstMidEndSepPunct{\mcitedefaultmidpunct}
{\mcitedefaultendpunct}{\mcitedefaultseppunct}\relax
\EndOfBibitem
\bibitem[Hohenberg and Kohn(1964)Hohenberg, and Kohn]{Hohenberg1964}
Hohenberg,~P.; Kohn,~W. Inhomogeneous electron gas. \emph{Phys. Rev.}
  \textbf{1964}, \emph{136}, B864\relax
\mciteBstWouldAddEndPuncttrue
\mciteSetBstMidEndSepPunct{\mcitedefaultmidpunct}
{\mcitedefaultendpunct}{\mcitedefaultseppunct}\relax
\EndOfBibitem
\bibitem[Bl{\"o}chl(1994)]{Blochl1994}
Bl{\"o}chl,~P.~E. Projector augmented-wave method. \emph{Phys. Rev. B}
  \textbf{1994}, \emph{50}, 17953\relax
\mciteBstWouldAddEndPuncttrue
\mciteSetBstMidEndSepPunct{\mcitedefaultmidpunct}
{\mcitedefaultendpunct}{\mcitedefaultseppunct}\relax
\EndOfBibitem
\bibitem[Kresse and Hafner(1993)Kresse, and Hafner]{Kresse1993}
Kresse,~G.; Hafner,~J. Ab initio molecular dynamics for liquid metals.
  \emph{Phys. Rev. B} \textbf{1993}, \emph{47}, 558\relax
\mciteBstWouldAddEndPuncttrue
\mciteSetBstMidEndSepPunct{\mcitedefaultmidpunct}
{\mcitedefaultendpunct}{\mcitedefaultseppunct}\relax
\EndOfBibitem
\bibitem[Kresse and Furthm{\"u}ller(1996)Kresse, and
  Furthm{\"u}ller]{Kresse1996}
Kresse,~G.; Furthm{\"u}ller,~J. Efficient iterative schemes for ab initio
  total-energy calculations using a plane-wave basis set. \emph{Phys. Rev. B}
  \textbf{1996}, \emph{54}, 11169\relax
\mciteBstWouldAddEndPuncttrue
\mciteSetBstMidEndSepPunct{\mcitedefaultmidpunct}
{\mcitedefaultendpunct}{\mcitedefaultseppunct}\relax
\EndOfBibitem
\bibitem[Kohn and Sham(1965)Kohn, and Sham]{Kohn1965}
Kohn,~W.; Sham,~L.~J. Self-consistent equations including exchange and
  correlation effects. \emph{Phys. Rev.} \textbf{1965}, \emph{140}, A1133\relax
\mciteBstWouldAddEndPuncttrue
\mciteSetBstMidEndSepPunct{\mcitedefaultmidpunct}
{\mcitedefaultendpunct}{\mcitedefaultseppunct}\relax
\EndOfBibitem
\bibitem[Lee \latin{et~al.}(2010)Lee, Murray, Kong, Lundqvist, and
  Langreth]{Lee2010a}
Lee,~K.; Murray,~{\'E}.~D.; Kong,~L.; Lundqvist,~B.~I.; Langreth,~D.~C.
  {Higher-accuracy van der Waals density functional}. \emph{Phys. Rev. B}
  \textbf{2010}, \emph{82}, 081101\relax
\mciteBstWouldAddEndPuncttrue
\mciteSetBstMidEndSepPunct{\mcitedefaultmidpunct}
{\mcitedefaultendpunct}{\mcitedefaultseppunct}\relax
\EndOfBibitem
\bibitem[Becke(1986)]{Becke1986}
Becke,~A. {On the large-gradient behavior of the density functional exchange
  energy}. \emph{J. Chem. Phys.} \textbf{1986}, \emph{85}, 7184--7187\relax
\mciteBstWouldAddEndPuncttrue
\mciteSetBstMidEndSepPunct{\mcitedefaultmidpunct}
{\mcitedefaultendpunct}{\mcitedefaultseppunct}\relax
\EndOfBibitem
\bibitem[Hamada(2014)]{Hamada2014}
Hamada,~I. {van der Waals density functional made accurate}. \emph{Phys. Rev.
  B} \textbf{2014}, \emph{89}, 121103\relax
\mciteBstWouldAddEndPuncttrue
\mciteSetBstMidEndSepPunct{\mcitedefaultmidpunct}
{\mcitedefaultendpunct}{\mcitedefaultseppunct}\relax
\EndOfBibitem
\bibitem[Huttmann \latin{et~al.}(2015)Huttmann, Mart{\'\i}nez-Galera, Caciuc,
  Atodiresei, Schumacher, Standop, Hamada, Wehling, Bl{\"u}gel, and
  Michely]{Huttmann2015}
Huttmann,~F.; Mart{\'\i}nez-Galera,~A.~J.; Caciuc,~V.; Atodiresei,~N.;
  Schumacher,~S.; Standop,~S.; Hamada,~I.; Wehling,~T.~O.; Bl{\"u}gel,~S.;
  Michely,~T. Tuning the van der Waals Interaction of Graphene with Molecules
  via Doping. \emph{Phys. Rev. Lett.} \textbf{2015}, \emph{115}, 236101\relax
\mciteBstWouldAddEndPuncttrue
\mciteSetBstMidEndSepPunct{\mcitedefaultmidpunct}
{\mcitedefaultendpunct}{\mcitedefaultseppunct}\relax
\EndOfBibitem
\bibitem[Perdew \latin{et~al.}(1996)Perdew, Burke, and Ernzerhof]{Perdew1996}
Perdew,~J.~P.; Burke,~K.; Ernzerhof,~M. {Generalized gradient approximation
  made simple}. \emph{Phys. Rev. Lett.} \textbf{1996}, \emph{77}, 3865\relax
\mciteBstWouldAddEndPuncttrue
\mciteSetBstMidEndSepPunct{\mcitedefaultmidpunct}
{\mcitedefaultendpunct}{\mcitedefaultseppunct}\relax
\EndOfBibitem
\end{mcitethebibliography}


\providecommand{\latin}[1]{#1}
\makeatletter
\providecommand{\doi}
  {\begingroup\let\do\@makeother\dospecials
  \catcode`\{=1 \catcode`\}=2 \doi@aux}
\providecommand{\doi@aux}[1]{\endgroup\texttt{#1}}
\makeatother
\providecommand*\mcitethebibliography{\thebibliography}
\csname @ifundefined\endcsname{endmcitethebibliography}
  {\let\endmcitethebibliography\endthebibliography}{}
\begin{mcitethebibliography}{5}
\providecommand*\natexlab[1]{#1}
\providecommand*\mciteSetBstSublistMode[1]{}
\providecommand*\mciteSetBstMaxWidthForm[2]{}
\providecommand*\mciteBstWouldAddEndPuncttrue
  {\def\EndOfBibitem{\unskip.}}
\providecommand*\mciteBstWouldAddEndPunctfalse
  {\let\EndOfBibitem\relax}
\providecommand*\mciteSetBstMidEndSepPunct[3]{}
\providecommand*\mciteSetBstSublistLabelBeginEnd[3]{}
\providecommand*\EndOfBibitem{}
\mciteSetBstSublistMode{f}
\mciteSetBstMaxWidthForm{subitem}{(\alph{mcitesubitemcount})}
\mciteSetBstSublistLabelBeginEnd
  {\mcitemaxwidthsubitemform\space}
  {\relax}
  {\relax}

\bibitem[N'Diaye \latin{et~al.}(2008)N'Diaye, Coraux, Plasa, Busse, and
  Michely]{N'Diaye2008}
N'Diaye,~A.~T.; Coraux,~J.; Plasa,~T.~N.; Busse,~C.; Michely,~T. {Structure of
  epitaxial graphene on Ir(111)}. \emph{New J. of Phys.} \textbf{2008},
  \emph{10}, 043033\relax
\mciteBstWouldAddEndPuncttrue
\mciteSetBstMidEndSepPunct{\mcitedefaultmidpunct}
{\mcitedefaultendpunct}{\mcitedefaultseppunct}\relax
\EndOfBibitem
\bibitem[Pieli{\'{c}} \latin{et~al.}(2020)Pieli{\'{c}}, Hall, Despoja,
  Raki{\'{c}}, Petrovi{\'{c}}, Sohani, Busse, Michely, and Kralj]{Pielic2020}
Pieli{\'{c}},~B.; Hall,~J.; Despoja,~V.; Raki{\'{c}},~I.~{\v{S}}.;
  Petrovi{\'{c}},~M.; Sohani,~A.; Busse,~C.; Michely,~T.; Kralj,~M. {Sulfur
  Structures on Bare and Graphene-Covered Ir(111)}. \emph{J. Phys. Chem. C.}
  \textbf{2020}, \emph{124}, 6659--6668\relax
\mciteBstWouldAddEndPuncttrue
\mciteSetBstMidEndSepPunct{\mcitedefaultmidpunct}
{\mcitedefaultendpunct}{\mcitedefaultseppunct}\relax
\EndOfBibitem
\bibitem[Bonilla \latin{et~al.}(2020)Bonilla, Kolekar, Li, Xin, Coelho, Lasek,
  Zberecki, Lizzit, Tosi, Lacovig, Lizzit, and Batzill]{Bonilla2020}
Bonilla,~M.; Kolekar,~S.; Li,~J.; Xin,~Y.; Coelho,~P.~M.; Lasek,~K.;
  Zberecki,~K.; Lizzit,~D.; Tosi,~E.; Lacovig,~P.; Lizzit,~S.; Batzill,~M.
  {Compositional Phase Change of Early Transition Metal Diselenide (VSe$_2$ and
  TiSe$_2$) Ultrathin Films by Postgrowth Annealing}. \emph{Adv. Mater.
  Interfaces} \textbf{2020}, \emph{7}, 2000497\relax
\mciteBstWouldAddEndPuncttrue
\mciteSetBstMidEndSepPunct{\mcitedefaultmidpunct}
{\mcitedefaultendpunct}{\mcitedefaultseppunct}\relax
\EndOfBibitem
\bibitem[van Efferen \latin{et~al.}(2021)van Efferen, Berges, Hall, van Loon,
  Kraus, Schobert, Wekking, Huttmann, Plaar, Rothenbach, Ollefs, Arruda,
  Brookes, Sch{\"{o}}nhoff, Kummer, Wende, Wehling, and
  Michely]{vanEfferen2021}
van Efferen,~C. \latin{et~al.}  {A full gap above the Fermi level: the charge
  density wave of monolayer VS$_2$}. \emph{Nat. Comm.} \textbf{2021},
  \emph{12}, 6837\relax
\mciteBstWouldAddEndPuncttrue
\mciteSetBstMidEndSepPunct{\mcitedefaultmidpunct}
{\mcitedefaultendpunct}{\mcitedefaultseppunct}\relax
\EndOfBibitem
\end{mcitethebibliography}

\end{document}


\maketitle

\newpage

\section*{Supplementary Note 1: Transition from mono- to multilayer growth mode}

\begin{figure*}[h!]
\centering
\includegraphics[width=\textwidth]{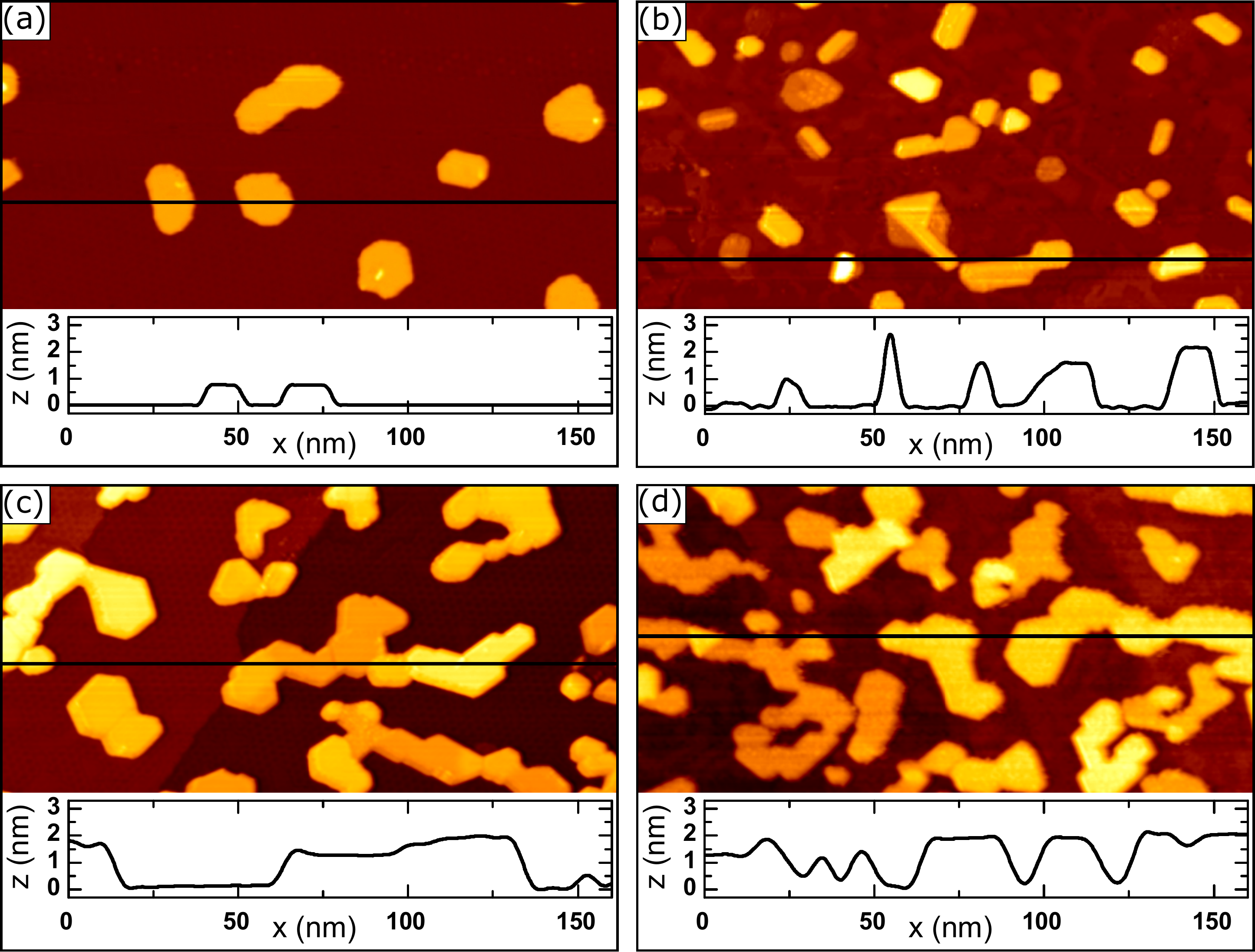}
\caption{\textbf{Transition from mono- to multilayer growth mode.} STM topographs after growth with identical synthesis parameters (growth temperature $\SI{300}{\K}$, S pressure during growth $\SI{5\times10^{-9}}{\milli\bar}$,  annealing temperature $\SI{800}{\K}$, S pressure during annealing $\SI{2\times10^{-9}}{\milli\bar}$), but variation of deposition time, which results in a change in growth mode. \textbf{a}~Deposition of 0.15 ML resulting in single-layer \va{} islands. \textbf{b}~Deposition of 0.45 ML results in a non phase-pure system of mixed single- and multilayer heights, which cover $20 \%$ of the surface area. \textbf{c}~After depositing 0.9 ML no single-layer heights can be found anymore, multilayer islands cover about $30\%$ of the surface. \textbf{d}~Depositing 1.35 ML results in a sample with only multilayer islands that cover $45\%$ of the Gr substrate.\\
Image information (image size, sample bias, tunnelling current): \textbf{a}~\SI{160 \times 80 }{\nano \meter^2}, \SI{1.3}{\V}, \SI{20}{\pico \ampere}, \textbf{b}~\SI{160 \times 80}{\nano \meter^2}, \SI{-1}{\V}, \SI{1}{\nano \ampere}, \textbf{c}~\SI{160 \times 80}{\nano \meter^2}, \SI{-1}{\V}, \SI{0.5}{\nano \ampere}, \textbf{d}~\SI{160 \times 80}{\nano \meter^2}, \SI{-0.7}{\V}, \SI{0.1}{\nano \ampere}.}
 \label{fig:VS2CoverageMonoMultiTransition}
\end{figure*}

\newpage

\section*{Supplementary Note 2: Formation of \vai{} at high S pressure}

\begin{figure*}[h!]
\centering
\includegraphics[width=\textwidth]{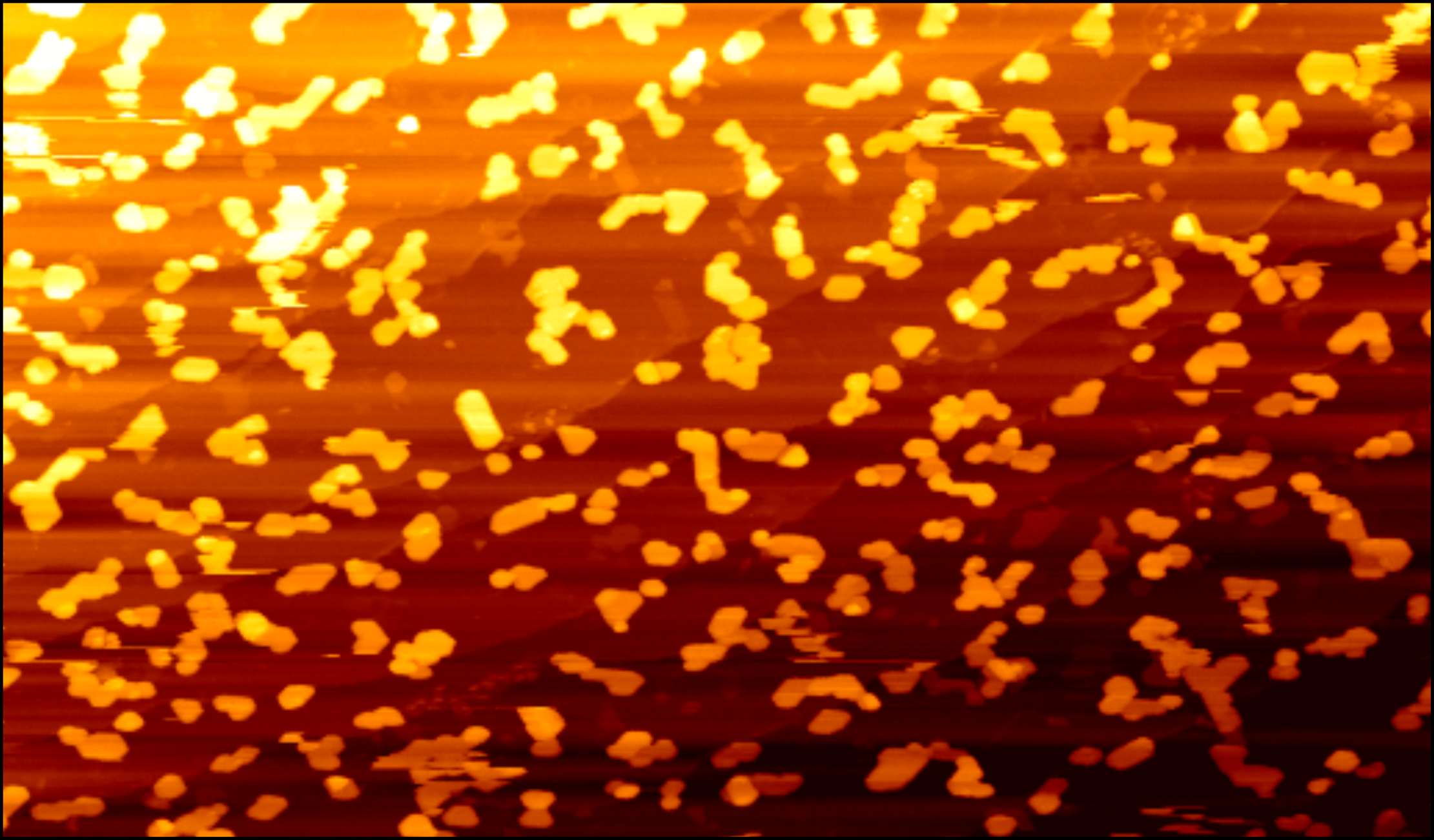}
\caption{\textbf{Formation of \vai{}-derived islands at high S pressure.}~Annealing to \SI{800}{\K} after deposition of 1.2 ML V in a high S pressure $p^\text{a}_\text{S} = \SI{8\times10^{-9}}{\milli\bar}$, \vai-derived islands in a variety of heights are formed.\\
Image information (image size, sample bias, tunnelling current): \textbf{a}~\SI{655 \times 385 }{\nano \meter^2}, \SI{-0.02}{\V}, \SI{8}{\pico \ampere}.}
 \label{fig:HighS}
\end{figure*}

\newpage

\section*{Supplementary Note 3: Line profiles of \vx{} islands}

\begin{figure*}[h!]
	\centering
		\includegraphics[width=1\textwidth]{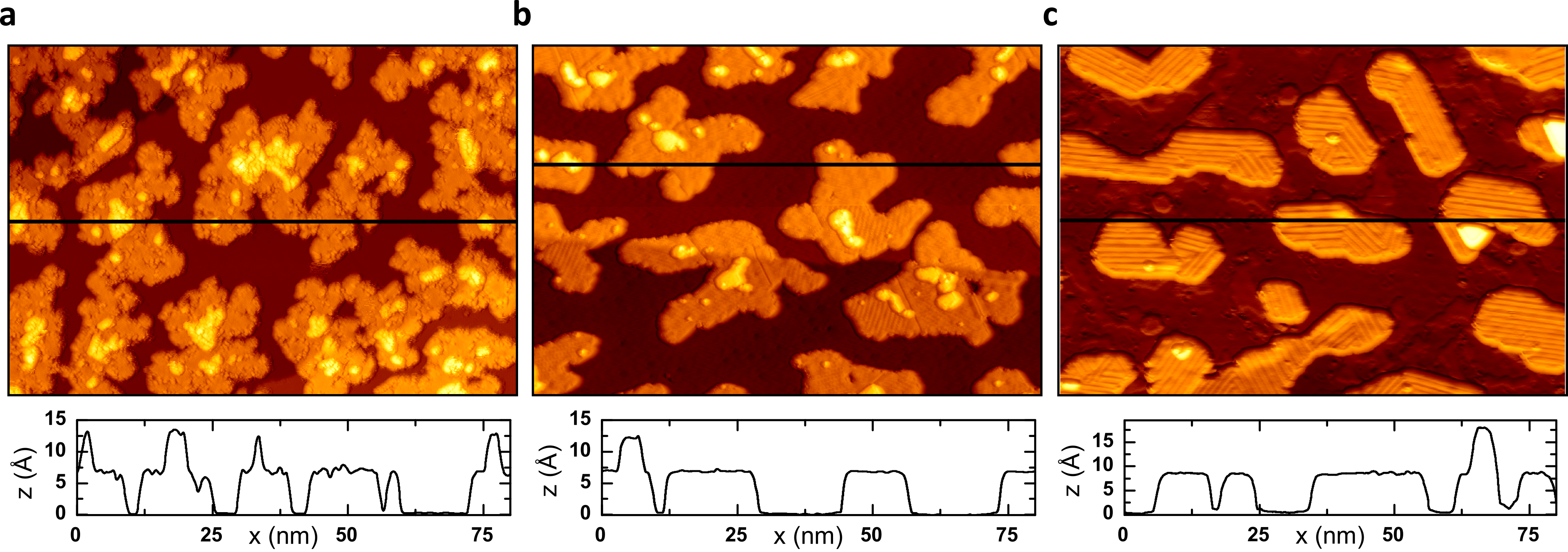}
			\caption{\textbf{Line profiles of \vx{} islands. a}~STM topograph of \va{} islands directly after growth at \SI{300}{\K}. \textbf{b}~STM topograph after annealing to \SI{600}{\K}. \textbf{c}~STM topograph after annealing to \SI{800}{\K}. All annealing was performed without S background pressure. Images taken at \SI{300}{\K}. Image information (image size, sample bias, tunneling current): main panels~\SI{80 \times 80 }{\nano \meter^2}, inset~\SI[parse-numbers=false]{20 \times 20}{\nano \meter^2}. \textbf{a}~\SI{-1.0}{\V}, \SI{210}{\pico \ampere}; \textbf{b}~\SI{-1.0}{\V}, \SI{100}{\pico \ampere}; \textbf{c}~\SI{-1.3}{\V}, \SI{10}{\pico \ampere}.}
 \label{fig:FigLineP}
\end{figure*}

\newpage

\section*{Supplementary Note 4: 1D crystallites of \vx}

\begin{figure*}[h!]
\centering
\includegraphics[width=\textwidth]{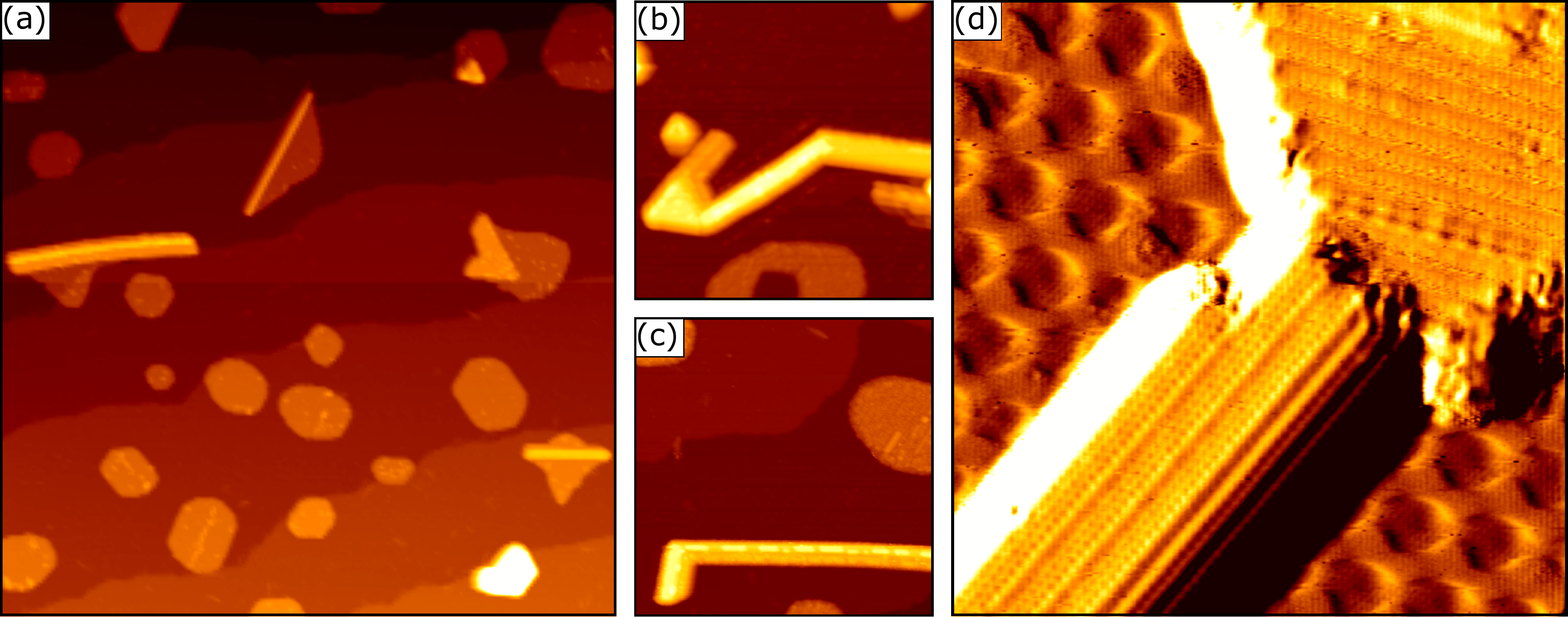}
\caption{\textbf{1D crystallites of \vx{}. a}~After annealing to \SI{900}{\K}, 1D crystallites form at the edges of formerly single-layer \vx{}. \textbf{b,c}~Solitary 1D structures which show apparent heights of several nm. \textbf{d}~At the atomic scale again striped reconstructions are visible, reminiscent of the striped surface of \vas.\\
Image information (image size, sample bias, tunnelling current): \textbf{a}~\SI{250 \times 250 }{\nano \meter^2}, \SI{-0.3}{\V}, \SI{20}{\pico \ampere}, \textbf{b}~\SI{80 \times 80}{\nano \meter^2}, \SI{-0.4}{\V}, \SI{10}{\pico \ampere}, \textbf{c}~\SI{80 \times 80}{\nano \meter^2}, \SI{-1}{\V}, \SI{10}{\pico \ampere}, \textbf{d}~\SI{20 \times 20}{\nano \meter^2}, \SI{-0.08}{\V}, \SI{0.4}{\nano \ampere}.}
 \label{fig:VS21DCrystallites}
\end{figure*}

The sulphur loss associated with elevated temperatures leads not only to vacancy superstructures but rather indicates the metastability of the \va{} phase by a quick transition to thermodynamically more stable phases. In that regard, the striped sulphur vacancy row phase \vas{} readily transforms into elongated crystallites as illustrated in Fig.~\ref{fig:VS21DCrystallites} when going to higher temperatures. In panel Fig.~\ref{fig:VS21DCrystallites}a, an overview of a formerly pure monolayer sample after annealing at 900 K in sulphur vapor is given. Besides the formation of small close to triangular-shaped bilayer islands, elongated bright structures form at the edge of an island strongly resembling rolled-up layers. Owing to their increased height of about \SIrange{2}{2.5}{\nm}, but low surface contact area to Gr, these structures are easily dragged over the surface or picked up by the STM tip. In Fig.~\ref{fig:VS21DCrystallites}b,c, single, free-standing crystallites, which are not connected to any residual monolayer are shown. In STM, these structures have an apparent height of up to 3 nm and are again covered in adsorbates, which arrange periodically. This indicates a lattice rearrangement similar to the one observed for \vas{} and may also be resolved atomically as shown in Fig.~\ref{fig:VS21DCrystallites}d. Here, the clean multilayer of \SI{16}{\nm} height shows striped ordering reminiscent of the monolayer analogue.

\newpage

\section*{Supplementary Note 5: DFT model of V$_3$S$_5$}
\begin{figure*}[h!]
	\centering
		\includegraphics[width=0.9\textwidth]{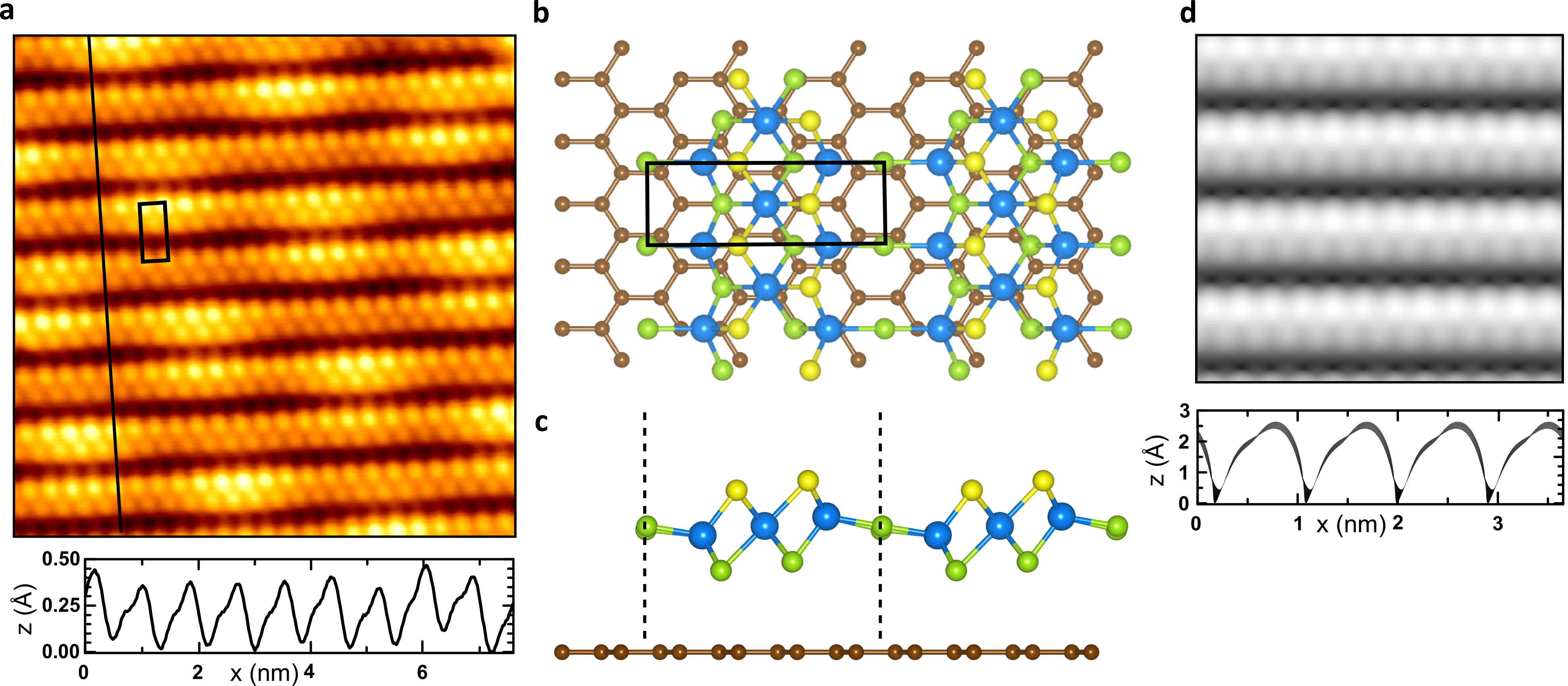}
			\caption{\textbf{Comparison of experimental stripe superstructure to DFT model of striped V$_3$S$_5$. a}~Atomically resolved STM topograph of the stripes. The unit cell of the dominant striped superstructure (black) is indicated. The black line indicates the position of the line profile. \textbf{b,c}~Top and side view of structural model of S-deficient \va{} on graphene obtained by DFT calculations. The black box indicates the unit cell. The bottom S atoms are colored green to allow for easy identification when viewed from the top. \textbf{d}~Simulated STM image (isosurface of the charge density) at \SI{-0.5}{\V}. Below, the isosurface is shown from the side, so the corrugation can be seen.
Image taken at \mbox{\SI{1.7}{\K}}. Image information (image size, sample bias, tunneling current): \textbf{a}~\mbox{\SI{7.5 \times 7.5}{\nano \meter^2}}, \mbox{\SI{-0.5}{\V}}, \mbox{\SI{1.4}{\nano \ampere}}
}
			
 \label{fig:SFigV3S5}
\end{figure*}

\newpage

\section*{Supplementary Note 6: S intercalation below Gr}
\begin{figure*}[h!]
	\centering
		\includegraphics[width=0.9\textwidth]{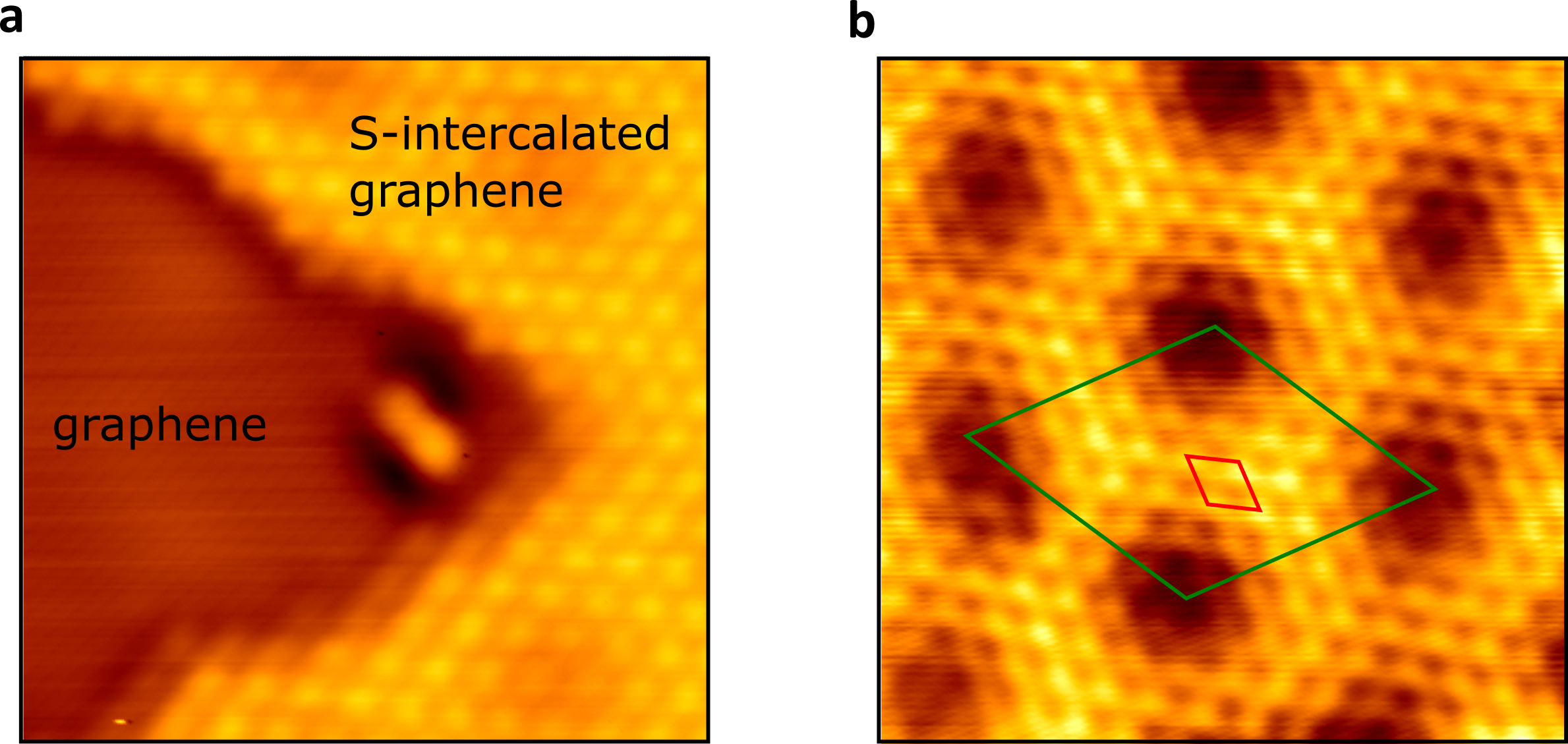}
			\caption{\textbf{S intercalation below Gr.}~After growth or annealing of \vx, patches of \ro{} S intercalation are visible below Gr. \textbf{a}~STM topograph of partially intercalated graphene. \textbf{b}~Patch of fully intercalated graphene. Visible are the moir\'e between graphene and Ir(111)~\cite{N'Diaye2008} and the \ro{} S intercalation~\cite{Pielic2020}. Both unit cells are drawn in the image, with the moir\'e unit cell in green ($a_\text{moir\'e} = \SI{25.3}{\angstrom}$) and the S intercalation in red ($a_\text{S\ro} = \SI{4.3}{\angstrom}$).
			Images taken at \SI{7}{\K}. STM parameters: \textbf{a}~\SI{6.2 \times 6.2}{\nano \meter^2}, \SI{-1.0}{\V}, \SI{50}{\pico \ampere}; \textbf{a}~\SI{6.4 \times 6.4}{\nano \meter^2}, \SI{-0.5}{\V}, \SI{100}{\pico \ampere}.
}
			
 \label{fig:SFigSInter}
\end{figure*}

\newpage
\section*{Supplementary Note 7: \SI{300}{\K} XPS spectrum of S 2p$_{3/2}$ and 2p$_{1/2}$ core-level spectra of \va{}}

\begin{figure*}[h!]
	\centering
		\includegraphics[width=0.6\textwidth]{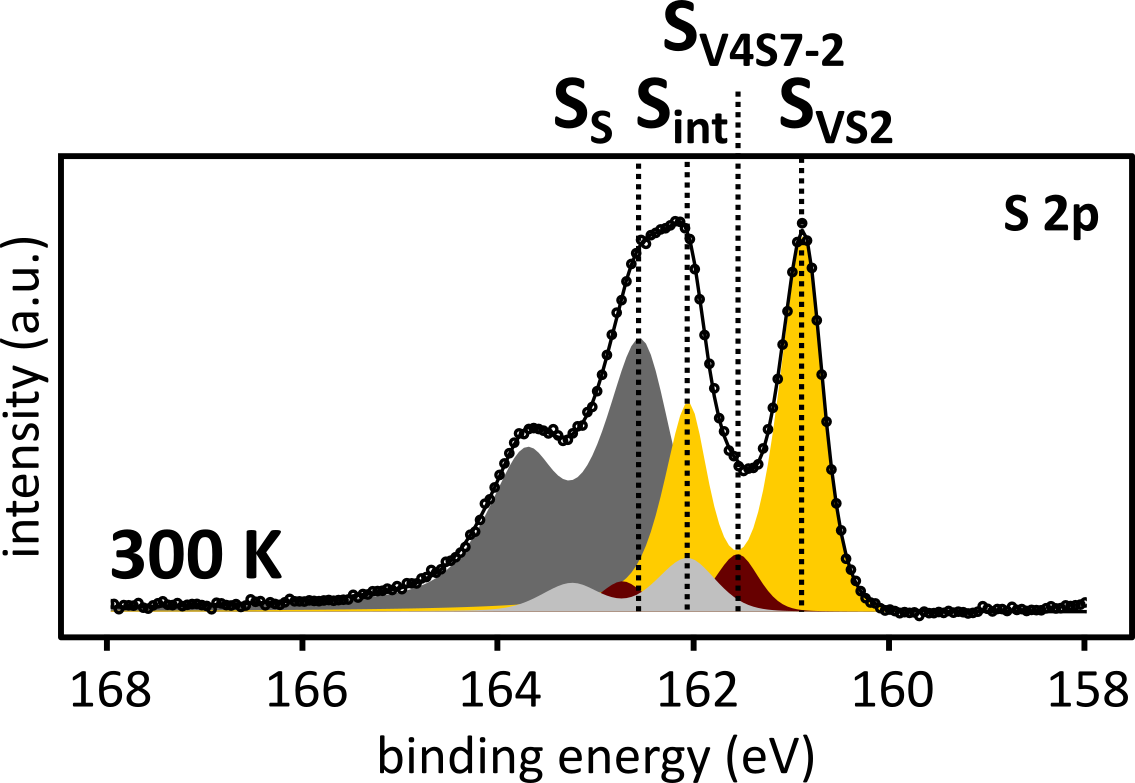}
			\caption{\textbf{\SI{300}{\K} XPS of S 2p$_{3/2}$ and 2p$_{1/2}$ core-level spectra of \va{}.} XP spectrum data measured at a photon energy of $hv = \SI{260}{\eV}$, fitted with 5 components, see text in main manuscript for discussion. The data points are represented by black circles and the overall fit by a solid black line. The S$_\text{V4S7-2}$ component is assumed to stem from edges and defects at this temperature.}
			
 \label{fig:SFigXPS300K}
\end{figure*}

\newpage

\section*{Supplementary Note 8:  Origin of S$_\text{V4S7-1}$ and S$_\text{V4S7-2}$ XPS components}
\begin{figure*}[h!]
	\centering
		\includegraphics[width=0.9\textwidth]{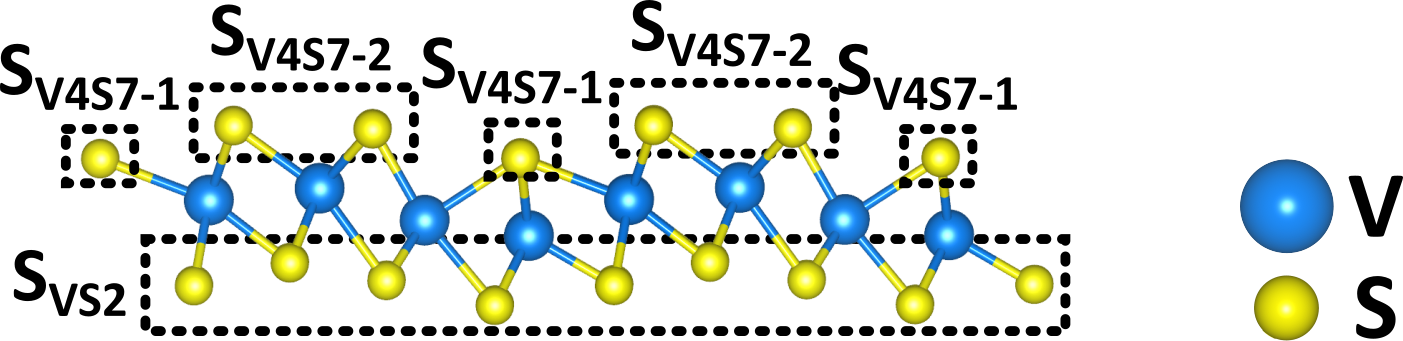}
			\caption{\textbf{Origin of S$_\text{V4S7-1}$ and S$_\text{V4S7-2}$ XPS components.}~Atomic structure models of \vas, with labeled boxes around three inequivalent atoms in the structure: 1) top S atoms that neighbour a missing S row, 2) top S atoms which do not sit directly next to a missing S row and 3) bottom S atoms. The box labels indicate which XPS components discussed in the main text correspond to the atoms. 
}
			
 \label{fig:SFigXPSV4S7}
\end{figure*}

\newpage

\section*{Supplementary Note 9: V 2p XP spectra of \vas}
\begin{figure*}[h!]
	\centering
		\includegraphics[width=0.6\textwidth]{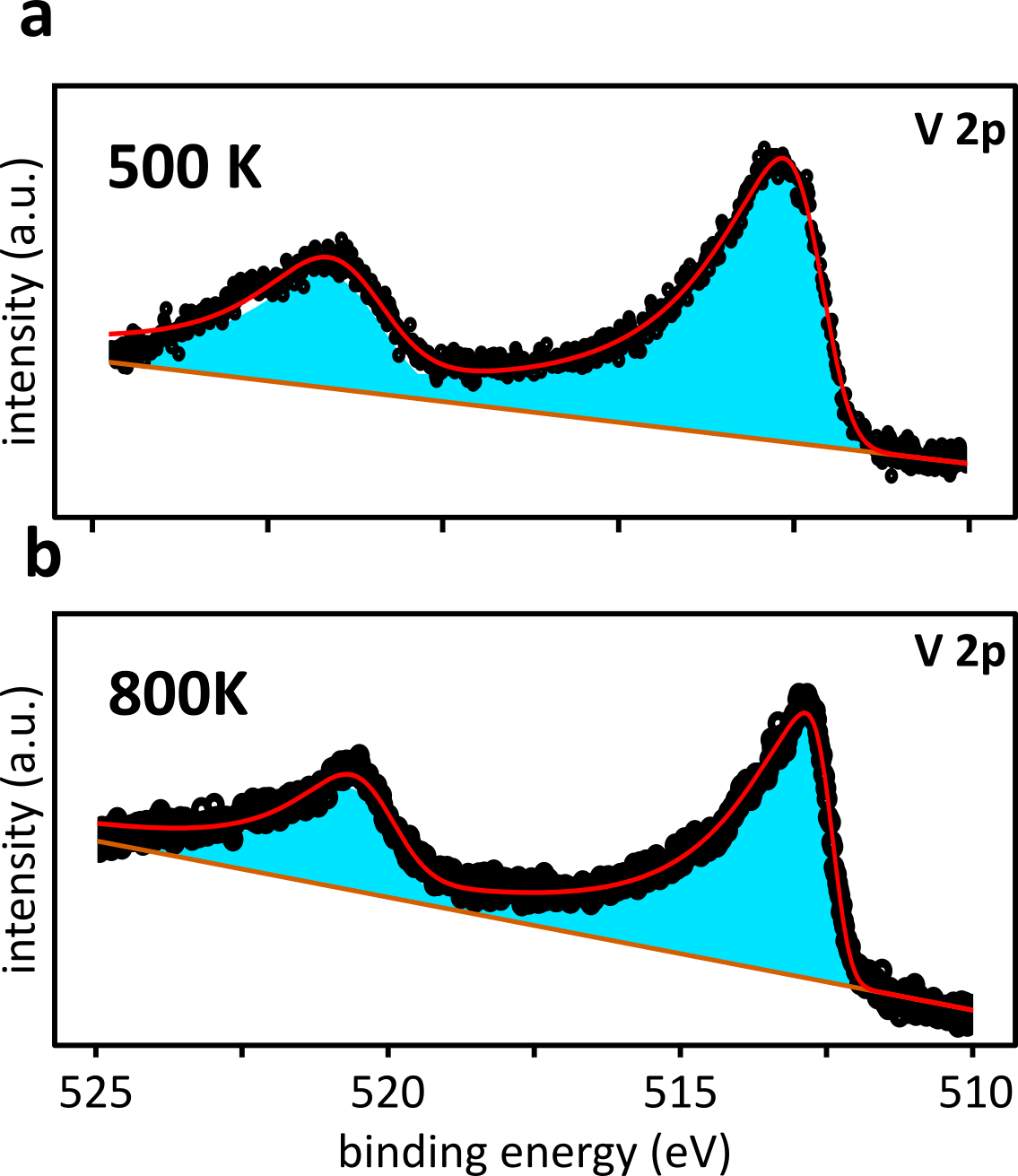}
			\caption{\textbf{XPS of V 2p$_{3/2}$ and 2p$_{1/2}$ core-level spectra of single-layer \va{} and \vas.}~ XP spectrum data measured at a photon energy of $hv = \SI{640}{\eV}$, fitted with 1 component. The data points are represented by black circles and the overall fit by a solid red line. The two spin-orbit split peaks are colored cyan.
}
			
 \label{fig:SFigXPSV2pV4S7}
\end{figure*}

All V 2p spectra of the single-layer sample can be fitted with the same doublet. Between the \va{} sample obtained at an annealing temperature of \SI{500}{\K}, see Fig.~\ref{fig:SFigXPSV2pV4S7}a, and \vas, which is created by annealing to \SI{800}{\K}, no additional components can be recognized. 

\newpage

\section*{Supplementary Note 10: LEED of phase-pure \vai-derived sample}

\begin{figure*}[h!]
	\centering
		\includegraphics[width=0.6\textwidth]{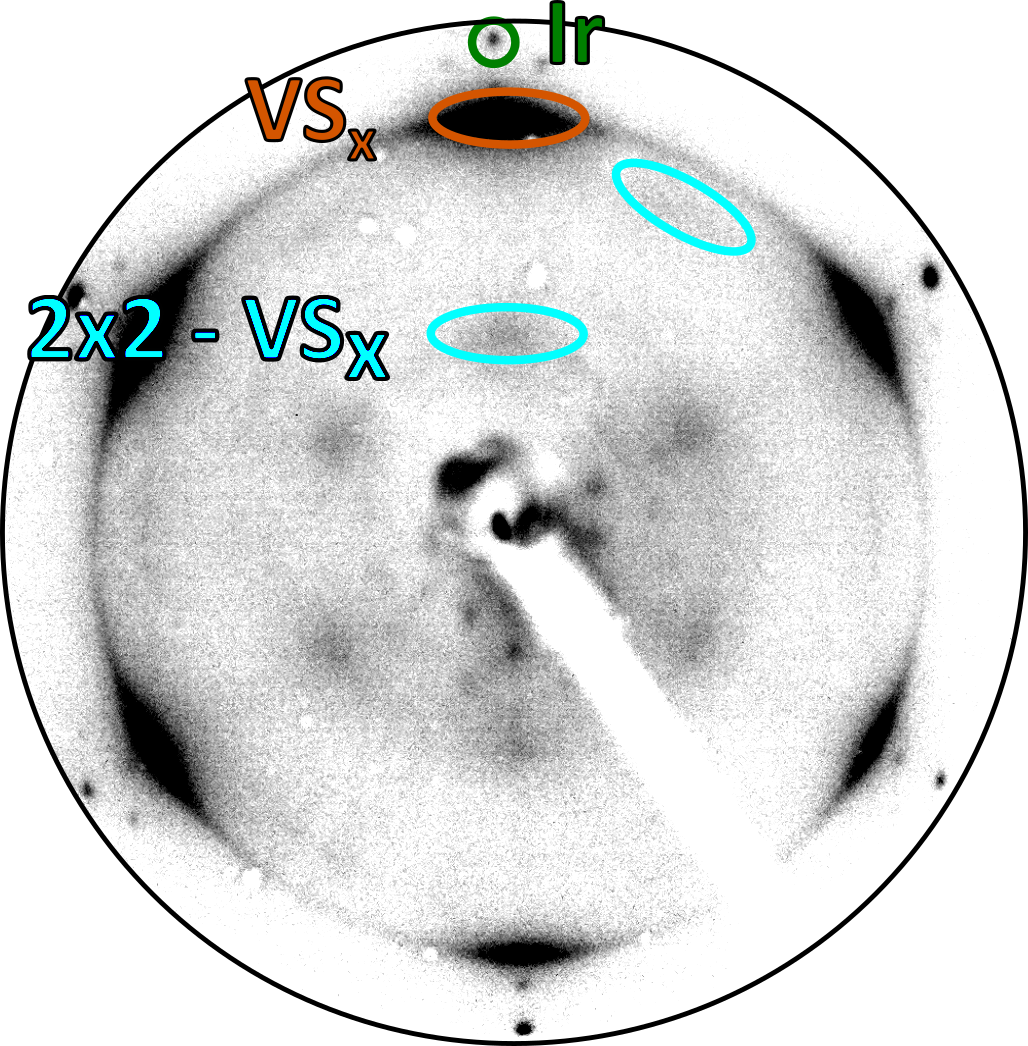}
			\caption{\textbf{LEED of phase-pure \vai-derived sample. a}~LEED pattern of \vx{} sample, where $25\%$ additional V atoms were deposited on a sample grown at \SI{300}{\K} and annealed to \SI{500}{\K}. After room temperature V deposition, the sample was subsequently annealed to \SI{800}{\K} in UHV (no S pressure). Indicated with colored circles are the Ir (green) first order reflections.  First order reflections of \vx{} (orange) and reflections of a $2 \times 2$ superstructure with respect to \vx{} (cyan) are visible. Note that unlike the other LEED images shown in this work, this one was not taken with a microchannel plate LEED, leading to a reduced gain power.
			LEED taken with \SI{85}{\eV}.}
			
 \label{fig:SFigLEEDBL}
\end{figure*}

\newpage

\section*{Supplementary Note 11: XP spectra of S 2p$_{3/2}$ and 2p$_{1/2}$ core-level spectra of multiheight VS$_x$, with $25\%$ extra V}

\begin{figure*}[h!]
	\centering
		\includegraphics[width=0.6\textwidth]{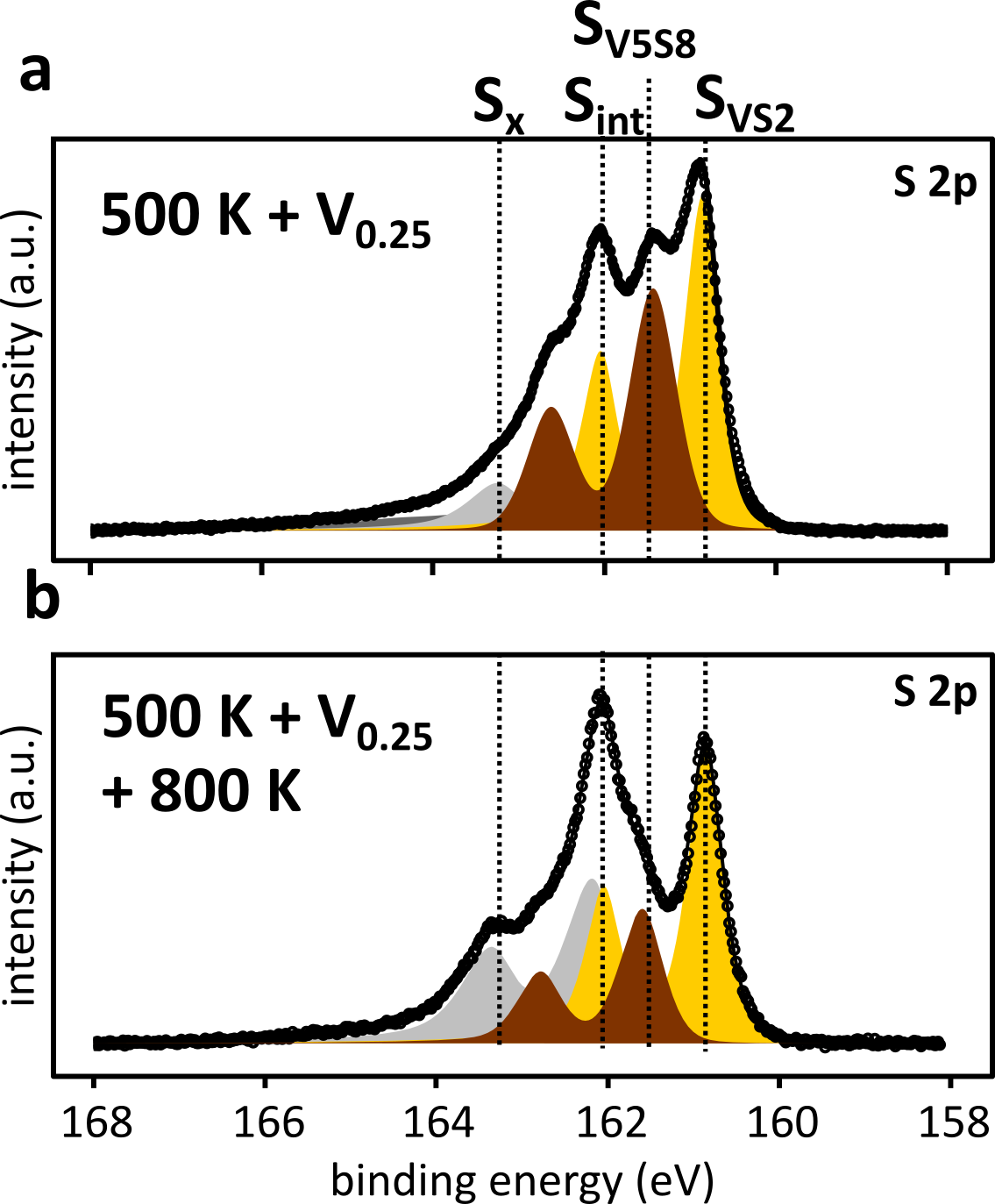}
			\caption{\textbf{XP spectra of S 2p$_{3/2}$ and 2p$_{1/2}$ core-level spectra of multiheight VS$_x$, with $25\%$ extra V. a}~XP spectrum of \vx{} sample, where $25\%$ additional V atoms were deposited on a \vx{} sample grown at \SI{300}{\K} and annealed to \SI{500}{\K}. The spectrum was taken immediately after room temperature V deposition. \textbf{b}~XP spectrum of the same sample, after a subsequent annealing step to \SI{800}{\K} in UHV (no S pressure). XP spectra measured at a photon energy of $hv = \SI{260}{\eV}$, fitted with 4 components, see text in main manuscript for discussion. The data points are represented by black circles and the overall fit by a solid black line. }	
 \label{fig:SFigXPSInter}
\end{figure*}

\newpage

\section*{Supplementary Note 12: V 2p XP spectra of multiheight sample}
\begin{figure*}[h!]
	\centering
		\includegraphics[width=0.6\textwidth]{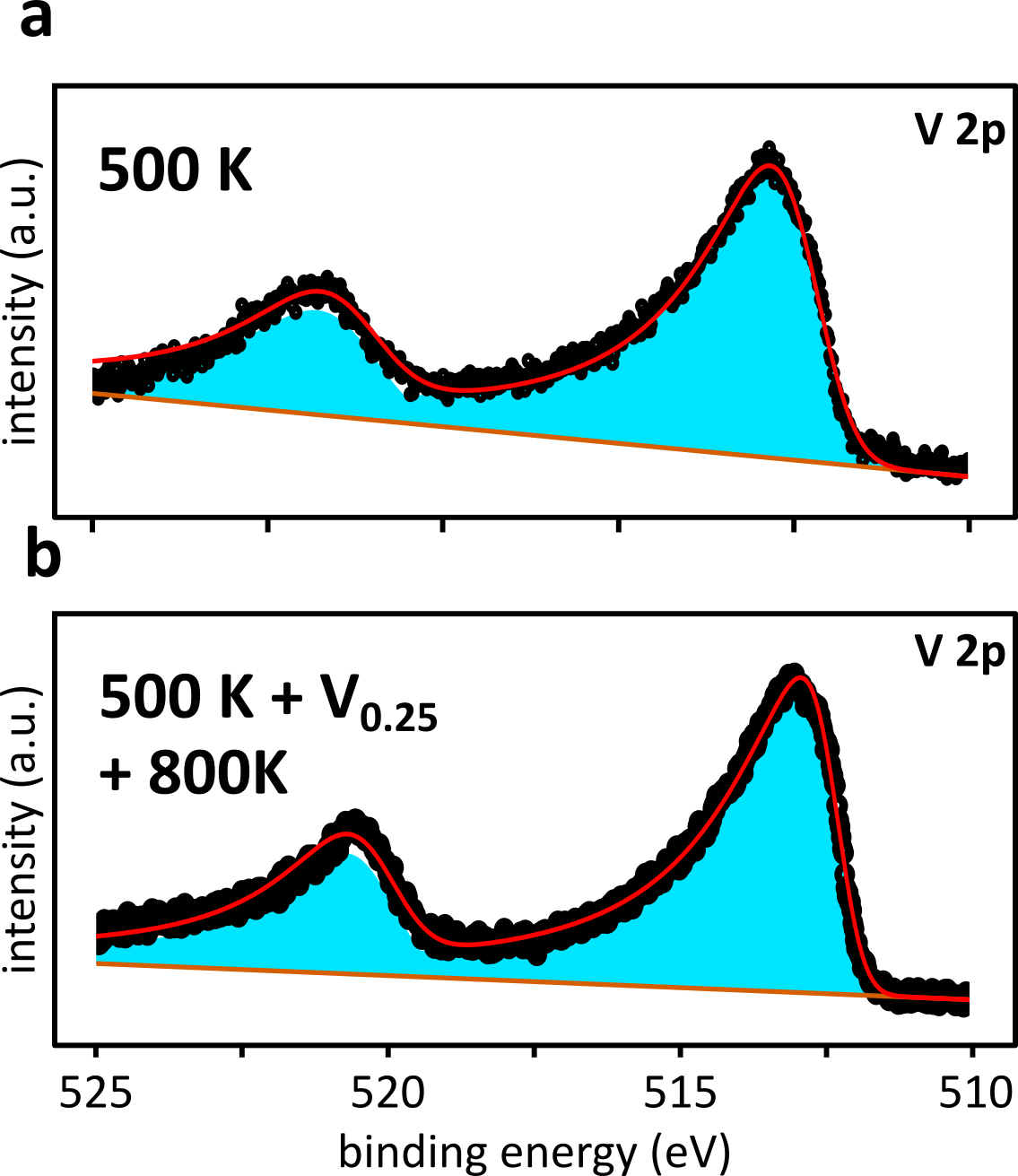}
			\caption{\textbf{XPS of V 2p$_{3/2}$ and 2p$_{1/2}$ core-level spectra of multiheight \va{}.}~ XP spectrum data measured at a photon energy of $hv = \SI{640}{\eV}$, fitted with 1 component. The data points are represented by black circles and the overall fit by a solid red line. The two spin-orbit split peaks are colored cyan.
			\label{fig:SFigXPSV2pV5S8}
}
			
 \label{fig:SFigXPSV2p}
\end{figure*}

All V 2p spectra of the multiheight sample can be fitted with the same doublet. Between the \va{} sample obtained at an annealing temperature of \SI{500}{\K}, see Fig.~\ref{fig:SFigXPSV2pV5S8}a, and the self-intercalated sample where extra V was deposited on the sample before annealing to \SI{800}{\K}, see Fig.~\ref{fig:SFigXPSV2pV5S8}b, no additional components can be recognized. This behavior was previously observed for self-intercalated \ve{} and attributed to the fact that the first coordination shell of the V atoms in the TMDC and intercalated layer are similar~\cite{Bonilla2020}.

\newpage

\section*{Supplementary Note 13: Single-layer \va{} and V$_9$S$_{16}$}

\begin{figure*}[h!]
	\centering
		\includegraphics[width=0.8\textwidth]{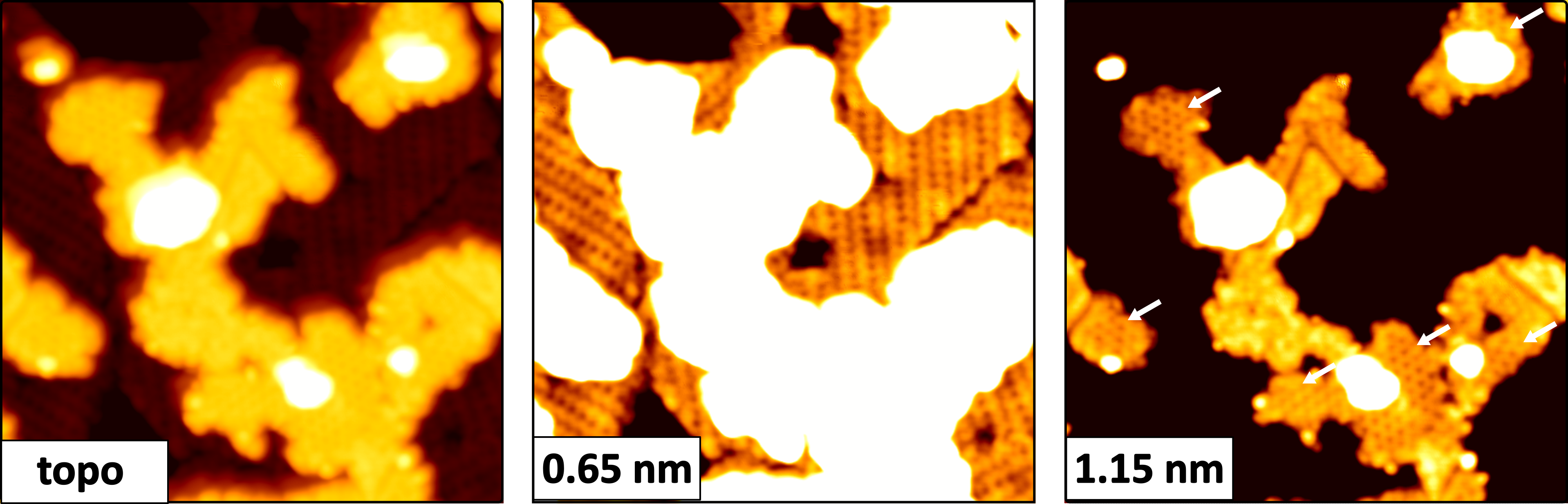}
			\caption{\textbf{Single-layer \va{} and V$_9$S$_{16}$}~Low-temperature STM topograph of a multiheight sample annealed to \SI{500}{\K}, see Fig.~4a of the main text. On the left, the complete topography is shown. In the middle the contrast is placed on single-layer \va{}, which has an unidirectional CDW phase~\cite{vanEfferen2021}. On the right, the same image is presented with the contrast placed on the higher islands. On most islands the \ro{} superstructure, characteristic for V$_9$S$_{16}$, has developed (highlighted by white arrows), while the island in the center looks disordered.
			Image taken at \SI{4}{\K}. STM parameters: \SI{20.5 \times 20.5}{\nano \meter^2}, \SI{-1.0}{\V}, \SI{50}{\pico \ampere}.
}
			
 \label{fig:SFigMLBL}
\end{figure*}
\newpage
\section*{Supplementary Note 14: Temperature dependence of \ro{} superstructure}

\begin{figure*}[h!]
	\centering
		\includegraphics[width=0.85\textwidth]{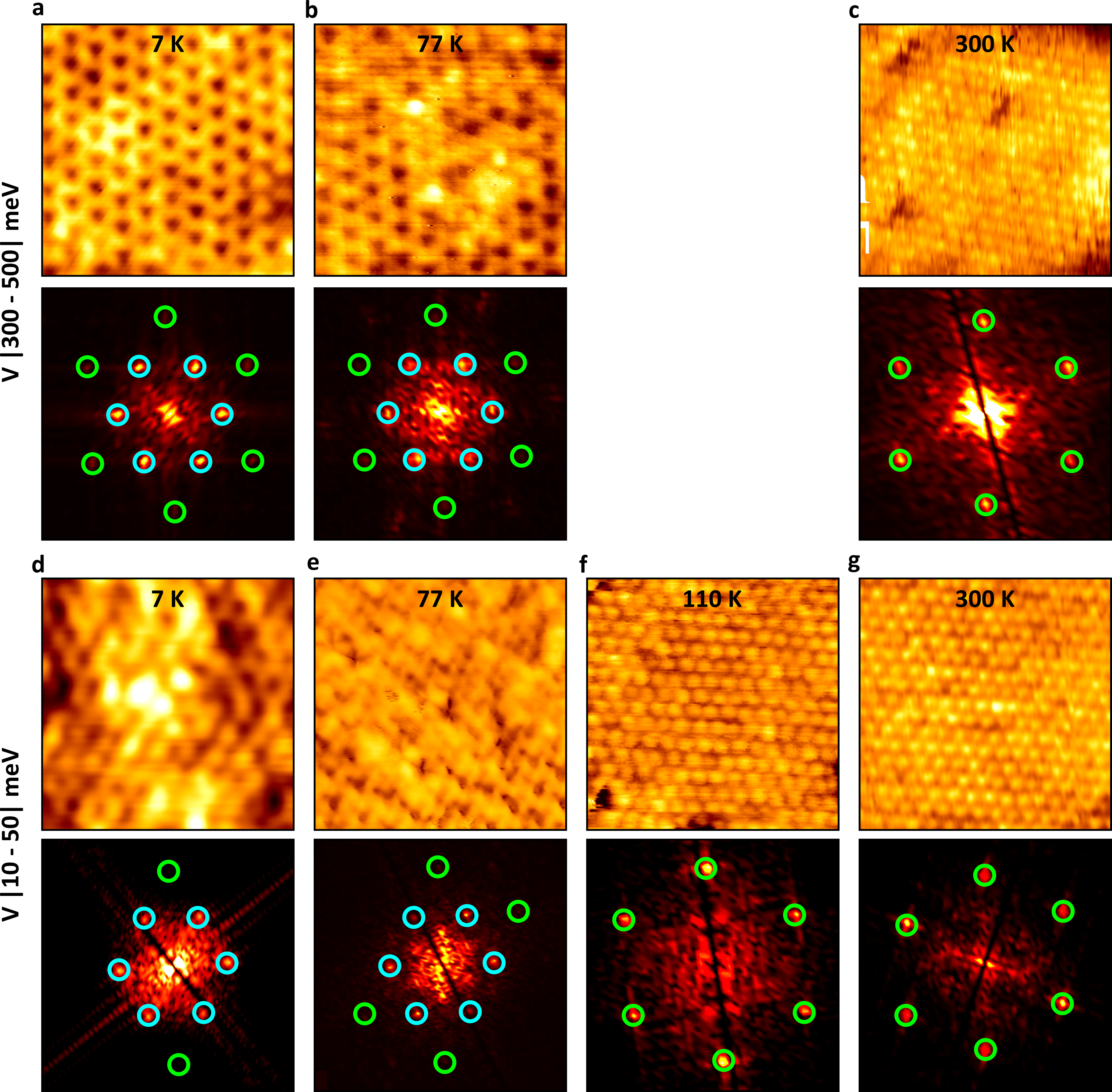}
			\caption{\textbf{Temperature dependence of \ro{} superstructure}~All topographs and FFTs used for graph in the top panel Fig. 6a of the main text. In the FFTs the Bragg peaks (green circles) and the \ro{} spots (cyan circles) are indicated. The FFTs are rotated for the purpose of comparison.
All topographs \SI{5 \times 5}{\nano \meter^2}.\\
STM parameters: \textbf{a}~\SI{-0.5}{\V}, \SI{50}{\pico \ampere}; \textbf{b}~\SI{-0.5}{\V}, \SI{200}{\pico \ampere} ;\textbf{c}~\SI{-0.3}{\V}, \SI{4.0}{\nano \ampere}; \textbf{d}~\SI{-0.05}{\V}, \SI{100}{\pico \ampere}; \textbf{e}~\SI{-0.01}{\V}, \SI{200}{\pico \ampere}; \textbf{f}~\SI{0.014}{\V}, \SI{500}{\pico \ampere}; \textbf{g}~\SI{0.01}{\V}, \SI{3.0}{\nano \ampere}.
}
			
 \label{fig:SFigFFTs}
\end{figure*}

\newpage
\section*{Supplementary Note 15: Location spectra and maps of multiheight sample}

\begin{figure*}[h!]
	\centering
		\includegraphics[width=0.9\textwidth]{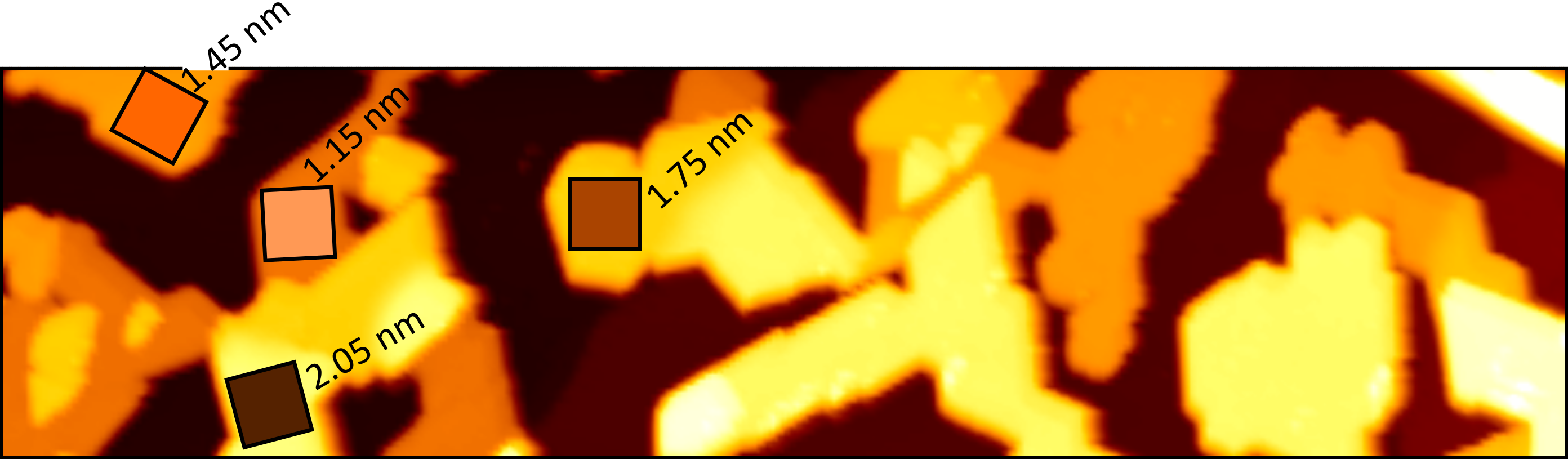}
			\caption{\textbf{Location of spectra and maps for multiheight sample.}~STM topograph of high-coverage sample with different islands heights. Indicated are the locations of the maps and spectra shown in Fig. 10 of the main text. All maps and spectra are taken with the same tip.
		Image taken at \SI{7}{\K}. STM parameters: \textbf{a}~\SI{200 \times 50}{\nano \meter^2}, \SI{-1.0}{\V}, \SI{100}{\pico \ampere}.
}
			
 \label{fig:SFigSpecLoc}
\end{figure*}



	\bibliography{./library}